\documentclass[twocolumn]{aastex62}

\graphicspath{{./}{figures/}}

\accepted{: The Astrophysical Journal}

\shorttitle{CGM around GRB Hosts at z = 2-6}
\shortauthors{Gatkine et al.}

\begin{document}

\title{The CGM-GRB Study. I. Uncovering The CircumGalactic Medium around GRB hosts at redshifts 2-6}

\correspondingauthor{Pradip Gatkine}
\email{pgatkine@astro.umd.edu}

\author[0000-0002-1955-2230]{Pradip Gatkine}
\affil{Dept. of Astronomy, University of Maryland, College Park, MD, USA}

\author{Sylvain Veilleux}
\affiliation{Dept. of Astronomy and Joint Space Science Institute, University of Maryland, College Park, MD, USA}
\affiliation{Institute of Astronomy and Kavli Institute for Cosmology, University of Cambridge, Cambridge CB3 0HA, United Kingdom}

\author{Antonino Cucchiara}
\affiliation{College of Science and Mathematics, University of the Virgin Islands, USVI, USA}



\begin{abstract}
Recent studies have revealed a dynamic interplay between the galaxy ecosystem and circumgalactic medium (CGM). We investigate the CGM at high redshifts ($z\gtrsim2$) by using bright afterglows of gamma-ray bursts (GRB) as background sources. We compiled a sample of medium-resolution ($\Delta v$ $<$ 50 km/s) and high signal-to-noise (typical SNR $\sim$ 10) spectra from 27 GRB afterglows covering $z\sim2-6$, with six of them at z $\gtrsim$ 4. 
We analyzed the high- and low-ionization absorption features within $\pm$400 km/s to extract the CGM and ISM kinematics. 
In the stacked spectra, high-ionization absorption profiles (e.g. C IV, Si IV) present significant absorption excess in their blue wings (v $<$ $-100$ km s$^{-1}$) relative to the red wings (v $>$ 100 km s$^{-1}$). The stronger blue wings in high-ionization species are indicative of the presence of ubiquitous warm outflows in the GRB hosts at high redshifts. We used simple toy models to kinematically distinguish the CGM and ISM absorption and estimate the CGM mass and outflow velocity. 
We find a tentative evidence of the evolution of the CGM metal mass by $\sim$0.5 dex between two redshift bins, each spanning 1 Gyr, $z1: 2-2.7$ and $z2: 2.7-5$. By comparing with past studies, we find that over the course of evolution of present-day galaxies with $M_{*} > 10^{10} M_{\odot}$, the ratio of C IV mass in the CGM to the stellar mass remains fairly uniform, with log$(M_{C IV}/M_{*}) \sim -4.5$ within $\pm$0.5 dex from $z\sim4$ to $z\sim0$, suggesting CGM-galaxy co-evolution.
   
\end{abstract}



\section{Introduction} \label{sec:intro}
The circum-galactic medium (CGM) is loosely defined as the multiphase material surrounding galaxies out to the virial radius \citep[typically spanning 10 to 300 kpc, depending on the mass and redshift of the galaxy;][]{tumlinson2017circumgalactic}. The CGM resides at the interface between the interstellar medium (ISM) of the galaxy and the intergalactic medium (IGM), and thus harbors galactic outflows, accretion flows, and recycling flows. The gas inflows fuel star formation while stellar winds and supernova explosions inject energy and metal-enriched matter at large distances into the ISM and CGM. Studying the CGM and its evolution in the early universe is key to understanding the feedback mechanisms in galaxies. The synergy between these processes is thought to shape galaxies and drive their evolution over cosmic timescales \citep{tumlinson2011large, schaye2014eagle, hopkins2014galaxies, voit2015regulation, angles2017cosmic, Nelson:2019jkf}. 

The history and mechanisms of metal enrichment of the universe remain poorly understood, primarily due to limited probes of galaxies, CGM, and IGM, at high redshifts ($z \ga 2$). The metal content of the
universe as a function of cosmic time can be estimated, given the cosmic star formation history and the models of stellar nucleosynthesis. Even with liberal estimates, current observations have only accounted for $\sim$50-70\% of the metals created in stellar processes \citep{campana2015missing, bouche2006missing}. As an example, the recent COS-Halos studies have inferred that at $z \sim 0$, only $\sim$20-25\% of the metals produced by the stars remain in the galaxy (ISM, stars, and dust), while $\sim$40\% of the metals reside in the CGM \citep{peeples2014budget}. 

At higher redshifts, the distribution of metals among the galaxy (ISM, stars, and dust), CGM, and IGM is even more uncertain due to limited observations. Simulations and observations at $z > 2$ suggest that the CGM could account for $\sim 30\%$ of the cosmic metal budget at that epoch (\citealt{schaye2014eagle, lehner2014galactic}). The transport of metals from their formation sites (i.e. galaxies) to the CGM and IGM is driven by galactic-scale outflows. The distribution of metals and baryons in the galaxy ecosystem provides critical constraints for galaxy evolution models and mechanisms of gas and metal transport \citep{rahmati2016cosmic, muratov2017metal}. Therefore, probing the CGM at high redshifts is essential to develop and test theories of galaxy evolution and metal enrichment. 

\subsection{Methods to Probe the CGM}
Various methods have been employed in past and ongoing observations to extract the diagnostic features of the multi-phased CGM. The most popular technique involves using a bright background quasar (QSO) to trace the CGM around an intervening galaxy. Various local CGM surveys ($z<0.5$) including COS-Halos \citep{tumlinson2013cos, werk2016cos, prochaska2017cos}, COS-Dwarfs \citep{bordoloi2014cos}, COS-GASS \citep{borthakur2015connection}, and others \citep{stocke2013characterizing, zhu2013calcium} utilize UV/optical absorption spectra to study CGM kinematics and physical properties through high ionization potential species (high-ion) such as O VI, N V, C IV, Si IV and low ionization potential species (low-ion) such as Fe II, Si II, C II, Ca II in the CGM. These observations allow matching the absorbers to their respective impact parameters from the galaxy with a precision of tens of kpc. The higher redshift surveys of QSO-galaxy pairings (eg: \citealt{fox2007hot, lehner2014galactic, turner2014metal, rudie2019column}), QSO-QSO pairings (QPQ; \citealt{hennawi2006quasars, prochaska2014quasars, lan2018circumgalactic}), and galaxy-galaxy pairings \citep{steidel2010structure, lopez2018clumpy} use rest-frame UV spectra for similar analysis with limited information about the associated galaxies and/or impact parameters.

In ``down-the-barrel" spectroscopy, a star-forming galaxy's own starlight is used as a background illumination to detect absorption from the intervening ISM and CGM \citep{martin2005mapping,steidel2010structure,bordoloi2011radial, rubin2012direct,kornei2012properties, heckman2015systematic, rubin2014evidence, rigby2018magellan}. This technique has been successful in tracing the galactic inflows and outflows in the CGM of star-forming galaxies with a caveat that the radial coordinate of the absorbing component remains unconstrained. 

\begin{figure}
\centering
\includegraphics[width=0.3\textwidth]{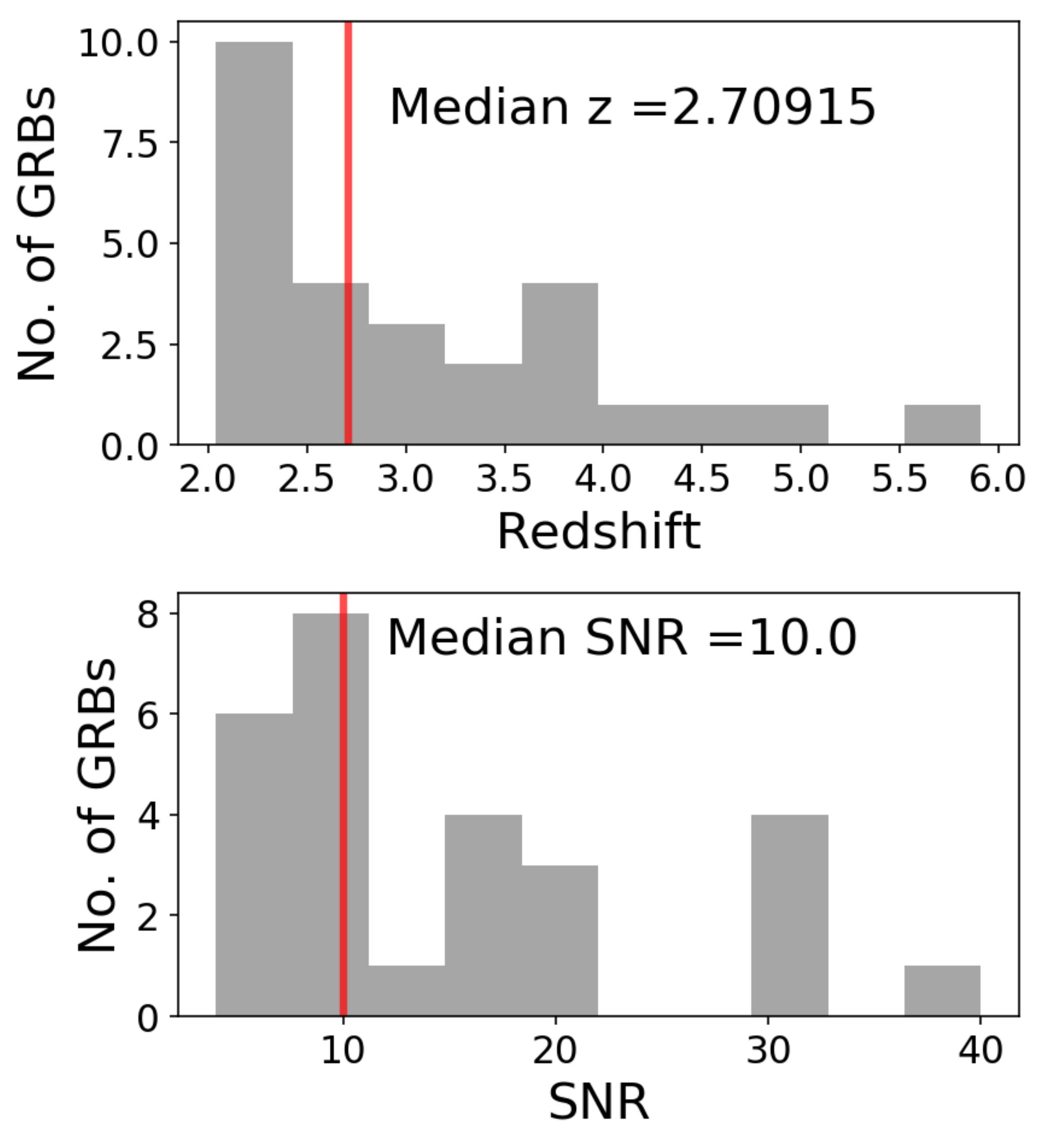}
\figcaption{\label{fig:Sample_properties} Upper panel: Distribution of redshifts in the GRB sample. The median redshift is 2.71. Lower panel: Distribution of the signal-to-noise ratios of the spectra used for this study. The median ratio is 10.}
\end{figure}

\subsection{The CGM-GRB Project}
 Our approach utilizes the spectra of bright afterglows of long Gamma-ray Bursts (GRBs) to  derive the kinematic properties of the CGM around their host galaxies and use a simple toy model to further constrain the physical properties of the CGM gas. This technique will further enable the investigation of possible relations between the CGM and the galaxy properties that may govern feedback processes and their evolution with redshift. Thanks to the nature of GRBs and the extensive follow-up effort over the past ten years, we were able to collect enough data to study the CGM of GRB hosts in the redshift range $2 \lesssim z \lesssim 6$.

Long-duration GRBs are the most powerful explosions in the universe, being several orders of magnitude brighter than typical supernovae. The prompt emission is followed by a rapidly fading ($\sim$1$-$2 days) X-ray, UV, and optical afterglow. GRB afterglows have been detected from low redshift $\sim$0.01 out to a redshift of 8.2 \citep{tanvir2009gamma,salvaterra2009grb, cucchiara2011photometric}, thus probing the first few billion years of the Universe, an era characterized by the formation and early evolution of galaxies that may have had a critical role in enriching the universe with metals.

Since GRB afterglows are bright background sources, their spectra harbor absorption features produced by the material along the line-of-sight including the host galaxy ISM as well as the CGM and intergalactic medium \citep{prochaska2007interstellar, fox2008high, prochaska2008resolving,  cucchiara2015unveiling}. Thus it provides an excellent opportunity to probe the chemical composition and physical conditions of the entire galaxy ecosystem. The GRBs also fade rapidly, making it possible to study the host galaxy component separately, in the absence of the bright GRB. 

Compared to QSO sightlines, GRB sightlines have key advantages: 1) GRBs happen within their host galaxy, thus probing the host galaxy ISM as well as the associated CGM, the main components of the galaxy ecosystem; 2) GRB discovery is based on gamma- and X-ray detection and hence, is largely independent with respect to host galaxy properties (further discussed in \S \ref{subsec:sample_prop}); 3) The optical afterglow fades in 1-2 days, clearing the path for future deep observations of the host galaxy to study its global properties and surroundings.

However, at the same time, the number of GRBs suitable for CGM investigation is small, despite the detection rate of 100 yr$^{-1}$ by dedicated space-based mission like the Neil Gehrels Swift Observatory (\emph{Swift}). Similarly, the fast-decaying nature of their afterglows requires opportune follow-up strategies in order to obtain high-SNR spectra. And finally, it requires separating the CGM and ISM contributions in the absorption spectra, which can be challenging. Our approach tackles this problem by using the kinematic information derived by absorption line spectroscopic data.

In this paper, we present the analysis of a dataset of 27 high-z (z $\gtrsim$ 2) GRBs, out of which 6 are at z $\gtrsim$ 4 (\S \ref{sec:data}, \S\ref{sec:methods}). We use column density line profiles to study the kinematics and line ratios of the absorbing gas to distinguish between the CGM and ISM (\S \ref{sec:infer_kinematics}). We designed a simple toy model (\S \ref{sec:toy_model}) to obtain an estimate of the outflow properties in the CGM. We estimate the CGM mass for a typical GRB host in this sample and summarize its possible evolution with redshift in \S\ref{sec:CGM_mass}. In \S\ref{sec:implications}, we discuss various implications our findings may have on our current understanding of the CGM kinematics, outflow rates, metal enrichment, and CGM-galaxy co-evolution. The key conclusions of this study are summarized in \S 8.

\section{The GRB Sample} \label{sec:data}
\subsection{Sample Properties}
Obtaining a good estimate of ionic column density as a function of velocity requires high signal-to-noise spectra (SNR $\gtrsim 5$) of medium resolution ($R\gtrsim 8000$, $\sim$ 50 km s$^{-1}$). The cut-off in resolution is selected to clearly distinguish the absorption systems at different velocities and minimize the errors in column density estimation due to saturation effects and blending (see \citealt{prochaska2006dissecting, cucchiara2013independent}). Our CGM-GRB sample consists of 27 long GRBs that satisfy these stringent criteria. 
The sample properties are summarized in Fig. \ref{fig:Sample_properties}. The median redshift of the sample is 2.71 and median SNR is 10. Notably, six of the GRBs are at z $>$ 4. Table \ref{tab:GRB_list} lists the GRBs in this sample along with the observational details.

As mentioned earlier, the transient nature of GRBs requires rapid-response facilities capable of observing their afterglow within a few minutes of their discovery. 
Our dataset comprises primarily of archival data acquired by the X-Shooter and UVES spectrographs on the Very Large Telescope (VLT). These spectra provide a wide wavelength coverage (from optical to NIR) and sufficiently high spectral resolution ($R\sim8,000-55,000$). In addition, we retrieve spectra from the archival dataset of the Keck telescope's HIRES and ESI spectrographs. 

The majority of these spectra were re-analyzed and normalized using the data analysis pipelines in \cite{cucchiara2013independent, cucchiara2015unveiling}. More recent data (from 2014) were acquired from the PHASE 3 VLT archive\footnote{\url{https://www.eso.org/sci/observing/phase3.html}}$^{,}$\footnote{\url{http://archive.eso.org/wdb/wdb/adp/phase3\_main/form}},
which provides fully reduced, research-ready one- and two-dimensional spectra. We utilized the flux-calibtrated one-dimensional spectra and normalized the GRB afterglow continuum using a spline function. Every spectrum is manually inspected and the overall continuum is determined using the python-based {\tt linetool} package \footnote{\url{https://github.com/linetools/linetools}}. The error from the continuum fit is propagated into the flux error spectrum.  

\subsection{Selection Effects} \label{subsec:sample_prop}
\noindent
Due to the high SNR and resolution requirement, this sample is biased towards the brighter end of GRB afterglows distribution. The afterglow magnitudes at the time of taking the spectra are listed in Table \ref{tab:GRB_list}, clearly indicating a limit at $m_{AB}$ $\sim$ 21.0 (with an exception of GRB 100219). This selection effect is complex since the magnitude at the time of observation depends on the intrinsic brightness and distance of the afterglow as well as the time elapsed since the prompt gamma-ray emission. Regardless, it can be said that this sample selectively avoids intrinsically faint afterglows. However, considering that the apparent brightness of the afterglow also depends on the host galaxy dust extinction, it can be said that this sample selectively avoids heavily dust-obscured ($A_{V}$ $>$ 0.5) sightlines \citep{perley2009host, kruhler2011seds, zafar2018x}.  

In general, long GRBs trace cosmic star formation \citep{greiner2015gamma, schady2017gamma}. At z $\sim$ 2.5, the typical star formation rate of GRB hosts is $\sim$10 $M_{\odot}$ yr$^{-1}$ \citep{kruhler2015grb}, the typical GRB host stellar mass is $\sim$ $10^{9.3}$ $M_{\odot}$ \citep{perley2016swift_2} and the typical gas phase metallicity is $\sim$ $0.05-0.5$ solar \citep{trenti2015luminosity, arabsalmani2017mass}. Thus, from a CGM perspective, this sample traces star-forming, low-mass galaxies at $z$ $>$ 2.   







\section{Methods} \label{sec:methods}

\subsection{Redshift Determination} \label{sec:redshift_det}
\noindent
In order to infer the kinematics of different chemical species, the redshifts of the GRB host galaxies in the sample need to be determined in a precise and uniform manner. Commonly, the galaxy redshifts are obtained using nebular emission lines (e.g. H$\alpha$) but this method is not viable for faint GRB hosts at high redshifts. Therefore, we use the local GRB environment (within few $\times$ 100 pc of the GRB) as a proxy for the systemic redshift of the GRB host galaxy. 

The fine structure transitions of the species such as such as Ni II*, Fe II*, Si II*, and C II* in the rest-frame UV \citep{bahcall1968fine} trace the ISM clouds in the vicinity of the GRB ($\sim$ few $\times$ 100 pc $-$ 1 kpc) due to UV pumping. This is further corroborated by temporal variations found in the strength of these lines in multi-epoch spectra of a few GRBs \citep{vreeswijk2007rapid, hartoog2013host, d2014vlt}. The strongest absorption components (i.e. velocity components) of these fine-structure transitions are therefore good proxies for the rest-frame velocity of the burst environment within a few hundred parsecs \citep{chen2005echelle, dessauges2006temporal, prochaska2006dissecting}. Therefore, we choose the redshift by visual inspection such that the strongest absorption components in the fine-structure transitions occur at $\sim$ 0 km s$^{-1}$, i.e. rest frame. We primarily use Si II* and C II* transitions for estimating the redshift. In case of saturation or confusion between Si II* and C II*, we use Ni II* lines due to their lower oscillator strength.   

In addition, we visually check for the presence of other low-ionization lines such as Fe II 1608, Si II 1526, and Al II 1670 (especially the weak transitions of Si II and Zn II) which are reliable tracers of the host galaxy ISM \citep{prochaska2006dissecting, cucchiara2013independent}, to visually confirm the redshift determination. With the use of the strongest components of fine-structure transitions to estimate the zero point, the GRB redshift is accurate to within 50 km s$^{-1}$.

\subsection{Spectroscopic Analysis} \label{sec:spec_analysis}
\noindent 
GRB spectra show a plethora of signatures, ranging from the circumburst and interstellar media to the galactic winds and circumgalactic medium. Prior studies have extracted the intervening systems along the GRB sightlines (at z $<$ z$_{GRB}$) to trace cosmic chemical evolution. These studies have used the doublets from Mg II, C IV, and Si IV, to determine the chemical enrichment of the universe similar to the analysis of quasars intervening systems \citep{prochaska2008resolving, fox2008high, fynbo2009low, simcoe2011constraints, thone2012grb, sparre2014metallicity, cucchiara2015unveiling, vergani2017chemical}. In this paper, we focus on the contrast between the high and low ion kinematics. The GRB spectra were normalized, binned, and fitted to extract the column density, Doppler parameter, and the line center (in velocity space) for each absorbing component within a velocity window of $\pm 400$ km s$^{-1}$ for key high- and low-ion species. The parameters of these species were then used to study the kinematics of the absorbing gas and estimate the likely origin of the absorbing component(s).  


\noindent 
Filtering: Various intervening systems previously reported in the literature (references in Table \ref{tab:GRB_list}) were identified and the regions where the absorption lines from the intervening system blend with rest-frame GRB absorption were flagged. In addition, the regions containing telluric absorption from the atmosphere were flagged. For lines with strong neighboring transitions (eg: Si II 1260 and S II 1259, etc), the velocity windows considered for fitting were carefully adjusted to minimize the confusion. 

\noindent 

\begin{figure}[!htb]
\centering
\includegraphics[width=0.45\textwidth]{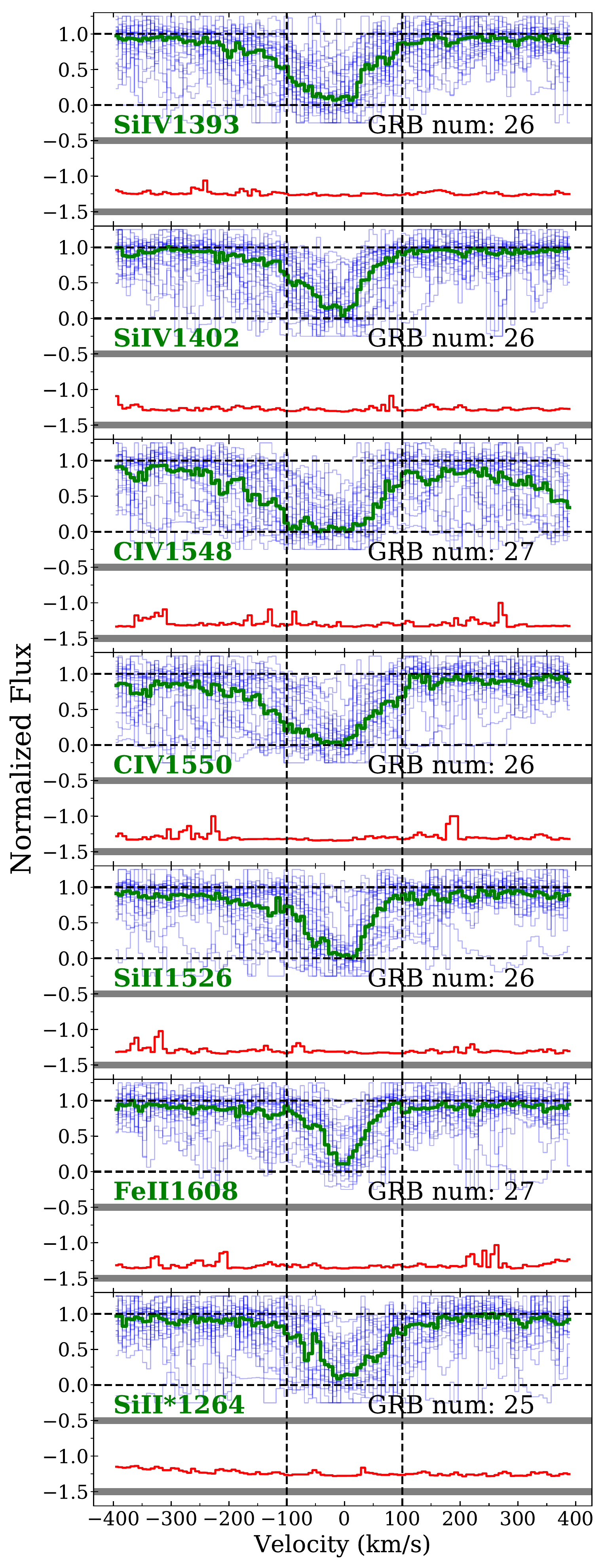}
\figcaption{\label{fig:Median_plot} The median  absorption line profile (in green) of the individual spectra (in blue) in the CGM-GRB sample. The red line shows the rms noise (shifted vertically by $-$1.5).}
\end{figure}

\noindent 
Voigt profile fitting: Thanks to the medium resolution spectra in this sample, it is possible to resolve the kinematics of  absorbing systems residing in the ISM and CGM into individual components in the velocity space. The GRB spectra were analyzed by fitting individual components, where the optical depth of each component is modeled as a Voigt profile.
Given the complex nature of GRB sightlines, typically more than 5 absorption components are observed in these spectra. While there are other $\chi^{2}$ grid-search-based codes (eg: VPFIT\footnote{Available at \url{https://www.ast.cam.ac.uk/~rfc/vpfit.html}} and MPFIT \citep{markwardt2009non}) for fitting the Voigt profiles, there are fundamental limitations of these methods for solving this particular problem. Obtaining useful results with a grid-search in such a large parameter space of a non-linear model is computationally expensive. Further, it is difficult to capture the degeneracy between various parameters in a quantitative manner, for instance, the degeneracy between Doppler parameter and column density for saturated components. The Bayesian approach provides a rigorous way to visualize the degeneracy and estimate the errors around the fit parameters in a systematic way for multi-component Voigt profiles.

For these reasons, we developed a Bayesian-inference-based, multi-component Voigt profile fitting code in Python to determine the best-fit values of the parameters for each component($i$), i.e. the column density ($N_{i}$), Doppler parameter ($b_{i}$), and line center ($v_{i}$). To sample the posterior probability distributions, the Markov Chain Monte Carlo (MCMC) method was used. The MCMC sampling was implemented using {\tt emcee} library in Python \citep{foreman2013emcee}. A detailed summary of the Bayesian Voigt-profile fitting method and error estimation is given in Appendix \ref{sec: Voigt_MCMC}. This code provides a robust and novel approach to fit the complex multi-component absorption systems such as GRB or QSO sightlines and obtain reliable estimates of the optimal parameters and the associated errors. 

\begin{figure}
\centering
\includegraphics[width=0.45\textwidth]{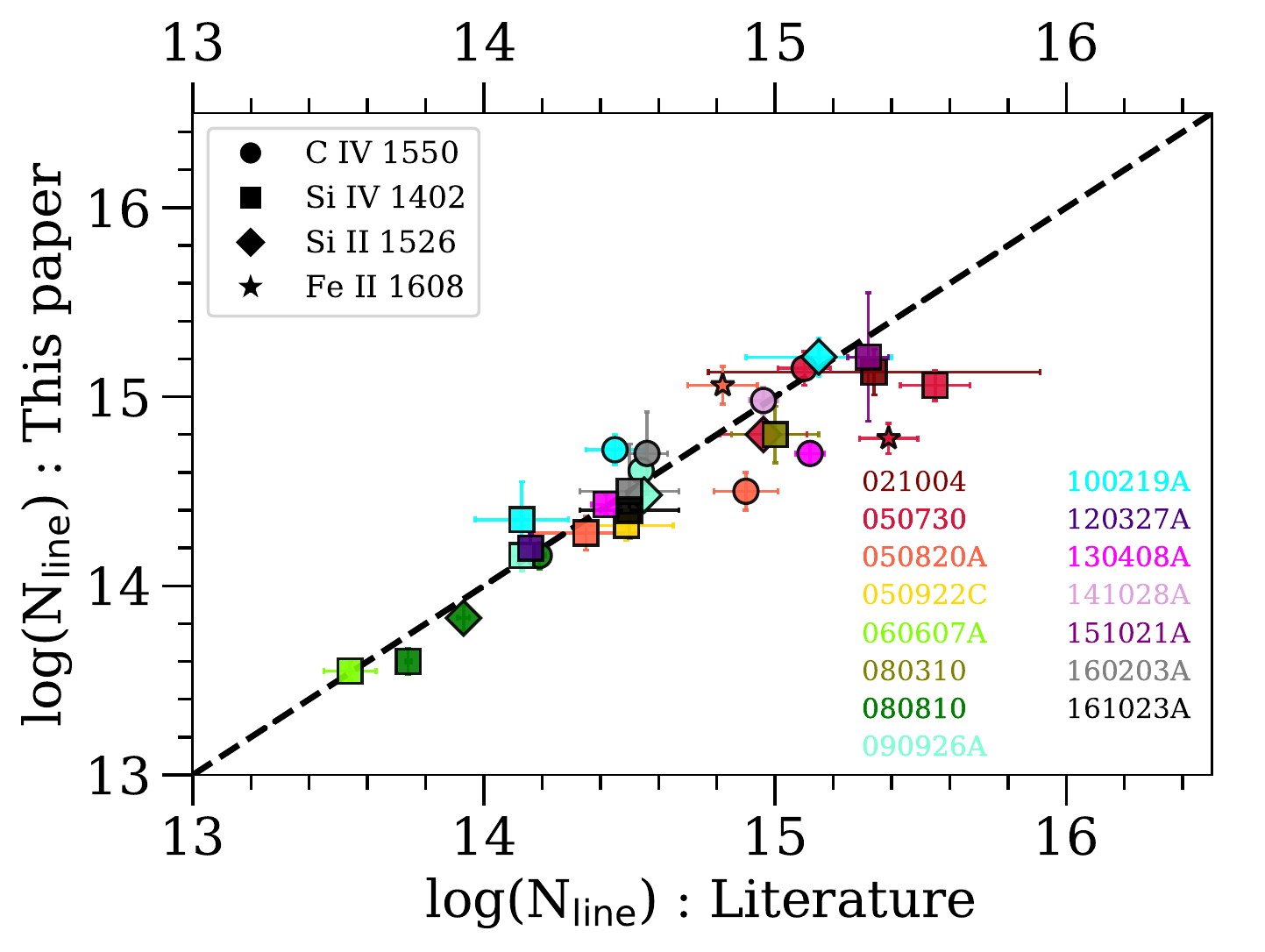}
\figcaption{\label{fig:Col_density_comparison} Comparison of total column densities partially compiled from the literature (references in Table \ref{tab:GRB_list}) with the total column densities derived in this paper. The dotted line represents the line of equality.}
\end{figure}

To fit multi-component Voigt profiles to a given transition, a velocity window spanning $\pm$400 km s$^{-1}$ around the GRB rest-frame is extracted. This velocity window enables fair comparisons with the simulations of CGM as well as previous observations of high- and low-redshift CGM. The spectra are binned by a factor of 2 to 4 depending on the noise level for easier visual inspection. The number of components to be fitted is determined through manual inspection of doublets (eg: C IV 1548 and 1550) and lines with similar ionization potential (eg: C IV, Si IV). The line spread function of the spectrometer is modelled as a Gaussian function and convolved with the synthesized multi-component Voigt profile to obtain the comparison spectrum for evaluating the residuals. The initial guesses of the parameters are manually provided to the MCMC routine (as priors) to find the optimal line-parameters and associated uncertainties corresponding to all the components ($N_{i}$, $b_{i}$, $v_{i}$). Also, doublets are fitted simultaneously. Our optimal parameters are consistent with the results from other references in Table \ref{tab:GRB_list} as shown in Fig. \ref{fig:Col_density_comparison}. The Voigt-profile fits are shown in figures \ref{fig:000926}$-$\ref{fig:170202A} and the line profile parameters are listed in Table \ref{tab:fit_profile}.  

\noindent
Caveats: The kinematic resolution of the analysis is linked to the spectral resolution of each observation, leading to a variance in the precision of the parameter estimation throughout the sample due to different spectral resolutions. For warm CGM (T $\sim$ 10$^{5-6}$ K), the thermal Doppler parameter is expected to be in the range $5-20$ km s$^{-1}$ for C and Si with further broadening expected due to turbulence \cite{lehner2014galactic}. For spectra with R $\sim$ 10000, we can marginally resolve b $\sim$ 15 km s$^{-1}$. In saturated regions, the determination of optimal parameters has a higher uncertainty due to degeneracy between the Doppler parameter and column density. In such cases, lines with similar ionization potential and weaker oscillator strength help provide an estimate without breaking the degeneracy. While the Voigt profile fitting works well in case of mildly saturated lines, it does not fully alleviate the uncertainty for strongly saturated lines (such as saturation spanning $\sim$ 100 km s$^{-1}$). In Table \ref{tab:fit_profile}, a quality flag of 0 indicates unsaturated or mildly saturated components while a quality flag of 1 indicates strongly saturated components. The tabulated profile parameters for strongly saturated components denote a likely but non-unique solution. 

\section{Inferring Kinematics}\label{sec:infer_kinematics}

\subsection{Median Plots}
In order to understand the overall kinematics of the sample, the normalized rest-frame spectra were plotted in velocity space. In Fig. \ref{fig:Median_plot}, various high- and low-ion transitions of the individual GRBs are shown in blue, while the median kinematic profiles are shown in green. The key qualitative results from the median plots are:\\ 
a) There is a significant blueward absorption excess at velocities $v \lesssim -100$ km s$^{-1}$, which is a clear signature of outflowing gas.\\
b) This blueward asymmetry is stronger in the high-ion transitions than in the low-ion transitions. The median profile for the fine structure transition of Si II* is, not surprisingly, more symmetric. \\
c) The median absorption for the low-ion transitions is fairly limited to within $\pm100$ km s$^{-1}$ unlike the high-ion lines, which extend much further (especially blueward).

These qualitative observations may indicate outflowing gas predominantly traced by the high-ion transitions. Thus, we expect at least two different phases that are kinematically distinct. In order to test this hypothesis in greater details we use a toy model to reproduce the observed kinematic behavior (Section $\ref{sec:toy_model}$).  
 

\subsection{Integrated Line Profiles} \label{subsec:integrated_profiles}
The absorption lines in the afterglow spectra from various high- and low-ion species were fitted with multi-component Voigt profiles as described in Section \ref{sec:spec_analysis}. In order to quantitatively measure and compare the kinematics of high- and low-ion species as well as compare the observations with models (as described in Section \ref{sec:toy_model}), the fitted continuous line profiles were converted into integrated column density profiles. The fitted Voigt profiles profiles (with $N_{i}$, $b_{i}$, and $v_{i}$) were converted to apparent optical depth profiles as a function of velocity ($\tau_{a}(v)$) such that $\tau_{a}(v)$ = $ln[F_{c}(v)/F(v)]$, where $F_{c}(v)$ is the normalized continuum flux level (i.e. 1) and $F(v)$ is the normalized flux from the fitted profile at velocity $v$ \citep{savage1991analysis}. The apparent column density is then evaluated as $N_{a}(v)$ $=$ $3.768 \times 10^{14} \tau_{a}(v)/f\lambda$, where $f$ is the oscillator strength and $\lambda$ is the rest-frame wavelength of the line in \AA. The integrated column density is the integral of $N_{a}(v)$ over bins of 100 km s$^{-1}$. 

\begin{figure}
\centering
\includegraphics[width=0.45\textwidth]{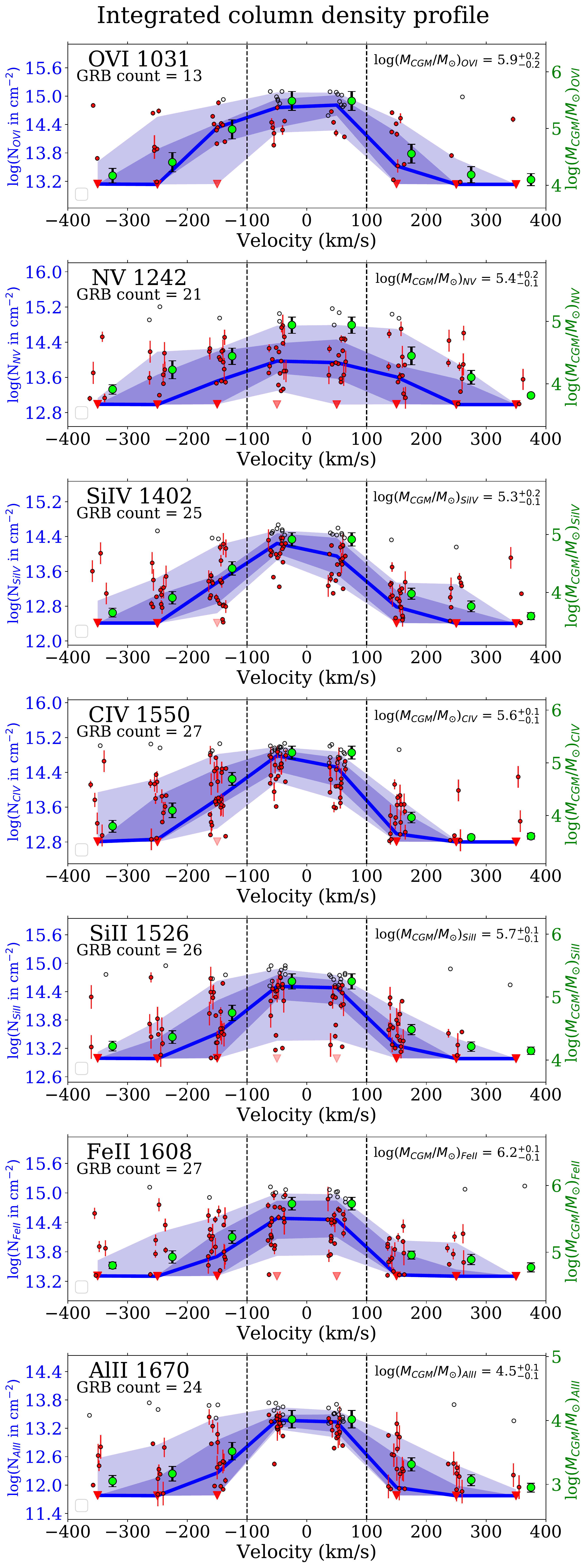}
\figcaption{\label{fig:CGM_mass} The integrated column density profiles for the high- and low-ion species of the CGM-GRB sample. The blue line and labels refer to the median values while the central 50 and 80 percentiles are shown in darker and lighter shades. The green line and labels indicate the estimated mass of the species in the CGM.}
\end{figure}

\begin{figure}
\centering
\includegraphics[width=0.45\textwidth]{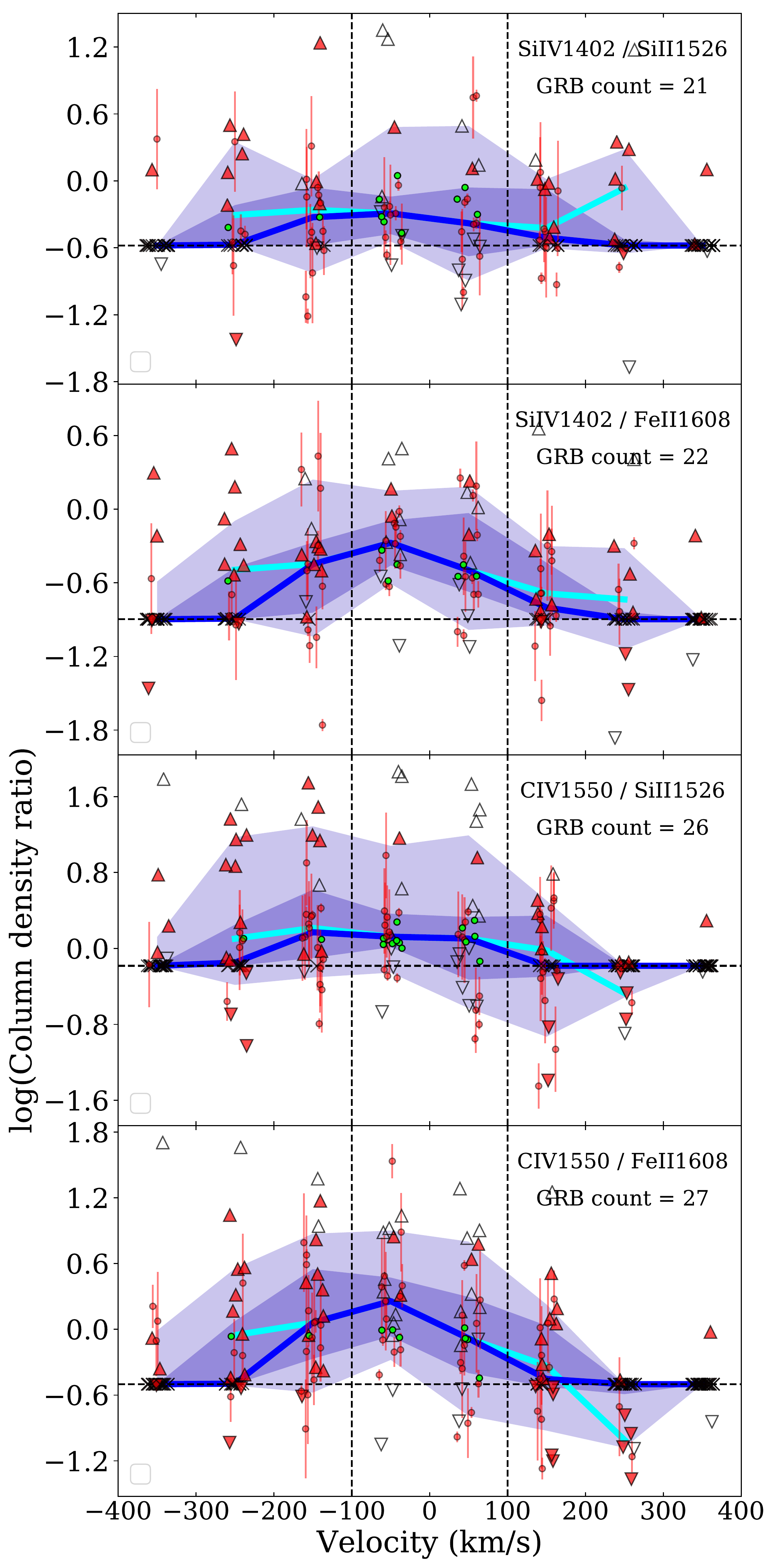}
\figcaption{\label{fig:Line_ratios}  
The high- to low-ion line ratios as a function of velocity. The line ratios are evaluated as the ratio of integrated column densities for the two lines as a function of velocity. The solid red symbols indicate that at least one of the two absorption lines is detected in that velocity bin. The empty symbols indicate that one of the two lines is saturated, while the green circles show the cases where both the lines are saturated in that velocity bin (double saturation points). A circle denotes detection of both lines, an upwards (downwards) triangle denotes lower (upper) limits. A cross symbol indicates points where both the lines have non-detections (double non-detection points). The blue line traces the median of the line ratios including double non-detection points at each velocity bin with the purple shades spanning the 50- and 75-percentile zones. The cyan line traces the median excluding the double non-detection points (i.e. requiring the detection of at least one of the two lines in the given velocity bin).
}
\end{figure}

The integrated column density profiles are shown in blue in Figures \ref{fig:CGM_mass}, \ref{fig:CGM_Mass_z1}, and \ref{fig:CGM_Mass_z2}. The strongly saturated components are treated as lower limits evaluated by imposing a maximum cap on $\tau_{a}(v)$ of 4.5 (equivalent to a lower cap of $\sim$0.01 on the normalized flux) and they are marked as open circles. The saturation issue does not significantly affect the velocity bins beyond $\pm$ 100 km s$^{-1}$. The downward triangles show integrated column density for a bin with no detected absorption and is evaluated using $\tau(v)$ $=$ 0.05 which denotes the typical detection limit of the sample. The error bars on the integrated column density are evaluated by calculating the difference between the Voigt profiles generated using the optimal parameters and the profiles generated using the 1-$\sigma$ deviation parameters (as shown in Fig. \ref{fig:MCMC_sample}). This captures both the fitting uncertainty as well as the noise spectrum. The median integrated column density profile for each line is shown in blue. The inner and outer purple-shaded regions show the central 50 and 80 percentiles of the integrated column density distribution for each bin, respectively. 



\subsection{Detection Fractions}
The integrated column density profiles help visualize the fraction of sightlines where various species are detected as a function of velocity. Broadly speaking, the detection fractions can be categorized in three kinematic regions: central ($|v| <$ 100 km s$^{-1}$), blue wing ($v <$ $-$100 km s$^{-1}$), and red wing ($v >$ $+$100 km s$^{-1}$). The detection fractions of various high- and low-ion species are calculated as the number of detected sightlines divided by the number of sightlines with the spectral coverage for that ion and are reported in Table \ref{tab:Det_fractions}. The O VI and N V absorption lines are redshifted to the low-sensitivity (blue) regions of the spectrographs for $z$ $\sim$ 2, which leads to a poor signal-to-noise ratio. In addition, the vicinity to the Lyman-alpha forest and possible doublets contamination can prevent reliable  detection of these species. Therefore, we only consider the detections where both doublets show absorption lines that can be reasonably fitted. The uncertainty on the detection fractions for these ions reflects the small number of cases where these lines are within the spectral coverage and are not affected by low SNR, saturation, or mismatch between the doublets. 

The central region within $\pm$100 km s$^{-1}$ shows the highest detection fraction for all the ions. The blue wings have a higher detection fraction compared to the red wings for all the ions except N V, as expected from the stacked spectra (Fig. $\ref{fig:Median_plot}$). We compare the detection fractions in this sample with KODIAQ sample $-$ a QSO-based survey of CGM absorbers at $z \sim 2 - 3.5$ \citep{lehner2014galactic}. The overall detection fractions in the KODIAQ survey for O VI (75\%), N V (55 $\pm$ 20\%), Si IV (95\%), C IV (90\%), and Fe II (78\%) are consistent with the detection fractions in blue wing region of the CGM-GRB sample within $\sim 10\%$.   


\begin{deluxetable}{cccc}
\tablecaption{Detection fractions in various kinematic regions\label{tab:Det_fractions}}
\tablehead{\colhead{Species } & \colhead{Blue wing } & \colhead{Central} & \colhead{Red wing} \\ 
\colhead{} & \colhead{(v $<$ -100 km s$^{-1}$)} & \colhead{($|v|$ $<$ 100 km s$^{-1}$)} & \colhead{(v $>$ 100 km s$^{-1}$)} } 

\startdata
O VI &   $0.65^{+0.2}_{-0.0}$ &   $0.8^{+0.2}_{-0.0}$ &   $0.6^{+0.2}_{-0.0}$ \\
\rule{0pt}{3ex}
N V &    $0.52^{+0.16}_{-0.0}$ &   $0.76^{+0.16}_{-0.0}$ &   $0.56^{+0.16}_{-0.0}$ \\
\rule{0pt}{3ex}
Si IV &  $0.92^{+0.04}_{-0.0}$ &   $0.96^{+0.04}_{-0.0}$ &   $0.69^{+0.04}_{-0.0}$ \\
\rule{0pt}{3ex}
C IV &   0.96 &   1.00 &   0.63  \\
Al III &  0.67 &  1.00 &   0.62 \\
Si II &  0.73 &   0.96 &   0.61 \\
Fe II &  0.67 &   0.92 &   0.55 
\enddata
\end{deluxetable}


\subsection{Kinematic Asymmetry} \label{subsec:kinematic_asymmetry}
It is evident from the stacked spectra in Fig. \ref{fig:Median_plot} that there is a clear excess of absorption in the blue wing ($v$ $<$ $-$100 km s$^{-1}$) relative to the red wing ($v$ $>$  100 km s$^{-1}$). The excess also appears to be stronger for the high-ionization species such as Si IV and C IV relative to the low-ionization species such as Si II and Fe II. We quantify this kinematic asymmetry using the median integrated column density profiles shown in Fig. \ref{fig:CGM_mass}. The median of the total integrated column densities are tabulated in three distinct kinematic regions in Table \ref{tab:Asymmetry}: central ($|v|$ $<$ 100 km s$^{-1}$), blue wing ( $v$ $<$ $-$100 km s$^{-1}$), and red wing ($v$ $>$ 100 km s$^{-1}$). The reported uncertainties are evaluated as 2-$\sigma$ intervals of the distribution of medians of total integrated column densities derived via a simple resample-with-replacement bootstrap technique. The degree of asymmetry is calculated as the ratio of blue- to red-wing column densities. The median asymmetry in the high ions C IV and Si IV are 0.36 and 0.52 dex, respectively, in contrast to $\sim$ 0.17 and 0.19 dex for the low ions Si II and Fe II, respectively. 

The blue absorption excess (or asymmetry) has been interpreted as a galactic outflow signature in previous GRB-afterglow sightline studies (eg: with 7 high-z GRBs in \citealt{fox2008high} and GRB 080810 in \citealt{wiseman2017gas}). From the kinematic distribution in the blue wing, the typical outflow velocity is $\sim$ 150 $-$ 250 km s$^{-1}$. Another important aspect is the higher degree of asymmetry in Si IV and C IV relative to Si II and Fe II. These high ionization species are more asymmetric by roughly a factor of $\sim$ 1.5 $-$ 2 than the low ionization species. This key observation implies that not only there is an excess of outflowing gas compared to infalling gas, but also the outflowing gas is more highly ionized than the infalling gas. Unlike all the other ions presented here, N V has no significant asymmetry and therefore, could be tracing a phase (or a combination of phases) that is kinematically distinct from the gas phase (or a combination of phases) traced by the other high- and low-ion lines. We return to this point in section \ref{sec:implications}.




\begin{deluxetable}{ccccc}
\tablecaption{Kinematic asymmetry in high and low ions \label{tab:Asymmetry}}
\tablehead{ \multicolumn{5}{c}{Integrated col. density in $cm^{-2}$} \\ 
\colhead{Species} & \colhead{Blue wing} & \colhead{Central} & \colhead{Red wing} & \colhead{Asymmetry} \\ 
\colhead{ } & \colhead{$log(N_{B})$} & \colhead{$log(N_{C})$} & \colhead{$log(N_{R})$} & \colhead{(dex)} } 

\startdata
O VI &    $14.39^{+0.15}_{-0.22}$ &   $14.98^{+0.08}_{-0.19}$ &   $13.80^{+0.13}_{-0.17}$ & 0.59 \\
\rule{0pt}{3ex}
N V &    $13.72^{+0.17}_{-0.25}$ &   $14.36^{+0.18}_{-0.24}$ &   $13.80^{+0.20}_{-0.31}$ & -0.08 \\
\rule{0pt}{3ex}
Si IV &  $13.46^{+0.18}_{-0.22}$ &   $14.49^{+0.15}_{-0.15}$ &   $13.10^{+0.14}_{-0.16}$ & 0.36 \\
\rule{0pt}{3ex}
C IV &  $13.87^{+0.09}_{-0.13}$ &   $14.99^{+0.14}_{-0.11}$ &   $13.35^{+0.07}_{-0.07}$ & 0.52 \\
\rule{0pt}{3ex}
Al III &  $12.48^{+0.14}_{-0.22}$ &   $13.70^{+0.09}_{-0.05}$ &   $12.32^{+0.06}_{-0.06}$ & 0.16 \\
\rule{0pt}{3ex}
Si II &  $13.75^{+0.19}_{-0.25}$ &   $14.82^{+0.10}_{-0.19}$ &   $13.58^{+0.09}_{-0.30}$ & 0.17 \\
\rule{0pt}{3ex}
Fe II &  $13.97^{+0.15}_{-0.17}$ &   $14.89^{+0.17}_{-0.23}$ &   $13.78^{+0.05}_{-0.10}$ & 0.19 
\enddata
\end{deluxetable}

\subsection{Line Ratios}
The ratio of column densities of various high- and low-ion species provides another perspective to learn about the physical conditions of the intervening gas in different kinematic regions. We select Si IV, C IV, Si II, and Fe II for this analysis since they have excellent spectral coverage in the sample and are fit reasonably well due to high signal-to-noise ratio in the corresponding observed wavebands. Si II 1526 and Fe II 1608 are taken as representative low-ion lines since a) they have a moderate line strength, thus preventing saturation, and b) there are no adjacent strong lines in the rest frame that could potentially blend/contaminate the $\pm$ 400 km s$^{-1}$ region. 

Figure \ref{fig:Line_ratios} shows the high- to low-ion line ratios as a function of velocity. The line ratios are evaluated as the ratio of the integrated column densities for the two lines as a function of velocity.  
The solid red symbols indicate that at least one of the two absorption lines is detected in that velocity bin. The empty symbols indicate that one of the two lines is saturated, while the green circles show the cases where both the lines are saturated in that velocity bin (double saturation points). The double saturation ratios are evaluated by taking the ratios of integrated column densities by putting a lower limit on the observed flux (as described in section \ref{subsec:integrated_profiles}). A circle denotes detection of both lines, an upwards (downwards) triangle denotes lower (upper) limits. A cross symbol indicates points where both the lines have non-detections (double non-detection points). In such cases, the ratio is taken as the ratio of their typical detection thresholds evaluated using $\tau(v)$ $=$ 0.05 (see section \ref{subsec:integrated_profiles}). The blue line traces the median of the line ratios including the double non-detection points with the purple shades spanning the 50- and 75-percentile zones. The cyan line traces the median excluding the double non-detection points (i.e. requiring the detection of at least one of the two lines in the given velocity bin).


To avoid large number of double non-detection points, we focus on velocity bins from $-250$ to $150$ km s$^{-1}$, where the double non-detection cases are limited to less than 40\%. In this region, the high- to low-ion ratio is higher in the blue wing relative to the red wing. This is more clearly noticeable in the Si IV / Fe II and C IV / Fe II ratios. The actual ratios in the blue wing are likely to be higher due to a large number of lower limits in the blue wings (i.e. detection of high ions and non-detection of low ions). This can also be seen by comparing the high- and low-ion lines in Table \ref{tab:Asymmetry}. The line ratios in the central region are more uncertain due to high occurrence of double saturation cases, but they appear to be commensurate with the blue wing ratios. Qualitatively, the line ratios highlight the distinct physical characteristics of the three kinematic zones and hint towards a general presence of high-ion rich outflowing gas at a projected speed of $\sim$ 150 $-$ 250 km s$^{-1}$.

\section{Toy Model} \label{sec:toy_model}
\noindent

In order to explain these observations, we simulate the sightlines through the ISM and CGM of the GRB hosts using the known characteristics of the GRB hosts at these redshifts. This modeling will help us disentangle the relative contributions of the host ISM and CGM to the observed column densities as a function of velocity. 

Unlike the detailed models that are available for the local multi-phase CGM \citep[e.g.][]{stern2016universal}, the CGM models for high-redshift galaxies are few and limited, especially for galaxies with $M_{*}  < 10^{10}M_{\odot}$. Therefore, we constructed a simple toy model to extract typical estimates of the physical properties of the CGM and thus help us interpret the observed kinematics. We adopt simple assumptions to derive CGM kinematics in terms of a few model parameters and compare the resulting column densities with the observations.  Ultimately, we aim to obtain a coarse estimate of typical kinematic properties of the CGM of GRB hosts at high redshifts. We will focus on the C IV kinematics since this feature is ubiquitously detected in all the sightlines, and the outflow component is prominent in the C IV kinematics. 

\begin{figure}
\centering
\includegraphics[width=0.45\textwidth]{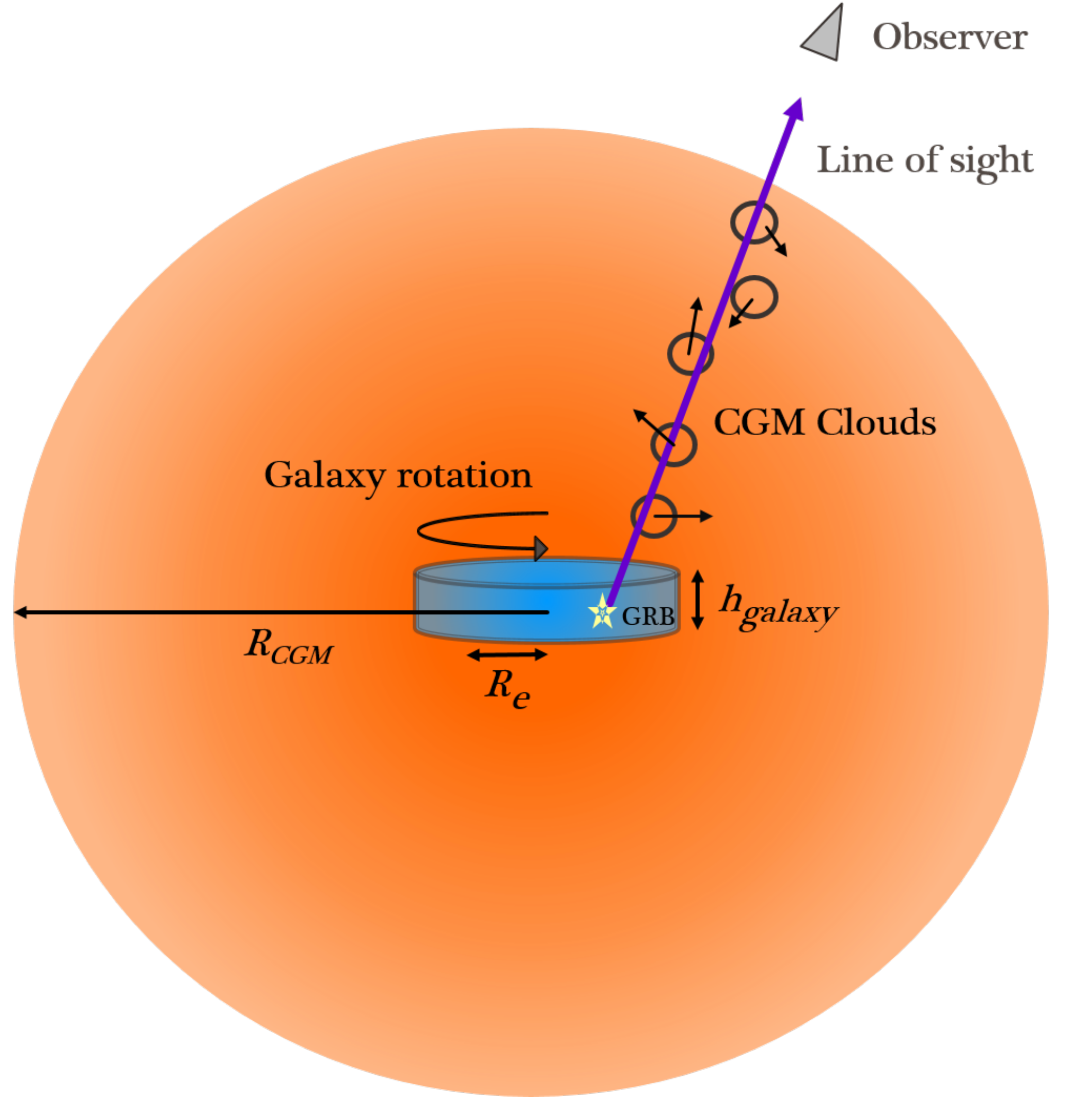}
\figcaption{\label{fig:CGM_model} A schematic of the toy model used for simulating a sightline that probes the kinematics of the ISM and CGM in a GRB host.}
\end{figure}

\subsection{Line-of-sight Simulations: Setup}
\label{sec:LOS_setup}
  
To simulate how the GRB sightlines sample the kinemaics of the CGM-galaxy system, we constructed a simple geometrical model of the system as shown in Figure \ref{fig:CGM_model}. The galaxy ISM is modeled as a disk with an exponential gas density distribution in the radial direction. The ISM parameters of the galaxy are chosen to reflect a typical, representative galaxy from the known population of GRB hosts in the redshift range $z$ $\sim$ 2 $-$ 3 \citep{wainwright2007morphological, perley2016swift_2, blanchard2016offset, arabsalmani2017mass}. The CGM is modeled as an isothermal sphere populated by clouds where the density of the cloud is inversely proportional to the square of the radial coordinate of the cloud (see Equation \ref{eqn:n_cloud}). This distribution assumes the clouds have originated from a mass-conserving outflow \citep{chisholm2017mass, steidel2010structure}. In addition to the isothermal velocity distribution, a certain fraction of clouds ($f_{out}$) are randomly selected to have an additional radial outward velocity component to simulate the outflows. The value of the additional radial velocity depends on the radial coordinate of the cloud and follows a ballistic velocity profile with radial launch velocity of $v_{out}$ at an outflow launching radius $R_{launch}$ of 2 kpc. The entire simulation setup is described in detail in Appendix \ref{appendix:LOS}. 

To create a complete sample of sightlines for each of these models, 200 GRB sightlines were synthesized by randomly sampling uniform distributions of a) GRB positions within the galactic disk and b) the 3D direction vectors of the sight line. The projected velocity of the intervening gas (from the ISM and CGM) along the GRB sightline was calculated with respect to the projected velocity of the gas in the immediate vicinity of the GRB. Setting this velocity reference is important to maintain consistency with the observations, where $v = 0$ is assigned to the strongest fine structure line absorption, a tracer of gas in the vicinity of the GRB as evident from the UV-pumping argument (\citealt{prochaska2006dissecting} and \citealt{vreeswijk2007rapid}).  


The most important model parameters affecting the observed CGM-ISM kinematics are listed in Table \ref{tab:Param_list}. We approximate the typical stellar mass of GRB hosts at $z > 2$ as $\sim$ $2\times10^{9}$ $M\odot$ \citep{perley2016swift_2}, thereby constraining the halo mass \citep{hopkins2014galaxies, wechsler2018connection} and thus the typical rotation and dispersion velocities. The volume filling fraction is approximated as 0.1 from prior CGM studies at lower redshifts \citep{werk2016cos, stocke2013characterizing}. With these assumptions, the free parameters of the model are: the CGM mass (traced by C IV), $M_{CGM}$, the launching velocity of the outflow, $v_{out}$, and the outflow fraction, $f_{out}$. Hence, we synthesize a matrix of 27 models with three distinct values of each of these model parameters, as stated in Table \ref{tab:Param_list}. 

\begin{figure*}
\centering
\includegraphics[width=0.9\textwidth]{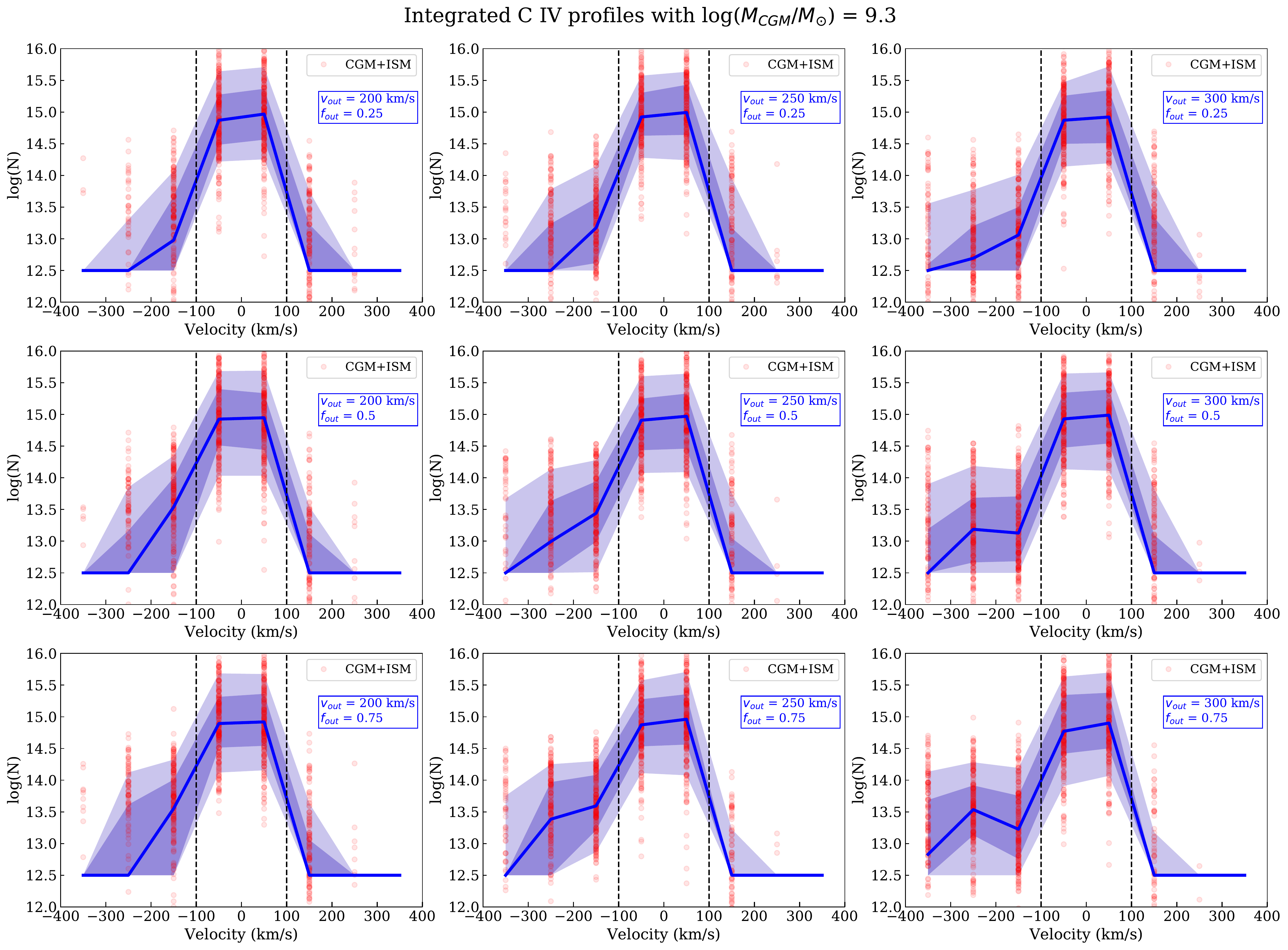}
\figcaption{\label{fig:Toy_CIV_9_0} Integrated C IV column density from the toy model assuming $M_{CGM}$ = $10^{9.3}M_{\odot}$. The panels show the results using an outflow launch velocity of 150 km s$^{-1}$, 220 km s$^{-1}$, and 300 km s$^{-1}$ (at 2 kpc) and an outflow fraction of 0.25, 0.5, and 0.75. }
\end{figure*}

\begin{figure*}
\centering
\includegraphics[width=0.9\textwidth]{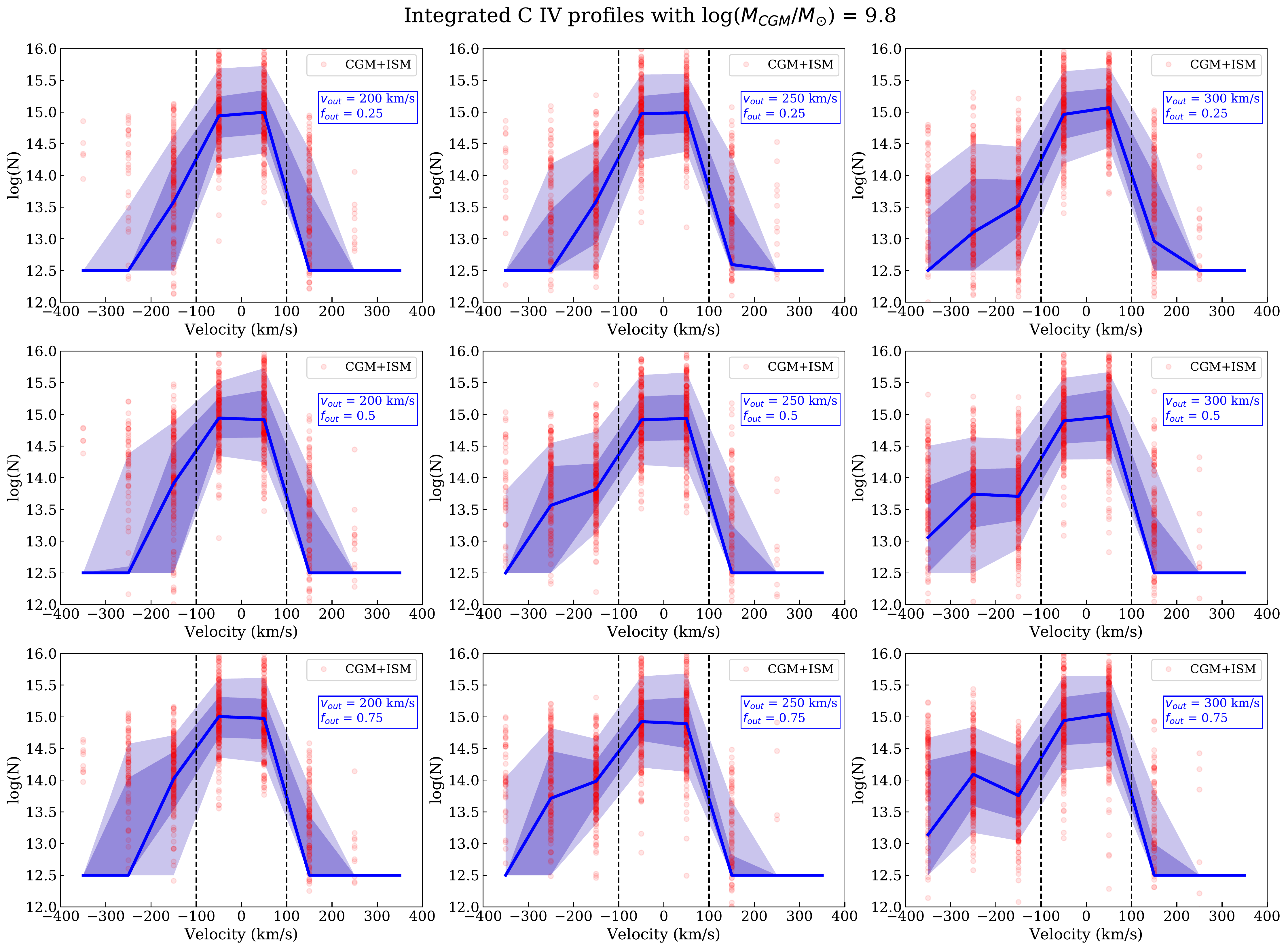}
\figcaption{\label{fig:Toy_CIV_9_5} Integrated C IV column density from the toy model assuming $M_{CGM}$ = $10^{9.8}M_{\odot}$. The panels show the results using an outflow launch velocity of 200 km s$^{-1}$, 250 km s$^{-1}$, and 300 km s$^{-1}$ (at 2 kpc) and an outflow fraction of 0.25, 0.5, and 0.75. }
\end{figure*}

\begin{figure*}
\centering
\includegraphics[width=0.9\textwidth]{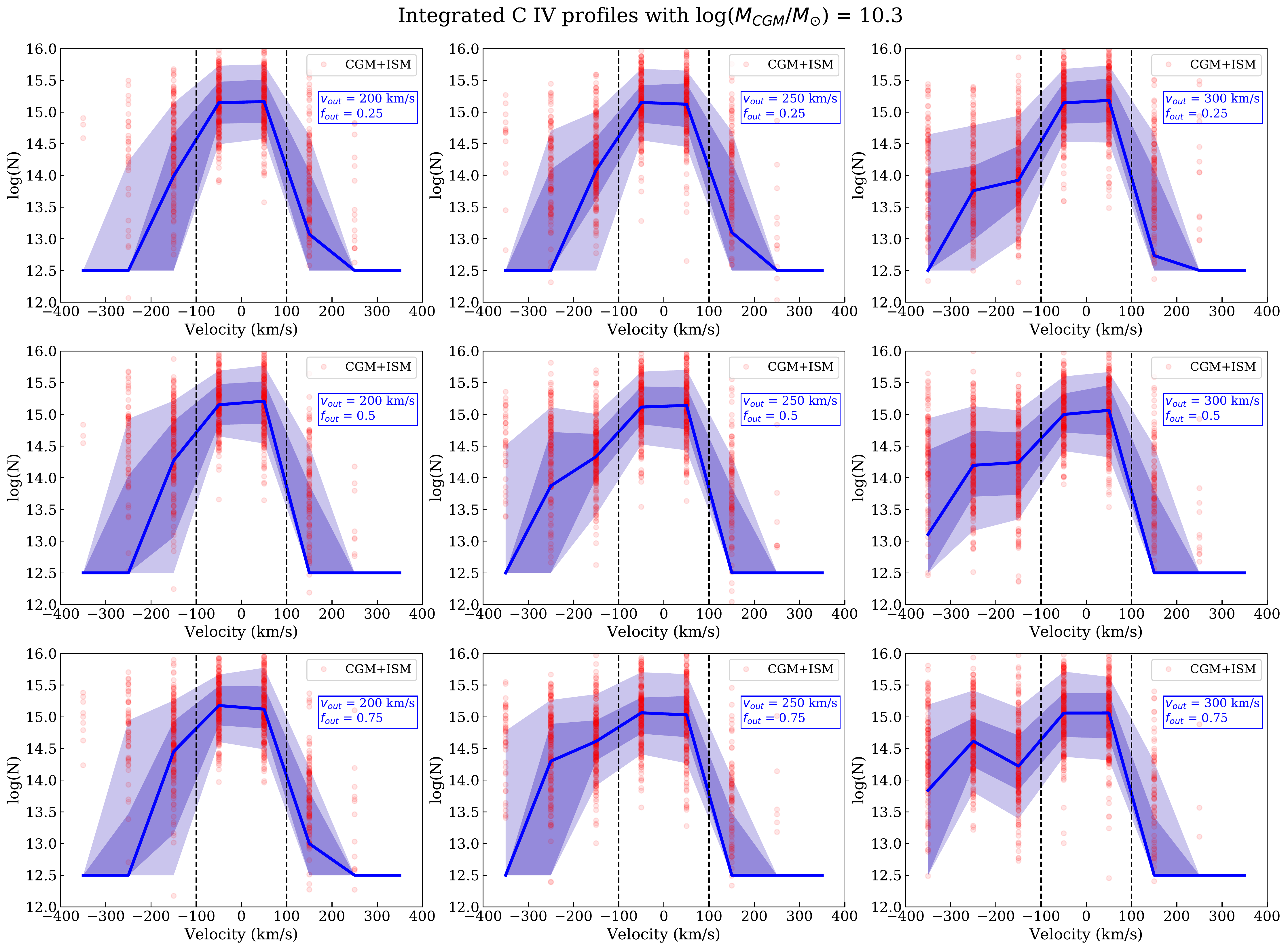}
\figcaption{\label{fig:Toy_CIV_10_0} Integrated C IV column density from the toy model assuming $M_{CGM}$ = $10^{10.3}M_{\odot}$. The panels show the results using an outflow launch velocity of 200 km s$^{-1}$, 250 km s$^{-1}$, and 300 km s$^{-1}$ (at 2 kpc) and an outflow fraction of 0.25, 0.5, and 0.75.}
\end{figure*}

The C IV kinematics are evaluated by making certain assumptions about the CGM and ISM metallicities and the ionization fraction for C IV, $f_{CIV}$. The C IV column density is evaluated as:

\begin{equation}
N_{CIV} = N_{total}\times \frac{f_{CIV}}{0.3} \times \frac{Z}{Z_{\odot}} \times (n_{C}/n)_{\odot}
\end{equation}

We conservatively assume $f_{CIV}$ = 0.3 as a maximal ionization fraction similar to \citet{bordoloi2014cos} and as derived in \citet{oppenheimer2013non} for $z > 2$. The CGM and ISM metallicities are assumed to be 0.25 and 0.15 solar, respectively \citep{trenti2015luminosity, arabsalmani2017mass, shen2013circumgalactic}. The solar ratio of carbon number density to total number density is 3.26 $\times$ 10$^{-4}$. 

\subsection{Line-of-sight Simulations: Results} \label{sec:LOS_results}

The kinematic separation of the CGM and ISM has been used in various previous studies (eg: to study the CGM of Andromeda galaxy) to distinguish the CGM and ISM contributions in the velocity profiles  \citep{fox2014cos, fox2015probing, lehner2015evidence}. In this paper, we performed kinematic simulations to infer the relative contributions of the CGM and ISM as a function of velocity using the characteristic parameters of GRB host galaxies of our sample. 

Figure 15 in Appendix \ref{appendix:LOS} shows the breakdown of the total column density into ISM and CGM components, with (left) and without an outflow (right). Two key inferences can be drawn from this illustration: a) an outflow component is necessary to explain the observed kinematic asymmetry, and b) the ISM component kinematically dominates the central region ($|v| < 100$ km s$^{-1}$), while the CGM component is the main contributor of column density in the red and blue wings ($v > 100$ km s$^{-1}$ and $v < -100$ km s$^{-1}$, respectively). With this insight, we use the kinematic separation to further estimate the physical properties of the CGM.

Figures \ref{fig:Toy_CIV_9_0}$-$\ref{fig:Toy_CIV_10_0} show the integrated CGM $+$ ISM column density for C IV as a  function  of  velocity  bins  of width 100 km s$^{-1}$ in a fashion similar to the observations. The results are shown for 27 synthesized models with different combinations of CGM mass (as traced by C IV), outflow launch velocity ($v_{out}$), and outflow fraction ($f_{out}$). The red dots indicate the integrated column density for each simulated GRB sightline. The blue line shows the median profile and the purple shades indicate the central 50 and 80 percentiles. The vertical dotted lines separate the central region dominated by the ISM from the blue and red wings dominated by the CGM.

We compare the blue and red wings with the observations since the central region is often saturated in the observations. As clearly seen in these figures, a higher outflow fraction increases the kinematic asymmetry while a higher outflow velocity shifts the asymmetry blueward, as expected. A higher CGM mass proportionally increases the median column densities in the red and blue wings, while the increment in the central region is slower since it is dominated by the ISM. 

Despite its simplicity, this kinematic model does an excellent job at reproducing the median kinematics profile as well as typical detection fractions and column density distribution as a function of velocity. These simulations rule out an outflow fraction of $\gtrsim 0.5$ and a CGM mass of $\gtrsim 10^{10.3}M_{\odot}$ by simple comparisons of the median profiles and the percentile distributions. However, it should be noted that a lower metallicity for the CGM would favor a higher CGM mass to explain the observations while a higher ionization fraction would favor a lower CGM mass. Nonetheless, the current assumptions are reasonable within a factor of 2. The model that best explains the observed kinematics is: $M_{CGM}$ = $10^{9.8}M_{\odot}$, $v_{out}$ = 300 km s$^{-1}$, $f_{out}$ = 0.25. Note that $v_{out}$ is the launch velocity at 2 kpc for a ballistic outflow. In section \ref{sec:CGM_mass}, we use these kinematic simulations to estimate the contribution of CGM in the central $\pm$ 100 km s$^{-1}$ velocity region, which is crucial in estimating the typical CGM mass in our sample. 


No extra inflow component has been added in these kinematic simulations. 
Various feedback simulations suggest that at $z \sim 2.5$, cold inflows are hardly detectable in metal lines due their low metallicity and low covering factors \citep{fumagalli2011absorption, goerdt2012detectability}. Given the observational challenges in detecting the inflows, we do not include an inflow component in our toy model. However, a plausible inflow scenario is explored in section \ref{subsec:inflows}.


\section{CGM Mass Estimate} \label{sec:CGM_mass}
\noindent
Estimating the mass of the CGM at high redshifts is a key step in determining the cosmic baryonic budget as well as the distribution of metals throughout the universe and the various mechanisms at play \citep{bouche2006missing, bouche2007missing, peeples2014budget, muratov2017metal}. 
In this section, we will estimate the typical mass of the CGM for high-z GRB hosts, explore its evolution with redshift, and compare the carbon mass in the CGM and ISM. The length scales considered here are proper distances.        
\subsection{Converting Column Densities into Mass} \label{subsec:Col_density_to_Mass}

Our kinematic toy models provide a reasonable handle on the CGM mass ($M_{CGM} \sim 10^{9.8} M_{\odot}$). In this section we provide further insights on the CGM mass from the observed column densities, adopting realistic geometrical assumptions, and compare the results with the toy models. For this, we assume the CGM-GRB sample as a set of random GRB sightlines probing the typical ISM and CGM of a GRB host at $z \sim$ 2 $-$ 5. 

\subsubsection{Method}

A key uncertainty in this formulation is the contribution of the CGM to the column density in the central region ($|v| < 100$ km s$^{-1}$). To resolve the CGM-ISM degeneracy in this region, we make certain assumptions based on the insights we gained from the toy models. Since the central region also suffers from saturation issue leading to the measurement of only the lower limit of the column density, we avoid making a direct use of the CGM to ISM ratio in this region. Instead, we use the column density in the $-$150 km s$^{-1}$ bin (i.e. $-200 < v < -100$ km s$^{-1}$) to extrapolate the central column density since the column density in the $-$150 km s$^{-1}$ velocity bin is measured more accurately. This ratio is in the range of 2 to 4 for the models that were found viable in the previous section. Hence, we use a factor of 3 to estimate the CGM-contributed column density in the observed spectra. To calculate the CGM mass, we will make use of C IV column density since it is ubiquitous and a good tracer of the outflow component.


To convert the integrated CGM column density of C IV, as approximated above, into the mass of C IV in the CGM, we assume that the CGM density profile is a power law given by:
\begin{equation} \label{eqn:n_cloud}
    n_{CGM}(r) = n_{0}\left( \frac{r}{R_{0}}\right)^{-\alpha}
\end{equation}
For convenience, we select the reference radius $R_{0}$ as the starting point of the CGM, equivalent to twice the half-light radius ($R_{e}$) of the galaxy. We assume a typical $R_{e}$ of 2 kpc for GRB hosts at $z > 2$, following previous observations of GRB hosts \citep{wainwright2007morphological, blanchard2016offset}, and thus $R_{0}$ = 4 kpc. Both the detailed simulations and our toy model indicate that the gas which leads to the observed absorption in C IV is spread out to about 2 $\times R_{vir}$. Beyond that, the gas density is too low to give rise to a detectable absorption component (assuming $n(r) \propto r^{-2}$). Given a constant volume filling fraction of $f_{vol}$, the mass of the CGM (for $\alpha \neq 3$) can be stated as: 

\begin{mathletters}
\begin{eqnarray} \label{eqn:M_CGM}
    M_{CGM} & = & 4\pi m_{H}f_{vol}\int_{R_{0}}^{R_{CGM}}n(r)\times r^{2}dr \\
     & = & 4\pi m_{H} f_{vol}n_{0}R_{0}^{\alpha}\left(\frac{1}{3-\alpha}\right)\left[R_{CGM}^{3-\alpha} - R_{0}^{3-\alpha}\right] 
\end{eqnarray}
\end{mathletters}\\

We further define the line covering fraction $f_{line}$ as the typical fraction of a sightline that passes through a CGM cloud. The typical line fraction does not strongly vary with radius if the volume filling fraction is constant. Given a constant $f_{line}$, the typical column density (for $\alpha \neq 1$) can be expressed as:
\begin{eqnarray} \label{eqn:N_CGM}
    N_{obs} & = & f_{line} \int_{R_{0}}^{R_{CGM}} n(r) dr \\
     & = & f_{line} n_{0}R_{0}^{\alpha}\left(\frac{1}{1-\alpha}\right)\left[R_{0}^{1-\alpha} - R_{CGM}^{1-\alpha} \right]
\end{eqnarray}

Combining the two equations, we get:

\begin{eqnarray}\nonumber \label{eqn:M_CGM_full}
    M_{CGM} & = & 4\pi m_{H}N_{obs}\left(\frac{f_{vol}}{f_{line}}\right)\left(\frac{1-\alpha}{3-\alpha}\right)\left[\frac{R_{CGM}^{3-\alpha} - R_{0}^{3-\alpha}}{R_{0}^{1-\alpha} - R_{CGM}^{1-\alpha}}\right] 
\end{eqnarray}
\\

In the following, we take $\alpha$ = 2, i.e.\ we consider the case where a significant fraction of the metal-rich CGM takes part in a mass-conserving outflow \citep{chisholm2017mass, pallottini2014circumgalactic, steidel2010structure}. Since $R_{0} \ll R_{CGM}$, the expressions can be simplified. Further, we evaluate the typical value of $f_{vol}/f_{line}$ by simulating various volume filling fractions in a spherical shell and measuring the distribution of line covering fraction for random sightlines. For small filling fractions ($\sim$ 0.05 $-$ 0.25, eg: as derived in \cite{werk2014cos}), the typical value of $f_{vol}/f_{line}$ ratio is found to be 1.2. Using this value, we estimate the mass of C IV in the CGM (in solar mass units) as:

\begin{equation} \label{eqn:M_CGM_CIV}
    M_{CGM, CIV} = 1.2 \times m_{C}N_{CIV}R_{0}R_{CGM} \times 10^{-13}
\end{equation}

\noindent 
where $m_{C}$ is the atomic mass number of carbon, $R_{0}$ and $R_{CGM}$ are in kpc, and M$_{CGM}$ is in solar masses. 

\subsubsection{Results}

As stated earlier, we estimate $R_{0}$ as 4 kpc and $R_{CGM}$ as 2$\times R_{vir}$ $\sim$ 100 kpc. We choose the geometric mean of all the sightlines in the sample as a representative value of $N_{CIV}$ since it best captures the large range in column density as a function of velocity whereas an arithmetic mean tends to be skewed to high values by the small number of large column densities. 
Similar mass estimates were performed for other species. The CGM mass estimates for various species as a function of velocity are shown in Fig. $\ref{fig:CGM_mass}$. For evaluating carbon mass in the CGM, the fraction of carbon in C IV phase ($f_{CIV}$) needs to be constrained which depends on the temperature, density, and ionization process. We assume a conservative maximal $f_{CIV}$ of 0.3 (see Fig. 7 in \cite{bordoloi2014cos}), which gives: 

\begin{eqnarray}\nonumber \label{eqn:M_CGM_C}
\mathrm{M_{CGM,C}}&\gtrsim& 4.8\times10^{4}\mathrm{M_{\odot}}\left(\frac{N_{CIV}}{\mathrm{10^{14} cm^{-2}}}\right)\\
&& \left(\frac{R_{0}R_{CGM}}{\mathrm{100\,kpc^{2}}}\right)\left(\frac{0.3}{f_{CIV}}\right)
\end{eqnarray}

Based on this formulation, the conservative lower limit on the carbon mass in the CGM of GRB hosts in our sample is $\sim$ $1.5 \times 10^{6} M_{\odot}$. 
The carbon mass in the CGM can be further extrapolated to derive the total mass of the CGM in the phase traced by C IV (T $\sim$ 4.5 $-$ 5.5 $\times$ $10^{5}$ K). For this, we assume that the metallicity in the CGM is roughly 0.25 solar. This assumption is informed by the detailed simulations of CGM at $z > 2$ \citep{shen2013circumgalactic} and the low-z CGM metallicity estimates (eg: \citealt{prochaska2017cos}). This is further supported by the observed gas-phase metallicity of $\sim$ 0.1 $-$ 0.2 solar for the ISM of the GRB hosts \citep{kruhler2015grb, arabsalmani2017mass}. Due to the metal-enriched outflows, the typical CGM metallicity tends to be $\sim$ 1.5 times the ISM metallicity \citep{muratov2017metal}. Thus, the total mass of the CGM gas traced by C IV can be expressed as:    
 
\begin{equation} \label{eqn:M_CGM_total}
    M_{CGM, total} \sim 10^{9}M_{\odot}\times \left(\frac{M_{CGM, C}}{10^{6}M_{\odot}}\right)\left(\frac{0.25}{Z_{CGM}}\right)
\end{equation}

Thus, the typical mass of the CGM gas traced by C IV in this sample is $\sim$ $10^{9.2}$ M$_{\odot}$, which is comparable to the typical stellar mass of the GRB hosts at these redshifts ($z \sim 2-4$).  This implies that the CGM is a very significant reservoir of baryons and metals in the galactic ecosystem at high redshifts. Thus, from a galaxy evolution standpoint, the CGM appears to be already in place at $z \sim 2-4$. Despite various uncertainties in the assumed parameters, we can say with high significance that the mass of the CGM in GRB hosts is at least as much as the mass that resides in the stars, and it can be higher by as much as $\sim$0.3 dex if the conservative assumptions are relaxed. 

The CGM mass estimates from the toy models (\S\ \ref{sec:LOS_results}) and column density profiles are complementary in nature, strengthening our CGM mass estimate of $\sim 10^{9.2-9.8} M_{\odot}$. It should be noted that the difference between the CGM mass of the optimal toy model ($10^{9.8} M_{\odot}$) and the CGM mass estimated here ($\sim 10^{9.2} M_{\odot}$) arises for two key reasons: a) the conservative estimate of the CGM contribution to the central $\pm$ 100 km s$^{-1}$ and b) the use of geometrical mean of column densities in the CGM mass measurement instead of arithmetic mean.  


\subsubsection{Caveats}
In this analysis, various simplifying assumptions have been made based on previous observations or simulation efforts. Here, we briefly discuss the sensitivity of the results to the assumed parameters. We assumed a CGM density profile of the CGM that follows $n(R) \propto R^{-2}$. A more compact profile, for instance $R^{-2.3}$, can reduce the CGM mass by 0.4 dex, whereas a more diffuse profile, eg $R^{-1.7}$, increases the CGM mass estimate by 0.4 dex. The product $R_{0}R_{CGM}$, which can be written as $4R_{e}R_{vir}$, marks the bounds we have defined for the CGM. We ascribe an uncertainty of $-0.3$ dex (reducing the $M_{CGM}$ estimate) to this product due to the gap in our knowledge associated with the faint nature and inherent variety of GRB hosts. 

On the other hand, $f_{CIV}$ = 0.3 is a conservative upper limit on the ionization fraction for the warm phase \citep{bordoloi2014cos}. This factor can be lower by a factor of $\sim 2-4$ in the range of temperatures and densities traced by C IV absorption in the CGM. Despite the evolution in the extragalactic background UV (ionizing) flux between $z \sim 0$  \citep[COS-Dwarfs;][]{bordoloi2014cos} and $z \sim 2.7$ \citep[this sample;][]{gilmore2009gev}, the ionization fraction $f_{CIV}$ does not exceed 0.3 for both collisional and photoionization models for a range of temperature ($10^{4.5} - 10^{5.5}$), number density ($10^{-2} - 10^{-5}$ cm$^{-3}$), and metallicity \cite[$0.1 - 1$ solar;][]{oppenheimer2013non}. The typical value expected in the warm phase traced by C IV is $\sim 0.1 - 0.2$. Thus, we attribute an uncertainty of $+ 0.3$ dex to $f_{CIV}$ (raising the $M_{CGM}$ estimate).

We have also assumed a constant typical metallicity for the CGM. While constraining the radial gradient of metallicity is an observational challenge, it is unlikely that the average metallicity of the CGM is significantly larger than 0.25 solar at these redshifts \citep[based on the observed metallicities of the DLAs asssociated with the GRB hosts at $z > 2$;][]{cucchiara2015unveiling, toy2016exploring}. As a limiting case, if the typical metallicity of the CGM gas traced by C IV is assumed to be the same as the typical ISM metallicity of these GRB hosts ($\sim 0.1$ solar), then we can expect an increment of $\sim 0.4$ dex in the CGM mass (from equation \ref{eqn:M_CGM_total}).      

Despite the simplifying assumptions and uncertainties stated here, it can be seen that the overall uncertainty in the CGM mass is small and would likely favor a higher warm CGM mass than calculated here. Thus, the robust mass estimates from our sample and the kinematic inferences drawn with the help of our toy model clearly indicate that the CGM is already a significant component of the GRB host galaxies at high redshifts, comparable to the mass of the host ISM. This has important implications on the evolution of the CGM and the distribution of metals and baryons throughout the galactic ecosystem as a function of time.

\subsection{CGM Mass Evolution with Redshift}\label{subsec:CGM_mass_redshift}

\begin{figure}
\centering
\includegraphics[width=0.45\textwidth]{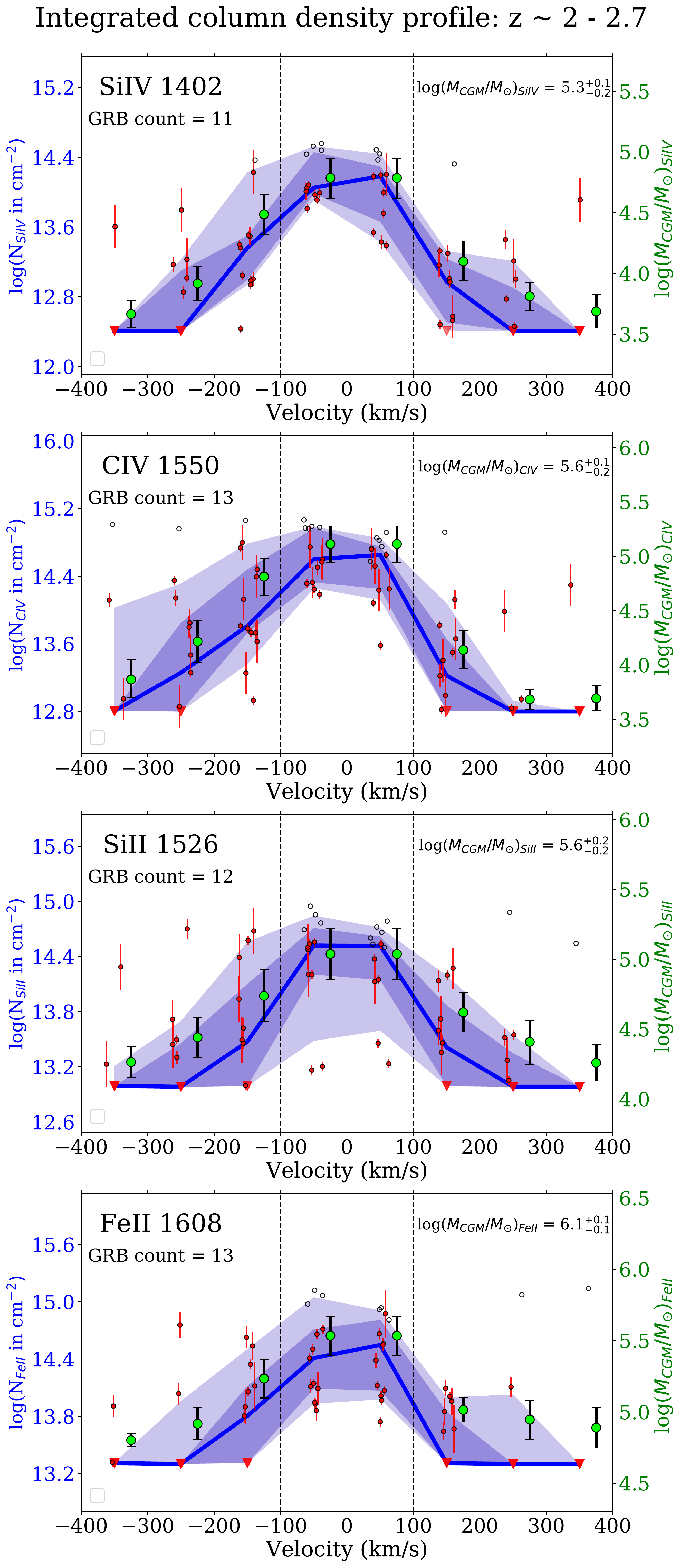}
\figcaption{\label{fig:CGM_Mass_z1} CGM column densities and mass estimates for high- and low-ion species in the redshift group \#1 (z $\sim$ 2 $-$ 2.7).}
\end{figure}

\begin{figure}
\centering
\includegraphics[width=0.45\textwidth]{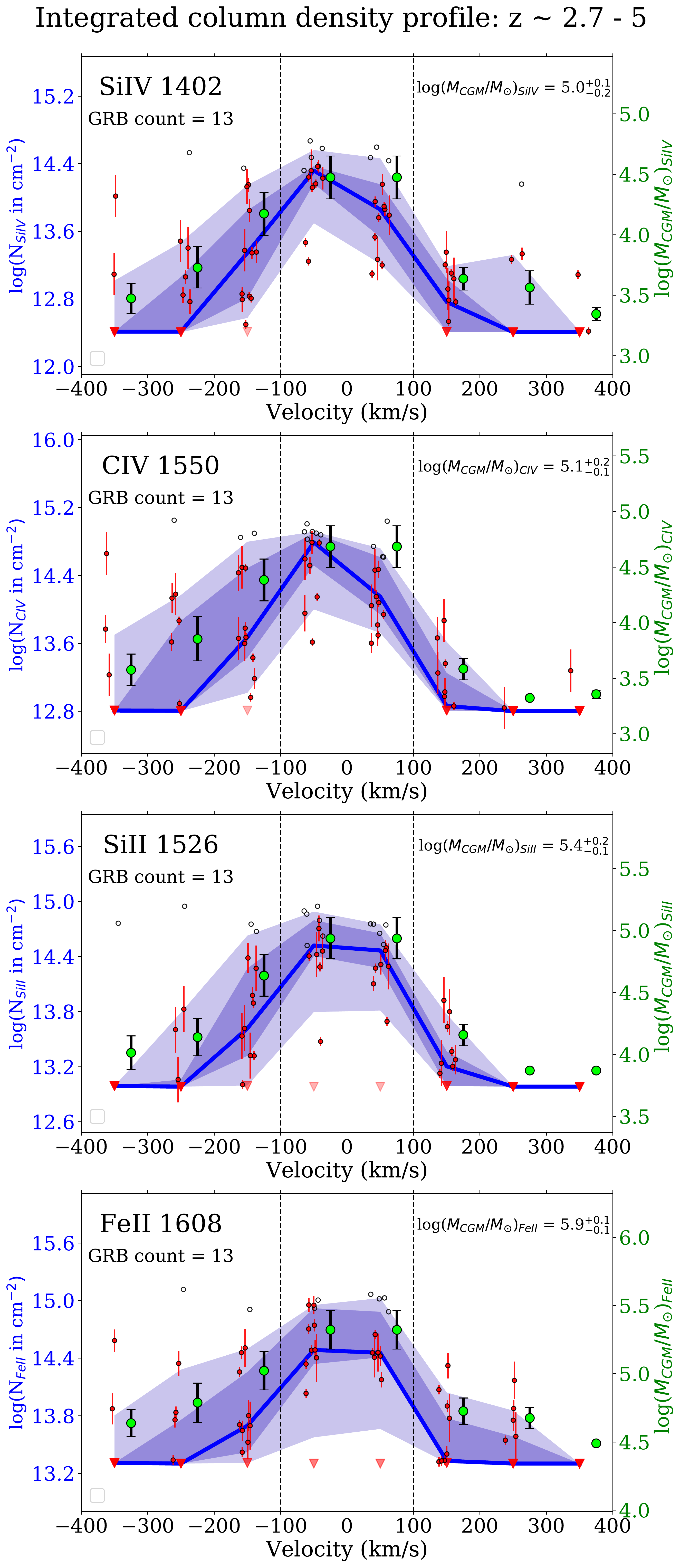}
\figcaption{\label{fig:CGM_Mass_z2} CGM column densities and mass estimates for high- and low-ion species in the redshift group \#2 (z $\sim$ 2.7 $-$ 5.0).}
\end{figure}

To study the evolution of the CGM mass with redshift, we divide the CGM-GRB sample into two roughly equal time bins of 1 Gyr $-$ group 1 ($z1 \sim 2 - 2.7$, midpoint: $z = 2.3$) and group 2 ($z2 \sim 2.7 - 5$, midpoint: $z = 3.6$). The number of GRBs in these two bins are also nearly equal. The integrated column density as a function of velocity for high- and low-ion species are plotted in Figures \ref{fig:CGM_Mass_z1} and \ref{fig:CGM_Mass_z2} in the same way as Figure \ref{fig:CGM_mass}. 

It can be seen in these figures that the high-ion kinematics in both redshift groups are quite similar with respect to the blue asymmetry. In the case of Si II, the blue asymmetry is weaker in the lower redshift bin (Fig. \ref{fig:CGM_Mass_z1}). There are two possible reasons for this lack of asymmetry in Si II: a) The ionization level of the outflows is different in at low z leading to a higher [Si IV / Si II] ratio relative to the high-z bin, or b) there is more inflowing gas in the low-z bin compared to the high-z bin which produces a relatively stronger Si II red wing. From the high-z simulations in \cite{shen2013circumgalactic}, it is clear that Si II is a much stronger tracer of the inflows compared to outflows by almost an order of magnitude. However, a more rigorous treatment using ionization modeling is required to distinguish between the two scenarios and constrain the physical state of the outflows and inflows. 

We follow the same procedure as the one described in section \ref{subsec:Col_density_to_Mass} for estimating the mass of the CGM in these two redshift bins. The key changes in the assumed parameters are: a) the typical value of $R_{vir}$ and b) the typical value of $R_{e}$ in these two redshift bins. The other parameters are not expected to change significantly. The virial radius is calculated as the radius within which the normalized density, $\rho/\rho_{cosmic}$ $>$ 200, using the NFW profile for dark matter distribution and standard cosmological parameters ($\Omega_{m}$ = 0.3, $\Omega_{rad}$ = 0, $\Omega_{\Lambda}$ = 0.7). Thus, the typical virial radii for the $z1$ and $z2$ redshift bins are 53 and 39 kpc, respectively. Therefore, $R_{CGM}$ for the two bins are 106 and 78 kpc. The typical half-light radius for the two bins are assumed to be 2 and 1.5 kpc, respectively, following the previous population study of the GRB hosts at these redshifts \citep{wainwright2007morphological, blanchard2016offset}. Hence, the typical $R_{0}$ for the two bins are taken as 4 and 3 kpc, respectively. With this setup, the masses of various species were calculated in the two redshift bins and are shown in Figures \ref{fig:CGM_Mass_z1} and \ref{fig:CGM_Mass_z2}. The C IV in the CGM was estimated to be $\mathrm{log(M_{\mathrm{CIV, z1}}/M_{\odot})} = 5.6^{+0.1}_{-0.2}$ and $\mathrm{log(M_{\mathrm{CIV, z2}}/M_{\odot})} = 5.1^{+0.2}_{-0.1}$. The total CGM masses in the two redshift bins (following the same procedure as described in section \ref{subsec:Col_density_to_Mass}) are estimated as $\mathrm{M_{CGM, z1} = 10^{9.2} M_{\odot}}$ and $\mathrm{M_{CGM, z2} = 10^{8.7} M_{\odot}}$. 

The C IV mass in the CGM, as shown in Figures \ref{fig:CGM_Mass_z1} and \ref{fig:CGM_Mass_z2}, is higher in the lower redshift bin by almost 0.5 dex. However, it is quite likely due to the combined effect of redshift evolution and the difference in typical stellar mass and star formation rate between the two redshifts. This will be explored in more depth by associating the CGM properties with the host properties in an upcoming paper on this sample. 

To convert the C IV mass to carbon mass, we make a conservative assumption for $f_{CIV}$. Despite the extragalactic ionizing UV background flux evolution (within a factor of 2) between the two redshift bins, $f_{CIV}$ does not exceed 0.3 for both photo- and collisional ionization models \cite{oppenheimer2013non}. Hence, we consider the maximal $f_{CIV}$ as 0.3 for both the redshift bins (same as in section \ref{subsec:Col_density_to_Mass}). The $M_{CGM, C}$ is thus estimated using Eqn. \ref{eqn:M_CGM_C} for both redshift bins and plotted in Fig. \ref{fig:CGM_mass_comparison} to compare with the $M_{CGM, C}$ estimates from COS-dwarfs study (at $z \sim 0$). The typical GRB host stellar masses in the redshift bins shown in the figure are median values from the SHOALS (Swift GRB Host Galaxy Legacy Survey) sample which is the largest systematic survey of long GRB hosts \citep{perley2016swift_2}. The stellar mass uncertainties are the bootstrapping $1\sigma$ intervals from the SHOALS survey. As described in \ref{subsec:Col_density_to_Mass}, our estimates for $M_{CGM, C}$ are evaluated within 2 virial radii. Hence, for a better picture of redshift evolution, the COS-dwarfs estimates within $\sim$ 1 virial radius are shown in green and the estimates within  $\sim$ 2 virial radii in light green. 

From Fig. \ref{fig:CGM_mass_comparison}, it can be seen that the CGM carbon mass (or by extension, metal mass) increased by only a factor of $\sim 2$ (comparing with the 2$R_{vir}$ estimates) from $z > 2$ to $z \sim 0$ for dwarf galaxies. Despite the fact that $f_{CIV}$ is assumed to be a conservative upper limit for all of these calculations, it can be clearly observed that most of the metal mass in the CGM of the low-mass galaxies represented by GRB hosts is already in place by $z \sim 2.5$. While the COS-dwarf galaxies are not the descendents of galaxies represented by GRB hosts at $z > 2$, this comparison has significant implications on the distribution of metals throughout the galaxy ecosystem as a function of redshift, as discussed in section \ref{subsec:Evolution_with_z}.  

\subsection{Comparison with Carbon Mass in the ISM}

\begin{figure}
\centering
\includegraphics[width=0.45\textwidth]{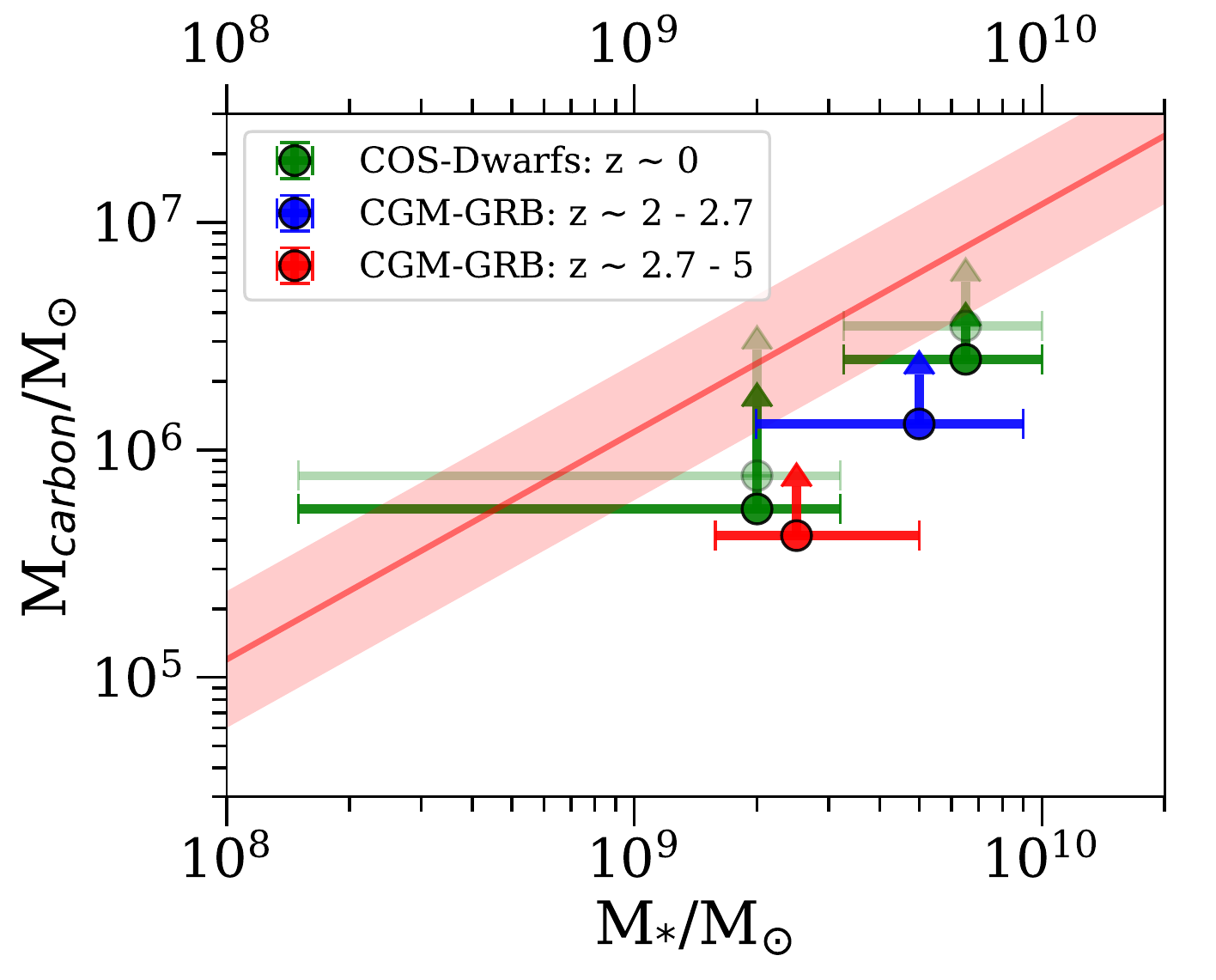}
\figcaption{\label{fig:CGM_mass_comparison} Estimates of the minimum carbon mass in the CGM of the GRB hosts in this sample. The typical minimum carbon mass in the CGM in the two redshift regions described in Figures \ref{fig:CGM_Mass_z1} and \ref{fig:CGM_Mass_z2} are compared with the z $\sim$ 0 dwarf-galaxy CGM from the COS-Dwarfs survey \citep{bordoloi2014cos}. The COS-dwarfs estimates within one virial radius are shown in dark green, while the estimates within two virial radii are shown in light green for easier comparison with the high-z values also derived within two virial radii.}
\end{figure}

By comparing the carbon mass in the ISM with that in the CGM, we can infer the level of CGM enrichment for GRB hosts at high redshifts. We estimate the gas phase carbon mass in the ISM of the GRB hosts by assuming a modest gas-phase metallicity of 0.15 solar \citep{arabsalmani2017mass, dave2017mufasa} with solar-like relative abundance pattern. Here, we assume a metallicity slightly lower than the median value reported in \cite{arabsalmani2017mass} since the UV/optical absorption- and emission-based metallicity estimates do not account for potentially lower metallicity gas in the ISM. The total ISM gas mass is estimated as 2$\times$ $M_{*}$ based on various molecular and neutral hydrogen measurements and simulations of high redshift main-sequence galaxies, where $f_{gas} = M_{gas}/M_{*}$ is reported to be $\sim$ 2 for $10^{9} < M_{*}/M_{\odot} < 10^{10}$ (\citealt{daddi2010very, carilli2013cool, lagos2014galaxies, dave2017mufasa,tacconi2018phibss}). We estimate an uncertainty of roughly $\pm$0.3 dex in the product of metallicity and the gas fraction ($f_{gas}$) to account for the uncertainty associated with the mass-metallicity relation for GRB hosts at $z > 2$ and the variation of $f_{gas}$ with redshift. Note that here we are considering total gas mass of the ISM to estimate the carbon mass in the ISM (and not just the warm phase of the ISM). With these reasonable assumptions, we plot the ISM mass of carbon in Figure \ref{fig:CGM_mass_comparison}. 

The carbon mass in the ISM is higher than the minimum carbon mass in the CGM by a factor of $\sim$ $3-5$. It should be noted again that the carbon mass in the CGM is a lower limit and considers only the warm phase traced C IV. In principle, the actual value of total carbon in the CGM can be higher by a factor of as much as $\sim2-3$ (see section \ref{subsec:Col_density_to_Mass}). This observation indicates that the carbon (metal) content of the CGM is $\sim 20-50\%$ of the carbon (metal) content of the ISM, indicating that a significant fraction of metals synthesized in the galaxy are able to escape into the CGM due to the galactic outflows. In \cite{bordoloi2014cos}, this fraction is $50-80\%$ for $z \sim 0$ galaxies of similar stellar mass. This finding hints towards a modest evolution in the carbon (metal) content, or in other words, a modest enrichment of the CGM over $\sim$ 11 Gyr span.   





\section{Implications} \label{sec:implications}
\noindent 
Using a sample of high SNR and medium resolution spectra of GRB afterglows at $z \sim 2 - 6$, we estimated various properties of the CGM for a typical GRB host at these redshifts. The observed blue asymmetry indicates a clear signature of an outflowing component predominantly traced by the high-ions (C IV, Si IV). Using the toy models and observed column density profiles, the typical mass of the host galaxy's CGM is estimated to be $\sim 10^{9.2-9.8} M_{\odot}$ (\S\ 5.2 and \S\ 6.1). In this section, we discuss the implications of our analysis on various important aspects of the CGM.

\subsection{Outflows and Metal Enrichment} \label{subsec:Outflows_metal_enrcih}
In \S \ref{subsec:CGM_mass_redshift}, the CGM-GRB sample is divided into two redshift bins each spanning 1 Gyr (bin 1: $z \sim 2 - 2.7$, bin 2: $z \sim 2.7 - 5$). Outflows are clearly detected in both redshift bins as shown in figures \ref{fig:CGM_Mass_z1} and \ref{fig:CGM_Mass_z2}. As described in \S \ref{subsec:kinematic_asymmetry}, the outflows are inferred from the blue asymmetry that is observed in both the high- and low-ion lines, but they are primarily traced by the high-ions (C IV and Si IV). Using the kinematic information from the observations and further clarity from the toy models, we can estimate various aspects of the outflows. From the toy models, the optimal model for explaining the observed kinematics has a typical outflow launch speed of $\sim 300$ km s$^{-1}$ (at 2 kpc), and $\sim$ 25\% of the CGM clouds in the model contribute to outflows in excess of the virial motion in the CGM of a typical dark matter halo mass $M_{h} \sim 10^{11.2} M_{\odot}$. With this insight, we estimate the key properties of the outflowing material in this section. We assume an NFW profile for the dark matter distribution with a concentration parameter of 4.5 (derived by forcing $R_{200} = R_{vir}$ and defining the mass enclosed within a virial radius as the halo mass).

\subsubsection{Outflow Mass}

The optimal toy model has $f_{out} = 0.25$, i.e. the outflow mass is 25\% the CGM mass within $R_{CGM}$. Thus, the estimates for the outflow masses in the two redshift bins are $M_{out, z1} \sim 10^{8.6} M_{\odot}$ and $M_{out, z2} \sim 10^{8.1} M_{\odot}$, where (z1: $z \sim 2 - 2.7$ and z2: $z \sim 2.7 - 5$). 
Assuming a halo mass of $10^{11.2} M_{\odot}$, a radial launch velocity of 250 km s$^{-1}$ (at 2 kpc) is just high enough for the outflowing gas to escape the CGM (i.e. 2$\times R_{vir}$).

The fraction of outflowing gas with a radial velocity $v_{radial} > 250$ km s$^{-1}$ at launch can be estimated using the velocity distribution of outflowing clouds in the toy model (see Appendix \ref{appendix:LOS}).
The velocity distribution is reasonably assumed to be an isotropic Gaussian with a standard deviation ($\sigma_{v, CGM}$) of 100 km s$^{-1}$ given by the virial velocity of the halo. The radially outward component of the isotropic standard deviation is $100/\sqrt{3}~(= 57.7)$ km s$^{-1}$. For the outflows, an extra radiallly outward component of $v_{out}$ is added to this distribution at launch, thus centering the Gaussian at $v_{out}$ (= 300 km s$^{-1}$ at launch for the optimal toy model). From this distribution, the fraction of clouds with a launch velocity $v_{radial} > 250$ km s$^{-1}$ is 80\% of the total outflowing gas (for the optimal toy model). Thus, the mass of outflowing gas in the CGM with sufficient initial kinetic energy to escape into the IGM is $M_{esc, z1} \sim 10^{8.5} M_{\odot}$ and $M_{out, z2} \sim 10^{8} M_{\odot}$. It should be noted that the time to traverse 2$\times R_{vir}$ in this dark matter halo is $\sim$ 0.5 Gyr.

\subsubsection{Mass Outflow Rate} \label{subsubsections: Mass-outflow}

We can now evaluate the mass outflow rate in these two redhsift bins. The outflow rate out of a spherical shell is given as: 
\vspace{-1ex}
\begin{equation}\label{eqn:outflow_rate}
    \dot{M}_{out} = \bar{m}n(R_{shell})\times 4\pi R_{shell}^{2} \times v_{out}(R_{shell}) \times f_{vol} \times{f_{out}}
\end{equation}

\noindent
where the average mass per atom or ion is $\bar{m}$ = 1.15 $m_{H}$. We consider $R_{shell}$ = 4 kpc (3 kpc) for the redshift bin $z1$ ($z2$) to estimate the outflows coming out of the galaxy disk. Note that we assume 4 kpc (3 kpc) as the boundary of the galaxy disk, $R_{0}$, in section \ref{subsec:CGM_mass_redshift}. Considering $M_{CGM}$ = $10^{9.2} M_{\odot}$ ($10^{8.7} M_{\odot}$), we get $n(R_{shell})$ = 0.028 (0.02) cm$^{-3}$ (following equation \ref{eqn:M_CGM}). The outflow launch velocity in the 2-4 kpc range is approximated as $\sim$ 300 km s$^{-1}$ from the optimal toy model and the observed kinematics for both redshift bins. Following the optimal toy model, we assume a volume filling fraction $f_{vol}$ = 0.1 and an outflow fraction $f_{out}$ = 0.25. Within this framework, we estimate the mass outflow rates (in the gas phase traced by C IV) as: \vspace{-1ex}

\begin{eqnarray}\nonumber
    \dot{M}_{out} &=& 0.27 M_{\odot}\mathrm{yr^{-1}} \left(\frac{n(R)}{\mathrm{10^{-2} \ cm^{-3}}}\right)\\ 
    && \left(\frac{R^{2}}{\mathrm{10 \ kpc^{2}}}\right)
     \left(\frac{v_{out}(R)}{\mathrm{300 \ km s^{-1}}}\right),
\end{eqnarray}

\noindent where $R$ is the radius of the shell, $R_{shell}$. Inserting the appropriate values for the two redshift bins, we get $\dot{M}_{out, z1}$ = 1.2 $M_{\odot}$ yr$^{-1}$ and $\dot{M}_{out, z2}$ = 0.5 $M_{\odot}$ yr$^{-1}$. This warm-phase mass outflow rate is comparable to the latest TNG simulation for galaxies with $M_{*} \sim 10^{9-10} M_{\odot}$ at $z = 2$ \citep[area under the curve for warm phase in Fig. 10 of][]{Nelson:2019jkf}. Note that the gas outflow rate in \cite{Nelson:2019jkf} is measured at 20 kpc, however. In our density profile assumption, $n(R)\times R^{2}$ is independent of $R$ and the cloud velocities at 20 kpc reduce to $\sim$70\% of their launch velocities for $v_{launch} = 200-300$ km s$^{-1}$. Hence, for comparison, $\dot{M}_{out, 20kpc}$ $\sim$ 0.7$\dot{M}_{out, launch}$ in our framework. 

As a corollary of the escaping mass ($M_{esc}$) calculation, we can now estimate the mass escape rate by modifying equation \ref{eqn:outflow_rate}. For this, we consider the boundary of the CGM as $R_{CGM} = 2R_{vir}$. At this radius, a ballistic outflow launched at 2 kpc with a velocity of 250 km s$^{-1}$ decelerates to 50 km s$^{-1}$ (from the NFW dark matter profile). Due to our assumption of mass conserving outflow (i.e. $n(R) \propto R^{-2}$), the product $n(R) \times R^{2}$ is invariant. The modified equation for the mass escape rate becomes:   
\begin{equation}
    \dot{M}_{esc} = \bar{m}n(R_{esc})\times 4\pi R_{esc}^{2} \times v_{out}(R_{esc}) \times f_{vol}{f_{out}}{f_{esc}}
\end{equation}

Using 50 km s$^{-1}$ for $v_{out}(R_{esc})$ and 80\% for $f_{esc}$ (the fraction of outflowing gas with a launch velocity $>$ 250 km s$^{-1}$ from Sec.\ 7.1.1), we get $\dot{M}_{esc} \sim 0.14 \times \dot{M}_{out}$. Thus, the estimates for the mass escape rate into the IGM for the two redshift bins are: $\dot{M}_{esc, z1}$ = 0.17 $M_{\odot}$ yr$^{-1}$ and $\dot{M}_{esc, z2}$ = 0.07 $M_{\odot}$ yr$^{-1}$. These estimates are robust within a factor of 2 over $\pm$30 km s$^{-1}$ variations in the launch velocity ($v_{out}$) in the optimal toy model. 

The calculations described in this section suggest that $\sim$ 80\% ($f_{esc}$) of the mass ejected from a GRB host galaxy of median halo mass ($M_{h} \sim 10^{11.2} M_{\odot}$) escapes to enrich the IGM while only 20\% ($f_{retain}$) stays within the CGM of a typical GRB host in this sample at $z > 2$. Combining this result with the 0.5 Gyr timescale to traverse 2$\times R_{vir}$, one can say that $\sim$ 20\% ($f_{out}\times$80\%) of the total CGM mass of a typical GRB host at $z \sim 2-5$ escapes out to the IGM over 0.5 Gyr. Conversely, the CGM mass in warm phase grows by $\sim$ 5\% ($f_{out}\times$20\%) every 0.5 Gyr if we assume that the non-escaping gas becomes a part of the virialized CGM. Such a growth rate, if steady, is sufficient to significantly grow the warm-phase CGM over a 5 Gyr period.  



\subsubsection{Mass Loading Factor and Energetics}

The mass loading factor is defined as $\eta = \dot{M}_{out}/SFR$ and serves as an important indicator of the mechanism driving the outflow. We consider a median SFR of 10 $M_{\odot}$ yr$^{-1}$ for GRB hosts from \cite{kruhler2015grb} at $z > 2$ (typical range $5-50 M_{\odot}$ yr$^{-1}$). The mass loading factors at the launch radii can then be estimated as: $\eta_{z1} = 1.2/10 = 0.12$ and $\eta_{z2} = 0.5/10 = 0.05$. Several simulations calculate the mass loading factors at some intermediate radius such as 20 kpc. The mass loading factors at 20 kpc can be evaluated by using decelerated velocities as (see section \ref{subsubsections: Mass-outflow}): $\eta_{z1, 20kpc} = 0.7 \times 0.12 = 0.084$ and $\eta_{z2, 20kpc} = 0.7 \times 0.05 = 0.035$.

While the outflow velocities in comparable mass ranges to our sample are always in the range of $200-400$ km s$^{-1}$, there is at least an order of magnitude variation in the mass-loading factors reported in various observational studies at high redshifts due to the diversity in probes, underlying assumptions, and the phases traced in the outflow. Therefore only an order-of-magnitude comparison can be done.

\cite{crighton2014metal} use a QSO sightline probe at $z = 2.5$ for a $M_{*} \sim 10^{9.1} M_{\odot}$ galaxy and infer a mass-loading factor of order 1. \cite{weiner2009ubiquitous} also infer an $\eta$ of order unity at launch for cold phase outflow (tracer: Mg II) in star-forming galaxies at $z \sim 1.4$ using rest-frame UV spectra of the galaxies. Along similar lines, \cite{martin2012demographics} report an $\eta$ of order 1 for Fe II-traced outflows in a redshift range $z \sim 0.4-1.4$ for star-forming galaxies over a wide range of stellar mass ($10^{9.5-11.5} M_{\odot}$). With a similar method, \cite{rubin2014evidence} conservatively estimate a cold gas mass loading factor $\eta_{5 kpc} \gtrsim 0.02-0.6$ for galaxies with $M_{*} \gtrsim 10^{9.6} M_{\odot}$ and SFR $\gtrsim 2 M_{\odot}$ yr$^{-1}$ in the redshift range $z \sim 0.3-1.4$. \cite{davies2018kiloparsec} use IFU spectra of star-forming galaxies with $z \sim 2-2.6$ and $M_{*} \sim 10^{9.5-11.5} M_{\odot}$ to estimate the mass loading factor $\eta \sim 0.05-0.5$ at launch. This sampling of the literature shows that overall, for galaxies of comparable mass with our sample, the mass-loading factor estimates range from $0.05-5$. It is likely that the higher mass loading in the down-the-barrel observations is due to the line-of-sight effects (down-the-barrel versus random sightlines of GRBs). This highlights the need for a multi-probe approach to trace the outflow process in various phases and for various orientations of a galaxy to capture the full picture of the CGM outflows. 


The mass loading factors estimated here are smaller than the estimates from various galaxy evolution and zoom-in simulations by 1-2 orders of magnitude. For example, cosmological zoom simulations such as \cite{angles2014cosmological} and \cite{shen2012origin} suggest an $\eta \sim 1$ at $z > 2$ to reproduce the morphological and dynamical properties of galaxies with $M_{h} \sim 10^{11-12} M_{\odot}$ at $z \sim 2$. However, it should be noted that these mass-loading factors encompass all the phases, and not just the warm phase traced by C IV or Si IV. The latest TNG simulations appear to resolve this issue by separating the phases in the outflows \citep{Nelson:2019jkf}. The total (all-phase) mass loading factor $\eta_{tot, 20kpc} \sim  4$ (Fig. 5 in \cite{Nelson:2019jkf}) for a main-sequence galaxy of $M_{*} \sim 10^{9.3} M_{\odot}$ at $z = 2$ whereas the loading factor for the warm phase is  $\eta_{warm, 20kpc} \sim  0.15$ (Fig. 10 in \cite{Nelson:2019jkf}). This warm phase mass loading factor is within a factor of 2 of the mass loading factor evaluated here. This also implies that a significant fraction of the outflowing mass is in other phases. It will be of further interest to carry out such comparisons at higher redshifts to synthesize a complete picture of the impact of outflows on galaxy evolution and vice versa. 

It should be noted that here we have assumed a ballistic outflow driven by star formation and no halo drag or outflow acceleration (eg: by cosmic rays, ram pressure, or radiation pressure) is considered (\citealt{murray2011radiation, hayward2016stellar, girichidis2016launching}). The comparative effects of the two can be non-trivial and need to be explored in the future.  
At the same time, the observational mass-loading factors have several uncertainties that need to be considered while comparing them to numerical simulations. The current analysis has been conducted for an ensemble, so only typical values are considered here. We will do this for individual hosts in an upcoming paper by connecting the CGM and galaxy properties. While comparing the outflow characteristics with simulations and other surveys, the ionization fraction, outflow fraction in various phases, and the dynamics of an outflow (various accelerations and drags) need to be treated more carefully. In addition, the relative fraction of the re-accretion of the enriched gas versus its virialization in the CGM needs to be explored in the simulations. 

The modest mass outflow rates estimated here can be entirely supernova-driven. The kinetic energy of the outflow can be estimated as $\frac{1}{2}\dot{M_{out}}{v^{2}}_{out} \sim 5.7 \times 10^{40}$ ergs/s. Following the formalism described in \cite{murray2005galaxies} (see equations 34 and 35), we assume a Salpeter IMF and that $\xi \sim 10\%$ of the supernova energy ($E_{SN} \sim 10^{51}$ ergs) is efficiently thermalized in the ISM and thus, driving the displacement of the gas. With a supernova rate ($f_{SN}$) of 10$^{-2}$ per $M_{\odot}$ of star formation, the energy deposition from supernovae becomes: 
\begin{equation}
    \dot{E}_{SN} \sim \xi f_{SN}E_{SN} \times SFR \sim 3 \times 10^{40} \frac{\xi}{0.1} \frac{\mathrm{SFR}}{1 \mathrm{M_{\odot} yr^{-1}}} \mathrm{ergs/s}
\end{equation}

Thus, SFR $\sim$ 2 $\mathrm{M_{\odot}~yr^{-1}}$ is energetically sufficient to drive the observed warm outflows. While a comparable mass outflow rate may be hidden in other gas phases (not traced by C IV), given the typical SFR $\sim$ 10 $\mathrm{M_{\odot}}$ yr$^{-1}$ for $z > 2$ GRB hosts, it is very likely that the outflows are predominantly driven by supernova energy injection.

\subsection{CGM-Galaxy Co-Evolution}\label{subsec:Evolution_with_z}

\begin{figure}
\centering
\includegraphics[width=0.5\textwidth]{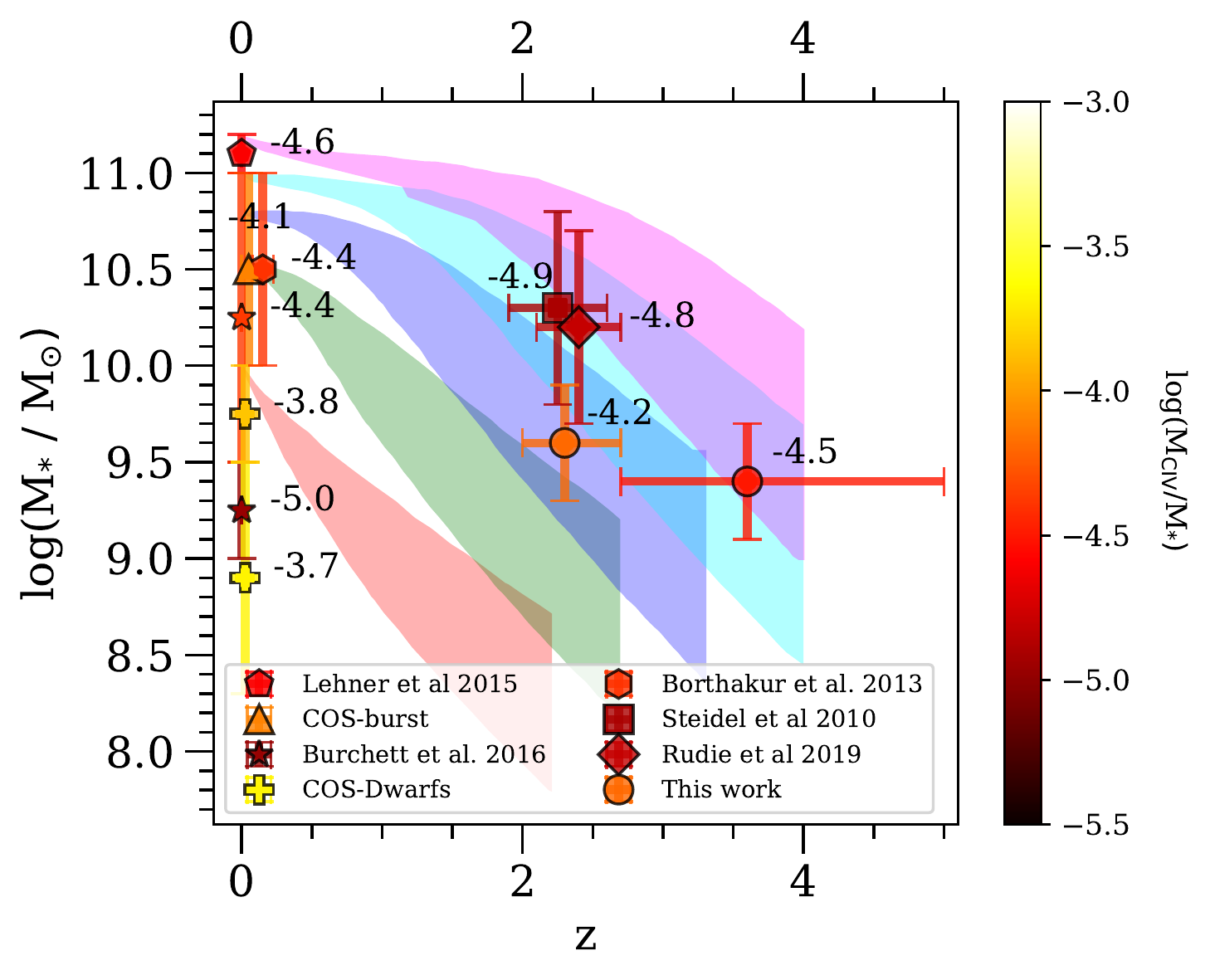}
\figcaption{\label{fig:M_CGM_vs_z_ratio} 
The evolution of $M_{\mathrm{C IV}}/M_{*}$ as a function of redshift. The background colored regions (from \citealt{wiklind2019evolution}) track the history of galaxies with current stellar mass $log(M_{*}/M_{\odot})$ = 10.0, 10.5, 11.0 and 11.2 (red, green, blue, magenta and purple) using the MEAM (multi-epoch abundance matching) selection method (\citealt{moster2012galactic}; \citealt{behroozi2013using}). The spread of the selection slices represents the 1$\sigma$ uncertainty as defined in \cite{behroozi2013using}. The foreground points show the ratio of C IV mass within $R_{vir}$ to the stellar mass of the galaxy. The color of a point indicates the ratio, scaled according to the colorbar on the right. For convenience, the log of the ratio is also written next to the points. The ratios have typical error bars of $\pm$ 0.5 dex due to uncertainties in the hosts stellar masses (thereby, virial radii) and C IV column density.}
\end{figure}

From the standpoint of galaxy evolution, it is interesting to compare the evolution of the CGM with the evolution of the host galaxy. We have combined published data from the literature (\citealt{steidel2010structure}; \citealt{borthakur2013impact}; COS-Dwarfs: \citealt{bordoloi2014cos}; KODIAQ: \citealt{lehner2015evidence}; \citealt{burchett2016deep}; COS-burst: \citealt{heckman2017cos}; \citealt{rudie2019column}) with our new observations to investigate the evolution of C IV mass in the CGM relative to the stellar mass ($\frac{M_{\mathrm{C IV}}}{M_{*}}$), as a function of redshift. For uniformity, we have only considered C IV column density within respective $R_{vir}$ (estimated using $M_{*} - M_{halo}$ relation; \citealt{wechsler2018connection}) and converted that into a C IV mass estimate using equation 4 in \cite{bordoloi2014cos}. The results are summarized in Figure \ref{fig:M_CGM_vs_z_ratio}. 

It can be seen that log$(M_{\mathrm{C IV}}/M_{*}) \sim -4.5$ throughout the evolution of current-day galaxies with $M_{*} > 10^{10}M_{\odot}$. This relatively uniform ratio within $\pm$ 0.5 dex indicates that typically the CGM co-evolves with the stellar mass of the galaxy all the way from $z \sim 4$ to $z \sim 0$. It is possible to explain this co-evolution with the cosmic star formation history. The star formation builds the stellar mass of the galaxy as well as the warm, ionized CGM traced by C IV. This important aspect of the CGM-galaxy co-evolution needs to be explored further in large-scale galaxy evolution simulations.   

Another possible explanation for the co-evolution of warm ionized CGM and the galaxy stellar mass can be the scaling of the virial radius with redshift and galaxy mass. This is especially important since the virial radius is an arbitrary choice to define the CGM. \cite{chen2012unchanging} studied the radial distribution of $L_{*}$ galaxies from $z \sim 2$ to $z \sim 0$ with similar halo mass ($M_{h} \sim 10^{12} M_{\odot}$) in that redshift range. An alternative definition of the extent of the CGM was used in this case to reveal that the radial distribution of column density (normalized to the `CGM radius') did not evolve substantially from $z \sim 2$ to $z \sim 0$. So it is very likely that the radial distribution of column density (normalized to the corresponding virial radius) remains fairly constant and the evolution of the CGM mass is observed due to the scaling of virial radius with redshift. From Figure \ref{fig:M_CGM_vs_z_ratio}, it could also be inferred that any major mergers leading to an increased halo and stellar mass grow the CGM metal mass in the same proportion as $M_{*}$. Further theoretical investigation of the CGM-galaxy co-evolution is necessary to corroborate these growth pathways. 

For current-day low-mass galaxies ($M_{*} < 10^{10} M_{\odot}$), there is a seemingly larger scatter in the log($M_{CIV}/M_{*}$) at $z = 0$, going from $-3.7$ to $-5.0$. The COS-Dwarfs sample (\citealt{bordoloi2014cos}; primarily made of star-forming dwarf galaxies) has a systematically higher ratio suggesting that the CGM of low-mass galaxies could be relatively more enriched in terms of either the metallicity or baryons. 
On the other hand, the low-mass sample (predominantly star-forming dwarfs) in a blind survey for C IV absorbers in QSO spectra \citep{burchett2016deep} shows a systematically lower ratio. However, one should note that the C IV column densities reported in \cite{burchett2016deep} refer only to high impact parameters ($> 0.3 R_{vir}$), which may substantially underestimate the total C IV mass. Therefore, further observations at a range of imapct parameters would be required to asses whether the $M_{\rm C IV}/M_*$ ratio is indeed higher for these low-mass galaxies. 

\subsection{Existence of Inflows?}\label{subsec:inflows}
The cold (T $< 10^{4}$ K) phase is almost exclusively traced by low-ions \citep{tumlinson2017circumgalactic}. If the relative proportions of warm (traced by C IV, Si IV) and cold (traced by Fe II, Si II) phases were the same for red and blue wings, we would expect to see a blue asymmetry in the low ions that resembles that of the high ions ($\sim$ 0.45 dex, see Table \ref{tab:Asymmetry}), but instead the asymmetry is only $\sim$ 0.18 dex. We interpreted this relatively stronger blue asymmetry in the high ions to be due to the outflows being more enriched in high ions. While this is our favorite explanation, another plausible explanation is the existence of cold inflows which make the red wings of low ions stronger compared to the red wings of high ions, thus leading to a lower `blue asymmetry' in low ions. In this section, we briefly explore this possibility. 


Under the aforementioned scenario, the inflow contribution of low ions can be evaluated as the fraction of red-wing column density of low ions that is leading to their relatively stronger red wings compared to the high-ions. 
From Table \ref{tab:Asymmetry}, we can estimate that  $N_{SiII} \sim 10^{13.2}$ cm$^{-2}$ and $N_{FeII} \sim 10^{13.4}$ cm$^{-2}$ are contributed by the inflows moving at $v \sim$ 150 km s$^{-1}$. This column density is consistent with the feedback simulations in \cite{shen2013circumgalactic} at $z \sim 3$. 

Since the characteristics of inflows at high-z are not well understood observationally, we rely on existing inflow models to infer a rough estimate of the inflow mass. Assuming a pristine inflow of 0.03 solar metallicity (\citealt{fumagalli2011absorption, glidden2016predominantly}) and an ionization fraction for Si II of $\sim$50\% for the cold phase (see Fig. 5 in \citealt{shen2013circumgalactic}), we derive $N_{H, inflow} \sim 10^{19.8}$ cm$^{-2}$. We get a similar estimate with Fe II. However, we note a caveat that, although we have considered a pristine metallicity of 0.03 solar, it is not clear how the circulation of metals enrich the inflows by the time they reach the galactic disk. In addition, the metal-enriched recycling flows also manifest itself as inflowing gas and could dominate the observational signature due to their high metal content.

Assuming a constant average density in the accretion stream and a typical area covering fraction of the inflow of $\sim 5\%$ (mostly concentrated along the direction of rotation of the galaxy; \citealt{fumagalli2011absorption, goerdt2012detectability}), we can estimate the inflowing mass as: 

\begin{equation}
\mathrm{M_{in}} = 10^{9}\mathrm{M_{\odot}}\left( \frac{R_{vir}}{\mathrm{50 kpc}}\right)^{2}\left( \frac{N_{H,in}}{10^{19.8} \mathrm{cm^{-2}}} \right)\left( \frac{f_{area}}{0.05} \right)\left( \frac{0.03}{Z/Z_{\odot}} \right)
\end{equation}

Note that we have considered $R_{vir}$ as the extent of the inflow instead of $2 \times R_{vir}$ since we are assuming a constant average density for the inflow stream. Thus, we get an inflowing mass of $\sim 10^{9}M_{\odot}$ within $R_{vir}$, which amounts to $\sim 60\%$ of the diffuse warm CGM traced by C IV that we calculated in section \ref{subsec:CGM_mass_redshift}. Note that this is a lower limit since we have only considered the high-velocity inflows ($v_{los} > 100$ km s$^{-1}$). This is an order-of-magnitude estimate owing to uncertainties in metallicity and covering fraction. 



The mass accretion flow rate can be approximated by considering the inflow motion as a free fall in the halo. Various theoretical models predict large radial streams for inflows until the inner few percents of the virial radius (eg: \citealt{fumagalli2011absorption}). Thus, the inflow timescale can be well approximated as free-fall timescale of the halo. From the typical stellar masses of GRBs at $z > 2$ we estimated a halo mass of $10^{11.2}$ M$_{\odot}$ (section \ref{subsec:Outflows_metal_enrcih} and Appendix \ref{appendix:LOS}). The free-fall timescale for such a halo is 500 Myr. This gives an overall inflow rate estimate of $\dot{M}_{in}$ = $M_{in}/t_{ff}$ = 2 M$_{\odot}$ yr$^{-1}$.    

How can this modest gas inflow rate drive a typical high-z GRB host SFR of the order of 10 M$_{\odot}$ yr$^{-1}$ ? 
It should be noted that the rate evaluated above is over a 500 Myr (free-fall) timescale, while the UV-derived star formation rates are fairly instantaneous in comparison ($<$ 100 Myr). 
GRBs typically take place in a transient (age $\sim 10-100$ Myr, see \citealt{erb2006stellar} and \citealt{levesque2010host}) high-SFR phase of its host. Therefore, a lower and steady cold gas accretion may be sufficient to support the typical long-term SFR history of a GRB host galaxy. In addition, large reservoirs of cold gas (from prior accretion) seem to be already present in these low-mass galaxies due to an order of magnitude lower star formation efficiency at $z > 4$ compared to $z < 3$ owing to the low metallicity \citep{reddy2012characteristic}. 

While this is not a direct observational evidence of cold inflows, these calculations provide an order-of-magnitude insight into the inflow rates of the low-mass galaxies at $z > 2$. Constraining the weak inflows will require high-quality spectra over a much larger sample. 



\subsection{Origin of O VI and N V: CGM vs Circumburst?}
The cases of O VI and N V in the CGM-GRB sample are interesting because of two key reasons: they trace similar regions in the temperature-density phase space \citep[$T \sim$ 10$^{5.5}$ K, $n$ $\sim$ 10$^{-4.5}$ cm$^{-3}$;][]{tumlinson2017circumgalactic}, but their kinematics appear to be distinct from each other. While O VI, like C IV and Si IV, shows a strong blue asymmetry of $\sim$ 0.6 dex, there is no such asymmetry in N V. The overall N V kinematics more closely resemble the low-ion kinematics. Despite the small number of sightlines (13) with O VI in the bandpass and often lower SNR in O VI due to lower sensitivity and Lyman-alpha forest, the kinematic differences are significant. It is important to understand the origin of N V and its distinct nature compared to O VI, in order to study the warm-hot CGM phase (T $\sim$ $10^{5.5}$K), as well the nature of photo- or collisional-ionized gas harboring N V.

In various low-$z$ CGM surveys using QSO sightlines such as COS-Halos \citep{werk2016cos}, N V is rare. In a similar survey at high $z$ (KODIAQ; \citealt{lehner2014galactic}), N V is more prevalent, but still with a detection rate of only $\sim$ 50\% and with a typical column density log(N(N V)/cm$^{-2}$) $\sim$ 13.6 (it is $\sim$ 14.0 in the CGM-GRB sample). On the other hand, O VI is ubiquitous in both high- and low-z surveys. Several sightlines through the Milky Way galactic halo and disk show N V absorption (\citealt{savage1997absorption, fox2015probing, karim2018probing}), but only $\sim$ 10\% show column densities N(N V) $\gtrsim$ $10^{14}$ cm$^{-2}$.       

The nature of the excess N V absorption in the GRB spectra is not well understood. Possible explanations for the N V absorption in the GRB spectra include: 1. photoionization of the circumburst medium within r $\sim$ 10 pc (\citealt{prochaska2008survey, fox2008high}), 2. recombination of the promptly ionized nitrogen (all electrons stripped) to N V within 10 pc \citep{heintz2018highly}, and 3. N V in the CGM (\citealt{heintz2018highly, fox2008high}). The kinematic similarity of N V absorption with the UV-pumped fine-structure lines (such as C II*, Si II*), which are associated with absorbers within a few hundreds of parsecs from the GRB \citep{vreeswijk2013time}, is considered an indicator for the circumburst origin of N V. 




For the GRB spectra presented here, the N V absorption within $\pm$400 km s$^{-1}$ typically comes from an ensemble of kinematically distinct absorbing components. Therefore, it is difficult to imagine that the absorption is produced solely by the circumburst medium. While most of the absorption components of N V (especially the strongest components) have a counterpart that is kinematically associated with fine-structure transitions (C II*, Si II*, as also seen in \cite{prochaska2008resolving}), there are also weaker components that are not associated with an excited transition within $\pm$ 30 km s$^{-1}$ (Fig.\ \ref{fig:000926}$-$\ref{fig:170202A}).  Therefore, we suggest that the low-$\vert v \vert$ N V absorption seen in the GRB spectra comes primarily from the highly ionized / recombining gas associated with the circumburst medium while the weaker absorption at higher $\vert v \vert$ comes primarily from the warm gas associated with the CGM. This would explain the relatively higher N V detection rate and column density along GRB sightlines. 

The typical O VI column densities derived in this survey are comparable to the typical O VI column densities in the high-z QSO sightline surveys \citep{lehner2014galactic} and the CGM of local star-forming galaxies \citep{tumlinson2011large}. While O VI absorption could also have a circumburst component since the recombination rates of N V and O VI are similar \citep{heintz2018highly}, the kinematics of N V are distinct. In fact, the O VI kinematics resemble the kinematics of C IV and Si IV more than those of N V. This strongly suggests that the ISM and CGM are the dominant contributors to the O VI column density, although a minor contribution from the circumburst medium cannot be ruled out. Detailed photoionization models addressing the circumburst O VI absorption are required to quantitatively ascertain this.

 As shown in \cite{prochaska2006dissecting} and \cite{chen2007absence}, C IV and Si IV absorption have no association with the circumburst medium up to several tens of parsecs. The ionizing photon flux from the GRB strips electrons from these species (with ionization potentials of 64.5 and 45.1 eV, respectively) and their recombination timescales are of the order of 1 year for a typical H II region \citep{chen2007absence}. Therefore, the C IV and Si IV lines primarily trace the typical ISM and CGM of the GRB host and not the circumburst medium. 

%
High-resolution rest-frame UV spectra of the afterglows at multiple epochs are required to probe their explosion sites and environments, constrain the ionization models, and thus understand the origin of peculiar N V absorption in the GRB afterglow spectra.

\section{Summary and Conclusions}
In this paper, we collected a large sample of medium resolution ($\delta v$ $<$ 50 km s$^{-1}$) and high SNR (typical $\sim 10$) spectra of 27 GRB afterglows in the redshift range $z \sim 2-6$ to systematically probe the kinematics and physical properties of the CGM at high redshifts using the absorption features of high- and low-ion species in the spectra. A simple toy model was constructed to aid this analysis through geometric and kinematic modeling of the CGM and the outflows. We further estimated the CGM mass and mass outflow rates in two different redshift regimes ($z1: 2-2.7$ and $z2: 2.7-5$). Finally, combining the results of past studies and this work, we investigated the CGM-galaxy co-evolution as a function of redshift. The key conclusions of this study are summarized as follows:

\noindent
1. Detection fractions: By inspecting the median plots of the rest-frame spectra for each species, three clear kinematic regions were identified: central region ($|v| < 100$ km s$^{-1}$), blue wing ($v < -100$ km s$^{-1}$), and red wing ($v > 100$ km s$^{-1}$). The high-ion species (C IV and Si IV) were found to have substantially higher detection fractions in the blue wings compared to the red wings (difference $\sim 30\%$). On the other hand, the low-ion species (Fe II, Si II, Al III) had a marginal excess in detection fraction in the blue wings (difference $\sim 10\%$). This shows that the blue wing component is substantially dominated by the high-ions compared to the red wings (Table \ref{tab:Det_fractions}). We interpret this as an evidence for warm phase (T $\sim$ $10^{4.5-5.5}$K) outflows. 

\noindent
2. Kinematic asymmetry: By stacking the spectra of individual absorption lines (Fig. \ref{fig:Median_plot}), a significant absorption excess was observed in the blue wings compared to the red wings, especially for the high-ions (C IV, Si IV). We further quantified this using multi-component Voigt-profile fitting and converting the Voigt profiles to integrated column densities in 100 km s$^{-1}$ windows (from $-$400 km s$^{-1}$ to $+$400 km s$^{-1}$). The blue-red kinematic asymmetry was stronger in high-ion species compared to the low-ion species by 0.2-0,3 dex in column density (Table \ref{tab:Asymmetry}). This key difference was further highlighted in the high- to low-ion line ratios (Fig. \ref{fig:Line_ratios}). This is a strong signature of warm outflows in the GRB hosts. 

\noindent
3. Kinematic simulations: To resolve the relative contributions of the CGM and ISM in terms of absorption line kinematics, we modeled a typical GRB host galaxy and the CGM around it using a simple toy model. The results from the toy model showed that the absorption produced by the host ISM is typically found within the central region ($\pm$100 km s$^{-1}$ in terms of line-of-sight velocity), while the CGM absorption dominates the absorption in the blue and red wings beyond $\pm 100$ km s$^{-1}$. An outflow component was further added to the virialized CGM to simulate star-formation driven outflows. We made various simplifying assumptions for the halo mass (M$_{h} \sim 10^{11.2}\mathrm{M_{\odot}}$), the CGM density profile ($r^{-2}$), the CGM/ISM metallicity, and the outflow itself (ballistic, no halo drag). By comparing the predictions of the toy models with the observed column density profiles, we estimated the physical properties of the CGM, including the outflow launch speed (300 km s$^{-1}$), the fraction of CGM clouds that are outflowing (25\%), and the CGM mass ($\sim$ 10$^{9.8} \mathrm{M_{\odot}}$). 

\noindent
4. CGM mass estimates: In \S\ \ref{subsec:Col_density_to_Mass}, we used the geometric mean of the integrated column density and a conservative extrapolation to estimate the CGM contribution in the central $\pm$ 100 km s$^{-1}$. From this, we deduced the mass of the warm-phase CGM traced by C IV ($\mathrm{M_{CGM}} \sim 10^{9.2} \mathrm{M_{\odot}}$). The CGM mass estimates from the toy models (\S\ \ref{sec:LOS_results}) and column density profiles (\S\ \ref{subsec:Col_density_to_Mass}) are complementary in nature, strengthening a CGM mass estimate of $\sim 10^{9.2-9.8} M_{\odot}$.
These estimates show that a) the mass contained in the CGM ($\sim$2 $\times$ $R_{vir}$) is comparable to the typical stellar mass of GRB hosts at $z > 2$ ($\mathrm{M_{*}} \sim  10^{9-10} \mathrm{M_{\odot}}$) and b) the CGM is already a significant component of the galaxy ecosystem for GRB host galaxies at $z > 2$.  

\noindent
5. Evolution of the CGM mass: The CGM-GRB sample was divided into two redshift bins each spanning $\sim 1$ Gyr ($z1: 2-2.7$ and $z2: 2.7-5$). Their CGM C IV masses were estimated to be $\mathrm{M_{CIV, z1} = 10^{5.6} M_{\odot}}$ and $\mathrm{M_{CIV, z2} = 10^{5.1} M_{\odot}}$ (Figs. \ref{fig:CGM_Mass_z1} and \ref{fig:CGM_Mass_z2}). 
\noindent
A comparison with the COS-Dwarfs survey for similar low-mass galaxies at low redshifts ($z < 0.3$) shows that the low-z galaxies are slightly more enriched by a factor of 2 relative to galaxies of similar masses at high z. This shows that the dwarf galaxies had metal-enriched environments as early as $z \sim 3-5$ and thus, most likely played a major role in the metal enrichment of the universe due to their shallow potential wells. 

\noindent
6. Outflow mass: The optimal toy model indicates that the fraction of outflowing clouds in outflow state is $f_{out} = 25\%$ and the outflow launch velocity at 2 kpc is $v_{launch} = 300$ km s$^{-1}$. The (warm) outflow mass in the two redshift ranges was estimated to be $M_{out,z1} \sim 10^{8.6} M_{\odot}$ and $M_{out,z2} \sim 10^{8.1} M_{\odot}$. Assuming no halo drag or outflow acceleration mechanisms, as much as 80\% of this outflowing gas has $v > v_{esc}$ at launch. Given a typical halo mass of $M_{h} \sim 10^{11.2} M_{\odot}$, the crossing time for $2 \times R_{CGM}$ is $\sim 0.5$ Gyr. This in turn implies that the CGM mass in the warm gas phase grows by $\sim 5\%$ ($f_{out} \times 20\%$) in 0.5 Gyr by retaining the slower outflows in the CGM. 

\noindent
7. Mass loading factor: We further estimated that the warm-phase mass outflow rates at a radius of 20 kpc radius is $\dot{M}_{out, z1} \sim 0.8~M_{\odot}$ yr$^{-1}$ and $\dot{M}_{out, z2} \sim 0.35~M_{\odot}$ yr$^{-1}$. The median SFR of GRB hosts at $z: 2-5$ is $\sim 10 M_{\odot}$ yr$^{-1}$. Therefore, the warm-phase mass loading factors in the two redshift bins are estimated to be $\eta_{z1, 20kpc} = 0.084$ and $\eta_{z2, 20kpc} = 0.035$. These mass-loading factors suggest that the outflows and thereby, the CGM metal enrichment, for these low-mass galaxies are exclusively driven by star formation. While these mass-loading factors are low, it is important to note that this only includes warm-phase outflows and thus highlights the need for a multi-probe approach to trace the outflows in various phases to produce a complete picture of the CGM outflows.

\noindent
8. CGM-galaxy co-evolution: 
We compared the evolution of C IV mass in the CGM with the stellar mass evolution as a function of redshift (Fig. \ref{fig:M_CGM_vs_z_ratio}). We find that log$(M_{\mathrm{C IV}}/M_{*}) \sim -4.5$ within $\pm 0.5$ dex throughout the evolution history of current-day galaxies with $M_{*} > 10^{10}M_{\odot}$. This relatively uniform ratio indicates that the CGM metal mass co-evolves with the stellar mass of the galaxy all the way from $z \sim 4$ to $z \sim 0$, for the progenitors of local galaxies with $M_{*} > 10^{10}M_{\odot}$. Therefore, the CGM-galaxy co-evolution is an important piece of the galaxy growth puzzle which needs to be explored further in large-scale galaxy simulations. 


In this paper, we systematically probed the CGM of high-z, low-mass, star-forming galaxies for the first time using a GRB host sample. We used typical stellar and dark matter halo masses to derive the CGM masses and outflow rates. However, the GRB host population is quite diverse and spans two order of magnitudes in stellar mass and SFR at $z >$ 2. In a future paper, we plan to examine the properties of individual GRB hosts and compare them with the properties of their CGM to help refine the CGM-galaxy connection. Detailed ionization modelling will be a crucial next step to derive better constraints on the physical properties of the CGM.   

\acknowledgments
P.G. was supported by NASA Earth and Space Science Fellowship (ASTRO18F-0085) for this research. S.V. acknowledges support from a Raymond and Beverley Sackler Distinguished Visitor Fellowship and thanks the host institute, the Institute of Astronomy, where this work was concluded. S.V. also acknowledges additional support by the Science and Technology Facilities Council (STFC) and Kavli Institute for Cosmology, Cambridge. This work was partly based on data obtained from the ESO Science Archive Facility. This research has also made use of the Keck Observatory Archive (KOA), which is operated by the W. M. Keck Observatory and the NASA Exoplanet Science Institute (NExScI), under contract with the National Aeronautics and Space Administration (NASA). The authors are grateful to Drs. Andrew Fox and S. Bradley Cenko for their useful comments in the early stages. 

\begin{deluxetable*}{cccccccc}[t]
\tablecaption{List of GRBs in the sample \label{tab:GRB_list}}
\tablehead{
\colhead{GRB} &
\colhead{z$_{GRB}$} &
\colhead{Instrument} & 
\colhead{Resolution} & 
\colhead{Acq.} &
\colhead{SNR} &
\colhead{log N$_\mathrm{H}$} &
\colhead{References} \vspace{-1ex}\\ 
\colhead{} & \colhead{} & \colhead{} & \colhead{} & 
\colhead{Mag$_\mathrm{AB}$} & \colhead{per pixel} & \colhead{ in $cm^{-2}$} & \colhead{} }
\startdata
000926 & 2.0385 & Keck/ESI     & 20,000    & 20.5  & 10	& 21.3 $\pm$ 0.25 & \cite{castro2003keck} \\
021004 & 2.3281 & VLT/UVES  & 40,000    & 18.8  & 6	 & 19.0 $\pm$ 0.2 & \cite{fiore2005flash}\tablenotemark{b} \\
050730 & 3.9672 & VLT/UVES  & 40,000    & 18.0  & 10 & 22.1 $\pm$ 0.1 & \cite{d2007uves}\tablenotemark{b}\\
050820A & 2.6137 & VLT/UVES & 40,000    & 21.0  & 12 &  21.1 $\pm$ 0.1 & \cite{prochaska2007interstellar}\tablenotemark{b}\\
050922C & 2.1996 & VLT/UVES & 45,000	& 19.5  & 10 &	21.55 $\pm$ 0.1 & \cite{prochaska2008survey}\tablenotemark{b} \\
060607A & 3.0738 & VLT/UVES & 55,000	& 16.5  & 30 &	 16.95 $\pm$ 0.03 & \cite{prochaska2008survey}\tablenotemark{b} \\
071031 & 2.6912 & VLT/UVES  & 55,000    & 18.5 & 10	&	22.15 $\pm$ 0.05 & \cite{fox2008high} \\
080310 & 2.4274 & VLT/UVES  & 55,000    & 17.5  & 30	&  18.7 $\pm$ 0.1 & \cite{fox2008high} \\
080804 & 2.205 & VLT/UVES   & 55,000    & 19.5  & 10	& 21.3 $\pm$ 0.1 & \cite{fynbo2009low}\\
080810 & 3.351 & Keck/HIRES & 50,000    & 17.0  & 30	&  17.5 $\pm$ 0.15 & \cite{page2009multiwavelength} \\
090926A & 2.106 & VLT/X-shooter (UVB) & 6000      & 17.9  & 20	&  21.73 $\pm$ 0.07 & \cite{d2010vlt} \\
100219A & 4.665 & VLT/X-shooter (VIS) & 10,000    & 22.2  & 4	& 21.13 $\pm$ 0.12 & \cite{thone2012grb} \\
111008A & 4.989\tablenotemark{c} & VLT/X-shooter (UVB) & 6500	&  21.0 & 10 & 22.3 $\pm$ 0.06  & \cite{sparre2014metallicity}\\
120327A & 2.813 & \multicolumn{1}{p{3.5cm}}{\centering VLT/X-shooter (UVB) \\ X-shooter (VIS)} & \multicolumn{1}{p{1.5cm}}{\centering 6250 \\ 10000} & 18.8  & 30	&  22.01 $\pm$ 0.09 & \cite{d2014vlt} \\
120815A & 2.358 & \multicolumn{1}{p{3.5cm}}{\centering VLT/X-shooter (UVB) \\ X-shooter (VIS)} & \multicolumn{1}{p{1.5cm}}{\centering 6000 \\ 11000} & 18.9  & 12	& 21.95 $\pm$ 0.1 & \cite{kruhler2013molecular} \\
120909A & 3.929 & VLT/X-shooter (VIS) & 10000 & 21.0 & 9	&  21.20 $\pm$ 0.10 & \cite{cucchiara2015unveiling} \\
121024A & 2.298\tablenotemark{c} & \multicolumn{1}{p{3.5cm}}{\centering VLT/X-shooter (UVB) \\ X-shooter (VIS)} & \multicolumn{1}{p{1.5cm}}{\centering 6000 \\ 12000} & 20.0  & 15 &  21.50 $\pm$ 0.10 & \cite{friis2015warm} \\
130408A & 3.757 & \multicolumn{1}{p{3.5cm}}{\centering VLT/X-shooter (UVB) \\ X-shooter (VIS)} & \multicolumn{1}{p{1.5cm}}{\centering 6000 \\ 12000} & 20.0  & 20 & 21.70 $\pm$ 0.10 & \cite{cucchiara2015unveiling} \\
130606A	& 5.911 & \multicolumn{1}{p{3.5cm}}{\centering VLT/X-shooter (VIS) \\ X-shooter (NIR)} & \multicolumn{1}{p{1.5cm}}{\centering 8000 \\ 6500} & 19.0  & 10 &  19.93 $\pm$ 0.2 & \cite{hartoog2015vlt} \\ 
130610A & 2.091 & VLT/UVES   & 40000      & 20.9  & 6 & $-$ & \cite{2013GCN.14848....1S} \\ 
141028A & 2.333 & \multicolumn{1}{p{3.5cm}}{\centering VLT/X-shooter (UVB) \\ X-shooter (VIS)} & \multicolumn{1}{p{1.5cm}}{\centering 5600 \\ 9600}  & 20.0  & 8 & 20.60 $\pm$ 0.15 & \cite{wiseman2017evolution} \\ 
141109A & 2.993\tablenotemark{c}  & \multicolumn{1}{p{3.5cm}}{\centering VLT/X-shooter (UVB) \\ X-shooter (VIS)} & \multicolumn{1}{p{1.5cm}}{\centering 6000 \\ 10000} & 19.2  & 5 & 22.10 $\pm$ 0.10 & \cite{heintz2018highly}\\ 
151021A & 2.329 & \multicolumn{1}{p{3.5cm}}{\centering VLT/X-shooter (UVB) \\ X-shooter (VIS)} & \multicolumn{1}{p{1.5cm}}{\centering 6000 \\ 10000} & 18.2  & 5 & 22.3 $\pm$ 0.2 & \cite{heintz2018highly} \\ 
151027B & 4.0633 & VLT/X-shooter (VIS)  & 9000 & 20.5  & 5 &  20.5 $\pm$ 0.2 & \cite{heintz2018highly} \\ 
160203A & 3.518 & VLT/X-shooter (VIS)  & 12000 & 18.0  & 5 &  21.75 $\pm$ 0.10 & \cite{heintz2018highly} \\
161023A & 2.709 & \multicolumn{1}{p{3.5cm}}{\centering VLT/X-shooter (UVB) \\ X-shooter (VIS)} & \multicolumn{1}{p{1.5cm}}{\centering 6000 \\ 10000} & 17.5  & 40 & 20.96 $\pm$ 0.05 & \cite{heintz2018highly} \\
170202A & 3.645 & \multicolumn{1}{p{3.5cm}}{\centering VLT/X-shooter (UVB) \\ X-shooter (VIS)} & \multicolumn{1}{p{1.5cm}}{\centering 6300 \\ 10500} & 20.8  & 8 & 21.55 $\pm$ 0.10 & \cite{selsing2018x} \\
\enddata
\tablenotetext{a}{Typical SNR in the bandpass.\vspace{0.5ex}} \tablenotetext{b}{Also previously analyzed in \cite{fox2008high}}
\tablenotetext{c}{Ni II* transition has been used in addition to Si II* and C II* for defining the redshift.}
\end{deluxetable*}

\begin{deluxetable*}{ccccccccccc}[t]
\tablecaption{Fit parameters of the spectra \label{tab:fit_profile}}
\tablehead{
\colhead{GRB} &
\colhead{z$_{GRB}$} &
\colhead{line} & 
\colhead{Component No.} & 
\colhead{log(N)} & 
\colhead{$\sigma_{\mathrm{log(N)}}$} &
\colhead{b (km s$^{-1}$)} &
\colhead{$\sigma_{\mathrm{b}}$} &
\colhead{center (km s$^{-1}$)} &
\colhead{$\sigma_{\mathrm{center}}$} &
\colhead{Flag \tablenotemark{a}}
}
\startdata
000926 & 2.0385 & C IV 1548 &	1 &	13.28 &	0.18 &	7 &	4 &	-368 &	4 &	0 \\
000926 & 2.0385 & C IV 1548 & 2 & 14.08 & 0.05	 & 29	& 6 & -332 & 19	& 0 \\
000926 & 2.0385 & C IV 1548 & 3 & 14.07 & 0.07 & 29 & 5 & -239 & 5 & 0 \\
000926 & 2.0385 & C IV 1548 & 4 & 14.35 & 0.21 & 20 & 5 & -181 & 19 & 0 \\
000926 & 2.0385 & C IV 1548 & 5 & 14.55 & 0.21 & 18 & 5 & -136 & 19 & 0\\
\enddata
\tablenotetext{a}{Flag of 0 indicates convergent fit, flag of 0 indicates saturated/degenerate fit.\vspace{0.5ex}}
\tablecomments{Table 4 is published in its entirety in machine-readable format. A small portion is shown here as an example.}
\end{deluxetable*}


\appendix 
\section{Voigt profile fitting using MCMC}
\label{sec: Voigt_MCMC}

As described in section \ref{sec:spec_analysis}, we developed an MCMC-based code for multi-component Voigt profile fitting. There are four key components in this Bayesian approach to Voigt-profile fitting: definition of Voigt function, prior distribution, log-likelihood of a set of parameters, and posterior function.\\

\noindent
\textbf{Definition of Voigt function}: In this analysis, we convert the spectrum in rest frame for each transition under consideration. The line properties are taken from \cite{morton2003atomic}, \cite{d2014vlt}, and the NIST database \footnote{\url{https://physics.nist.gov/PhysRefData/ASD/lines\_form.html}}. For given properties of a transition (wavelength $\lambda_{0}$, oscillator strength $f_{osc}$, and damping constant $\gamma$), the normalized flux at a wavelength $\lambda$ is given by:

\begin{equation}
F(v)_{N, b, \lambda_{0}} = exp[-\tau(\lambda)]
\end{equation}

where $\tau(\lambda)$ is the optical depth as a function of wavelength in rest frame and $\lambda_{0}$ is the line center. The optical depth is further parametrized as: 

\begin{equation}
\tau(\lambda)_{N, b, \lambda_{0}} = \frac{Nf_{osc}\sigma_{0}\lambda}{b} \:\: V(\lambda)_{b, \lambda_{0}} \:\: , with \:\: \sigma_{0} = \frac{\sqrt{\pi}e^{2}}{m_{e}c}
\end{equation}

where $V(\lambda)$ is the Voigt function defined by a convolution of the Doppler-broadened Gaussian profile (depends on $N, b, f_{osc}$) and pressure-broadened Lorentzian profile (depends on the damping coefficient, $\gamma$). For the functional form of the Voigt profile, see eq. 6 in \cite{petitjean1995qso}. The Voigt function in our code is evaluated using the {\fontfamily{qcr}\selectfont Voigt1D} routine in {\fontfamily{qcr}\selectfont astropy} implemented using a high accuracy analytical approximation described in \cite{mclean1994implementation}). Since the continuum is separately fitted and normalized (see \ref{sec:spec_analysis}), we do not add a continuum model in this treatment. The modeled $F(v)$ is further filtered using a Gaussian kernel of width equal to the line spread function of the instrument used to obtain the spectrum. \\

\noindent
\textbf{Posterior distribution function:}
The posterior distribution function represents the probability distribution of a set of model parameters given the observed data. In this case, the model of the optical depth as a function of velocity ($\tau(v)$) is defined as a sum of Voigt profiles of multiple absorbing components for a given transition. The model parameters are denoted by $\Theta = [N_{i}, b_{i}, v_{0,i}]_{i = 1,2,..,n}$ where $n$ is the number of absorbing components to be fitted to the line transition. According to Bayes' theorem,
\vspace{-2ex}
\begin{equation}
p(\Theta|D) = \frac{p(D|\Theta)p(\Theta)}{p(D)} 
\end{equation}

where $D$ is the observed data. The distributions $p(\Theta|D)$, $p((D|\Theta)$, $p(\Theta)$ are the posterior distribution, likelihood distribution, and prior distribution. The normalization $p(D)$ is independent of model parameters here and therefore, the parameter estimation can be achieved by maximizing the product of likelihood and prior distributions. We use a python package called {\fontfamily{qcr}\selectfont emcee} which implements the ensemble sampling algorithm as described in \cite{foreman2013emcee} to efficiently sample the posterior distribution in a high-dimensionality and often correlated parameter space in this particular problem.\\

\noindent
\textbf{Definition of likelihood function:}
To evaluate the likelihood of a particular set of parameters, a noise model is required. We use a Gaussian noise model assuming large photon number limit. This model is suitable for fitting the non-saturated absorption components. Although it may not serve as the best approximation for saturated parts of the spectrum, it is well suited for fitting the wings (due to large number of photons) of such saturated components. The likelihood function is defined as follows:

\begin{equation}
p(D|\Theta) = \frac{1}{\sqrt{2\pi\epsilon_{i}^{2}}} exp \left[ -\frac{(D_{i}-F_{i})^{2}}{2\epsilon_{i}^{2}} \right]
\end{equation}

where $D_{i}$, $\epsilon_{i}$, $F_{i}$ are the observed flux, error in the flux, and model flux at a certain wavelength $i$.\\

\noindent
\textbf{Prior distribution:}
 A uniform prior distribution is defined over the range of possible values for each parameter to ensure its non-informative nature over this range (eg: log(N$_{Si IV}$) from 9 to 18 and b-parameter from 5 to 70 km s$^{-1}$). The number of absorbing components to be fitted to a line transition are manually selected by validating their presence and strength in another transition of the same species or a doublet/triplet system and/or another transition of similar ionization state. This approach is particularly helpful for saturated lines where the number of absorbing components can be constrained by identifying the components in weaker transitions of the same species or the same class (high-/low-ion). The priors for line centers are manually provided as 50-90 \% of the extent of the particular component in velocity. 

The region very close to sky lines is not considered while defining components. In addition, we mark all the lower-z intervening systems identified in the literature for each GRB. This way, any undesirable contamination is avoided. The initial guesses provided to the algorithm are randomly sampled from the range defined in the prior distribution. In case of doublets or multiplets, the spectra are fit simultaneously. The likelihood distribution for these systems is defined as the product of likelihood distributions of individual transitions. This ensures that the resulting posterior distribution represents the complete doublet/multiplet system.
\vspace{1ex}

\noindent
\textbf{Parameter estimation:}
The number of steps required to establish a reasonable convergence of the Markov chains is estimated using the autocorrelation time method as described in the emcee package \citep{goodman2010ensemble, foreman2013emcee}.  We increase the number of steps (i.e. length of a Markov chain) in powers of 4 (capped at 10$^5$ $\times$ no. of parameters due to computational limit) until the autocorrelation time converges within 10\% of the previous step. The optimal parameters are obtained by maximizing the marginalized posterior distribution function for each parameter. The 1-$\sigma$ confidence interval is estimated as the range of parameter value that covers the central 68\% of the marginalized posterior distribution function.  

\begin{figure*}
\centering
\includegraphics[width=0.9\textwidth]{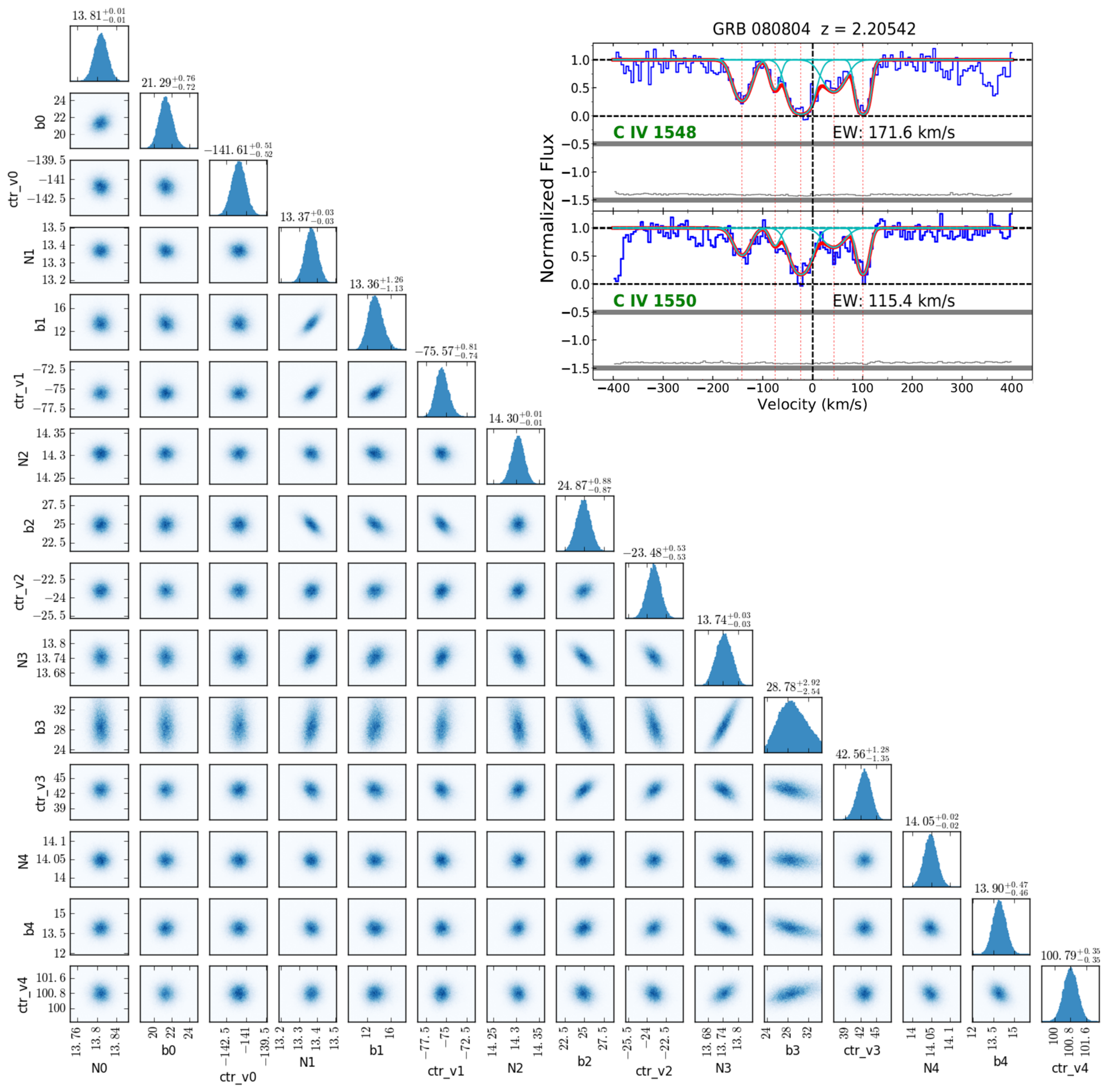}
\figcaption{\label{fig:MCMC_sample} A sample plot to show the parameter estimation using the MCMC-based method. N, b, and ctr\_v indicate the column density, Doppler parameter and central velocity of the respective component. In this example, the C IV doublet is fitted together. The corner plot shows the marginalized posterior distribution function. The 1-$\sigma$ parameter uncertainty is estimated as the range of parameter value that covers the central 68\% of the marginalized posterior distribution function.}
\end{figure*}

\begin{figure*}
\centering
\includegraphics[width=0.9\textwidth]{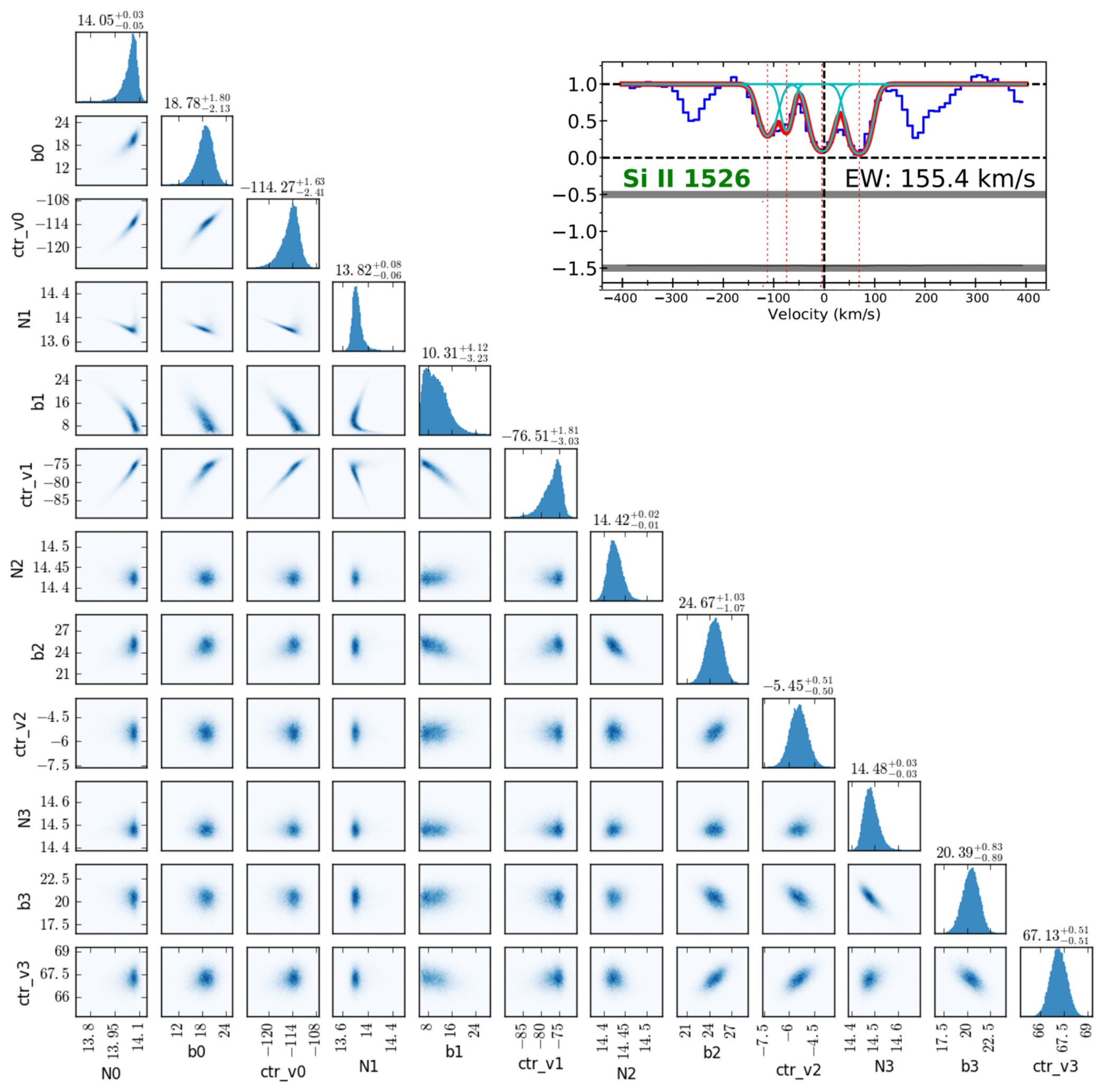}
\figcaption{\label{fig:MCMC_sample_161023A} Same as Fig. \ref{fig:MCMC_sample} for GRB 161023A with an R $\sim$ 8000 spectrum (X-shooter). The corner plot shows the marginalized posterior distribution function for Si II 1526. The 1-$\sigma$ parameter uncertainty is estimated as the range of parameter value that covers the central 68\% of the marginalized posterior distribution function.}
\end{figure*}





\section{Line-of-sight simulation with toy model}\label{appendix:LOS}

\noindent
The GRB sightline through the galaxy ISM and CGM is simulated using a simple toy model to understand the kinematic distinction between the ISM and CGM in a statistical sample and use it to better estimate the physical properties from the observed data. A large number of GRB  explosions are simulated in a representative galaxy with randomly chosen burst locations in the galactic disk and pointing directions. We assume a typical stellar mass of the GRB host galaxy of $\sim 10^{9.3}$ $M_{\odot}$ at z $>$ 2 for these simulations (\citealt{perley2016swift_2}). The halo mass ($M_{h}$) is taken as $10^{11.2}$ $M_{\odot}$ as described in previous simulations in the literature \citep{hopkins2014galaxies, wechsler2018connection}. The virial radius is calculated using a NFW profile for the dark matter distribution and standard cosmological parameters ($\Omega_{m}$ = 0.3, $\Omega_{rad}$ = 0, $\Omega_{\Lambda}$ = 0.7). The virial radius for the simulated galaxy at z $\sim$ 2.5 is 50 kpc. The corresponding virial velocity is $\sim$ 100 km s$^{-1}$. The assumed setup for the simulation is summarized in Table \ref{tab:Param_list}. \vspace{-1ex}

\begin{figure*}
\centering
\includegraphics[width=0.9\textwidth]{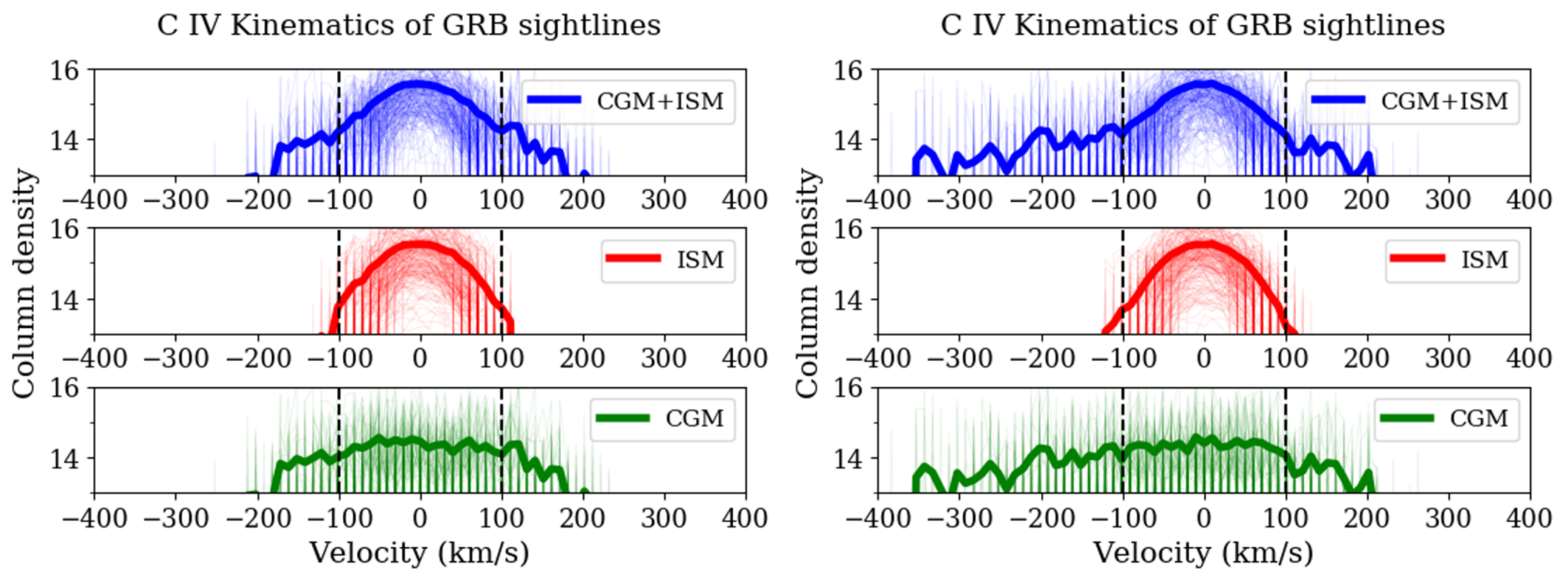}
\figcaption{\label{fig:GRB_sightlines} Overall simulation CIV kinematics with log($M_{CGM}/M_{\odot}$) = 9.8 are decomposed into ISM and CGM components and shown with (right) and without (left) an outflow component. The simulations are run for 200 GRB sightlines, which are shown in faint traces and the average column density profiles are shown in dark traces. The figure on the right has $v_{out}$ = 250 km s$^{-1}$ and $f_{out}$ = 0.25.}
\end{figure*}

\begin{deluxetable}{lccc}[htbp]
\tablecaption{Typical values of the toy model parameters, z $\sim$ 2.5 \label{tab:Param_list}}
\tablecolumns{4}
\tablenum{5}
\tablewidth{0pt}
\tablehead{
\colhead{Parameter} &
\colhead{Symbol} &
\colhead{Value} &
\colhead{References (as applicable)} \\
}
\startdata
Galaxy stellar mass &  $log(M_{*})$ & 9.3  & \cite{perley2016swift_2}\\
Galaxy halo mass & $log(M_{h})$  & 11.2
& \cite{hopkins2014galaxies}, \cite{wechsler2018connection}  \\
Mass in the ISM\tablenotemark{a} & $log(M_{ISM})$  & 9.3  &   \\
Mass in the CGM\tablenotemark{a,b} & $log(M_{CGM})$  & 9.3, {\bf 9.8}, 10.3  &   \\
\hline\\
Galaxy half-mass radius & $R_{e}$ & 2 kpc & \cite{wainwright2007morphological}, \cite{blanchard2016offset}\\
Galaxy radius & $r_{galaxy}$ & 4 kpc & \\
Galaxy height & $h_{galaxy}$ & 3 kpc & \\
Radial range of GRB location & $R_{GRB}$ & $0.4 - 4$ kpc & \cite{blanchard2016offset}\\
Max. height of GRB location  & $h_{max}$ & $\pm$ 1 kpc & \\
Virial radius  & $R_{vir}$ & 50 kpc  & \\
CGM cloud radius & $R_{cloud}$ & 0.4 kpc & \\
Volume filling fraction & $f_{vol}$ & 0.1 & \cite{stocke2013characterizing}, \cite{werk2016cos} \\
Simulation region  & $-$ & 2$\times R_{vir}$   & \cite{shen2013circumgalactic}\\
\hline\\
Flat rotation velocity & $v_{flat}$ & 100 km s$^{-1}$ & \cite{arabsalmani2017mass}\\
ISM dispersion velocity & $\sigma_{v, ISM}$ & 50 km s$^{-1}$ & \cite{lan2018circumgalactic}\\
CGM dispersion velocity & $\sigma_{v, CGM}$  & 100 km s$^{-1}$ & \cite{lan2018circumgalactic}\\
Outflow launch velocity\tablenotemark{b} & $v_{out}$ & 200, 250, {\bf 300} km s$^{-1}$ & \\
Outflow launching radius & $R_{launch}$ & 2 kpc & \\
Outflow fraction\tablenotemark{b} & $f_{out}$ & {\bf 0.25}, 0.5, 0.75 & \cite{ford2014tracing}, \cite{muratov2015gusty} \\
\hline
Number of sightlines & $-$ & 200 & \
\enddata
\tablenotetext{}{The boldfaced values are for the optimal toy model which best explains the observed C IV column density profile}
\tablenotetext{a}{The mass of warm ionized CGM traced by the C IV ion (T $\sim$ $10^{4.5-5.5}$K)}
\tablenotetext{b}{These parameters are modified to obtain a model that closely matches the observations}
\end{deluxetable}
\vspace{5mm}


\noindent
\textbf{Geometrical Setup:} The galaxy is modeled as a cylinder with radius $r_{galaxy}$ and height $h_{galaxy}$. This volume is treated as the extent of the ISM with an exponential density distribution such that the total enclosed mass equals $M_{ISM}$ and half of the mass is enclosed within the half-light radius ($R_{e}$). For simplicity, we assume that the mass of the ISM traced by high-ion species ($M_{ISM}$) is equal to the stellar mass of the galaxy ($M_{\star}$). $R_{e}$ is assumed to be 2 kpc following previous observations of GRB hosts \citep{wainwright2007morphological, blanchard2016offset}. The CGM is defined as a sphere of radius $R_{CGM}$ surrounding the galaxy. This sphere is populated with a uniform probability distribution by clouds of radius $R_{cloud}$ and each cloud has a gas number density that is given as: 
\begin{equation} \label{eqn:n_cloud}
    n_{cloud} = n_{0}\left( \frac{r}{r_{0}}\right)^{-2}
\end{equation}
where $r$ is the radial coordinate of the cloud and $r_{0}$ is a reference radius. In this simulation, $r_{0}$ is taken to be the same as $r_{galaxy}$. The gas number density in the CGM clouds at $r_{0}$ is $n_{0}$. The radial density variation is modeled as inverse square law assuming the clouds originate from a mass conserving galactic outflow (\citealt{chisholm2017mass}), consistent with other CGM modeling efforts at z $\sim$ 2.5 (eg: equation 22 in \cite{steidel2010structure}).
\vspace{1ex}

\noindent
A constant volume filling fraction (i.e.\ the fraction of CGM volume filled by the clouds) of 0.1 is assumed, in line with the volume filling fraction estimates in local CGM studies \citep{werk2016cos, stocke2013characterizing}. The value of $n_{0}$ is selected such that the integrated mass in the CGM (within 2 $\times R_{vir}$) equals $M_{CGM}$ for a given model. $M_{CGM}$ is the mass of CGM phase traced by C IV ion and is considered a variable quantity among different models. $M_{CGM}$ is selected to be roughly within an order of magnitude of the stellar mass. This assumption is in line with the observations and models presented in the literature (see Fig. 8 in \citealt{tumlinson2017circumgalactic} and \citealt{shen2013circumgalactic, peeples2018figuring}). 
\vspace{1ex}

\noindent
\textbf{GRB location:} The location of the GRB is randomly selected with a uniform spatial probability distribution in the region within $r_{inner}$ and $r_{galaxy}$, where $r_{inner}$ is chosen to be 0.4 kpc since the GRB host imaging surveys indicate that $\sim$80\% of the long GRBs to have an offset greater than 0.4 kpc \citep{blanchard2016offset}. The vertical location of the GRB is constrained within $\pm$ $h_{max}$, where $h_{max}$ is chosen to be 1 kpc to constrain the occurrence of GRBs in the region of active star formation. The GRB sightline is randomly chosen with a uniform probability distribution.  
\vspace{1ex}

\noindent
\textbf{Kinematics:}  For the ISM kinematics, a flat rotation curve is assumed with $v_{circ}$ = 100 km s$^{-1}$, in accordance with the stellar mass of the GRB host galaxy \citep{arabsalmani2017mass}. A 3-dimensional velocity dispersion ($\sigma_{v, ISM}$) of 50 km s$^{-1}$ is added to this. The individual CGM clouds move at random velocities with an isothermal distribution described by a Gaussian centered at 0 km s$^{-1}$ with a standard deviation ($\sigma_{v, CGM}$) of 100 km s$^{-1}$, given by the virial velocity of the halo. The dispersion speed is informed by the assumption of the galaxy mass and prior estimates from local and high-z CGM observations (\citealt{steidel2010structure, lan2018circumgalactic}). In this construct, although the column density depends on the value of $M_{CGM}$, the kinematic extent of the CGM is fairly insensitive. 
\vspace{1ex}

\noindent
\textbf{Outflows:} In order to simulate galactic outflows, a radially outward component is added to a fraction $f_{out}$ of the CGM clouds. The outflow velocity varies with the radial coordinate of the cloud to model the ballistic motion under the gravity of the dark matter halo. The  outflow launch velocity at a radial distance ($R_{launch}$) of 2 kpc is defined as $v_{out}$. The fraction $f_{out}$ is assumed to be $25-75\%$, in line with circumgalactic outflow simulations (\citealt{muratov2015gusty, ford2014tracing}). The  outflow launch velocity is varied between $200-300$ km s$^{-1}$ with a step of 50 km s$^{-1}$ (\citealt{muratov2015gusty, lan2018circumgalactic}). The outflow component is only added up to the radius where it decelerates to zero, i.e., no inflow component is added. The kinematics obtained from the observed data can be used to constrain $f_{out}$ and $v_{out}$. 
\vspace{1ex}


\noindent
\textbf{Simulation scheme:} A GRB location and line of sight (LOS) is selected. Within the ISM portion of the LOS, the length is divided into 100 sections. For each section, the component of velocity along the LOS and the column density is evaluated. In the CGM portion, the column density contributed by intersecting clouds and the LOS velocity is evaluated. The zero velocity for kinematics is defined as the LOS velocity of the material in the immediate vicinity of the GRB location (similar to the use of strongest fine structure transitions to define the zero velocity for the observed spectra). The results are compiled to synthesize the column density as a function of velocity for the entire LOS, as well as for the separate portions from the ISM and CGM . This helps in visualizing the kinematic distinction between the ISM and CGM. The column density is further binned in 100 km s$^{-1}$ velocity bins in the same fashion as the observed data. The simulation is repeated 200 times with new GRB locations and LOS to infer the variation in kinematics. The velocity-binned column density profiles are further stacked to plot the kinematics for the sample which can then be compared with the observations. 



%

\vspace{5mm}
\facilities{VLT X-shooter, Keck-HIRES, Keck-ESI, VLT-UVES}


\software{{\fontfamily{qcr}\selectfont astropy} \cite{robitaille2013astropy},  ~{\fontfamily{qcr}\selectfont emcee} \cite{foreman2013emcee} }

\bibliographystyle{apj}
\bibliography{ref}

\begin{thebibliography}{}
\expandafter\ifx\csname natexlab\endcsname\relax\def\natexlab#1{#1}\fi

\bibitem[{Angl{\'e}s-Alc{\'a}zar {et~al.}(2014)Angl{\'e}s-Alc{\'a}zar,
  Dav{\'e}, {\"O}zel, \& Oppenheimer}]{angles2014cosmological}
Angl{\'e}s-Alc{\'a}zar, D., Dav{\'e}, R., {\"O}zel, F., \& Oppenheimer, B.~D.
  2014, The Astrophysical Journal, 782, 84

\bibitem[{Angl{\'e}s-Alc{\'a}zar {et~al.}(2017)Angl{\'e}s-Alc{\'a}zar,
  Faucher-Gigu{\`e}re, Kere{\v{s}}, Hopkins, Quataert, \&
  Murray}]{angles2017cosmic}
Angl{\'e}s-Alc{\'a}zar, D., Faucher-Gigu{\`e}re, C.-A., Kere{\v{s}}, D.,
  {et~al.} 2017, Monthly Notices of the Royal Astronomical Society, 470, 4698

\bibitem[{Arabsalmani {et~al.}(2018)Arabsalmani, M{\o}ller, Perley, Freudling,
  Fynbo, Le~Floc'h, Zwaan, Schulze, Tanvir, Christensen,
  {et~al.}}]{arabsalmani2017mass}
Arabsalmani, M., M{\o}ller, P., Perley, D., {et~al.} 2018, Monthly Notices of
  the Royal Astronomical Society, 473, 3312

\bibitem[{Bahcall \& Wolf(1968)}]{bahcall1968fine}
Bahcall, J.~N., \& Wolf, R.~A. 1968, The Astrophysical Journal, 152, 701

\bibitem[{Behroozi {et~al.}(2013)Behroozi, Marchesini, Wechsler, Muzzin,
  Papovich, \& Stefanon}]{behroozi2013using}
Behroozi, P.~S., Marchesini, D., Wechsler, R.~H., {et~al.} 2013, The
  Astrophysical Journal Letters, 777, L10

\bibitem[{Blanchard {et~al.}(2016)Blanchard, Berger, \&
  Fong}]{blanchard2016offset}
Blanchard, P.~K., Berger, E., \& Fong, W.-f. 2016, The Astrophysical Journal,
  817, 144

\bibitem[{Bordoloi {et~al.}(2011)Bordoloi, Lilly, Knobel, Bolzonella, Kampczyk,
  Carollo, Iovino, Zucca, Contini, Kneib, {et~al.}}]{bordoloi2011radial}
Bordoloi, R., Lilly, S.~J., Knobel, C., {et~al.} 2011, The Astrophysical
  Journal, 743, 10

\bibitem[{Bordoloi {et~al.}(2014)Bordoloi, Tumlinson, Werk, Oppenheimer,
  Peeples, Prochaska, Tripp, Katz, Dav{\'e}, Fox, {et~al.}}]{bordoloi2014cos}
Bordoloi, R., Tumlinson, J., Werk, J.~K., {et~al.} 2014, The Astrophysical
  Journal, 796, 136

\bibitem[{Borthakur {et~al.}(2013)Borthakur, Heckman, Strickland, Wild, \&
  Schiminovich}]{borthakur2013impact}
Borthakur, S., Heckman, T., Strickland, D., Wild, V., \& Schiminovich, D. 2013,
  The Astrophysical Journal, 768, 18

\bibitem[{Borthakur {et~al.}(2015)Borthakur, Heckman, Tumlinson, Bordoloi,
  Thom, Catinella, Schiminovich, Dav{\'e}, Kauffmann, Moran,
  {et~al.}}]{borthakur2015connection}
Borthakur, S., Heckman, T., Tumlinson, J., {et~al.} 2015, The Astrophysical
  Journal, 813, 46

\bibitem[{Bouch{\'e} {et~al.}(2007)Bouch{\'e}, Lehnert, Aguirre, P{\'e}roux, \&
  Bergeron}]{bouche2007missing}
Bouch{\'e}, N., Lehnert, M.~D., Aguirre, A., P{\'e}roux, C., \& Bergeron, J.
  2007, Monthly Notices of the Royal Astronomical Society, 378, 525

\bibitem[{Bouch{\'e} {et~al.}(2006)Bouch{\'e}, Lehnert, \&
  P{\'e}roux}]{bouche2006missing}
Bouch{\'e}, N., Lehnert, M.~D., \& P{\'e}roux, C. 2006, Monthly Notices of the
  Royal Astronomical Society: Letters, 367, L16

\bibitem[{Burchett {et~al.}(2016)Burchett, Tripp, Bordoloi, Werk, Prochaska,
  Tumlinson, Willmer, O’Meara, \& Katz}]{burchett2016deep}
Burchett, J.~N., Tripp, T.~M., Bordoloi, R., {et~al.} 2016, The Astrophysical
  Journal, 832, 124

\bibitem[{Campana {et~al.}(2015)Campana, Salvaterra, Ferrara, \&
  Pallottini}]{campana2015missing}
Campana, S., Salvaterra, R., Ferrara, A., \& Pallottini, A. 2015, Astronomy \&
  Astrophysics, 575, A43

\bibitem[{Carilli \& Walter(2013)}]{carilli2013cool}
Carilli, C., \& Walter, F. 2013, Annual Review of Astronomy and Astrophysics,
  51

\bibitem[{Castro {et~al.}(2003)Castro, Galama, Harrison, Holtzman, Bloom,
  Djorgovski, \& Kulkarni}]{castro2003keck}
Castro, S., Galama, T., Harrison, F., {et~al.} 2003, The Astrophysical Journal,
  586, 128

\bibitem[{Chen(2012)}]{chen2012unchanging}
Chen, H.-W. 2012, Monthly Notices of the Royal Astronomical Society, 427, 1238

\bibitem[{Chen {et~al.}(2005)Chen, Prochaska, Bloom, \&
  Thompson}]{chen2005echelle}
Chen, H.-W., Prochaska, J.~X., Bloom, J.~S., \& Thompson, I.~B. 2005, The
  Astrophysical Journal Letters, 634, L25

\bibitem[{Chen {et~al.}(2007)Chen, Prochaska, Ramirez-Ruiz, Bloom,
  Dessauges-Zavadsky, \& Foley}]{chen2007absence}
Chen, H.-W., Prochaska, J.~X., Ramirez-Ruiz, E., {et~al.} 2007, The
  Astrophysical Journal, 663, 420

\bibitem[{Chisholm {et~al.}(2017)Chisholm, Tremonti, Leitherer, \&
  Chen}]{chisholm2017mass}
Chisholm, J., Tremonti, C.~A., Leitherer, C., \& Chen, Y. 2017, Monthly Notices
  of the Royal Astronomical Society, 469, 4831

\bibitem[{Crighton {et~al.}(2014)Crighton, Hennawi, Simcoe, Cooksey, Murphy,
  Fumagalli, Prochaska, \& Shanks}]{crighton2014metal}
Crighton, N.~H., Hennawi, J.~F., Simcoe, R.~A., {et~al.} 2014, Monthly Notices
  of the Royal Astronomical Society, 446, 18

\bibitem[{Cucchiara {et~al.}(2015)Cucchiara, Fumagalli, Rafelski, Kocevski,
  Prochaska, Cooke, \& Becker}]{cucchiara2015unveiling}
Cucchiara, A., Fumagalli, M., Rafelski, M., {et~al.} 2015, The Astrophysical
  Journal, 804, 51

\bibitem[{Cucchiara {et~al.}(2011)Cucchiara, Levan, Fox, Tanvir, Ukwatta,
  Berger, Kr{\"u}hler, Yolda{\c{s}}, Wu, Toma,
  {et~al.}}]{cucchiara2011photometric}
Cucchiara, A., Levan, A., Fox, D.~B., {et~al.} 2011, The Astrophysical Journal,
  736, 7

\bibitem[{Cucchiara {et~al.}(2013)Cucchiara, Prochaska, Zhu, M{\'e}nard, Fynbo,
  Fox, Chen, Cooksey, Cenko, Perley, {et~al.}}]{cucchiara2013independent}
Cucchiara, A., Prochaska, J., Zhu, G., {et~al.} 2013, The Astrophysical
  Journal, 773, 82

\bibitem[{Daddi {et~al.}(2010)Daddi, Bournaud, Walter, Dannerbauer, Carilli,
  Dickinson, Elbaz, Morrison, Riechers, Onodera, {et~al.}}]{daddi2010very}
Daddi, E., Bournaud, F., Walter, F., {et~al.} 2010, The Astrophysical Journal,
  713, 686

\bibitem[{Dav{\'e} {et~al.}(2017)Dav{\'e}, Rafieferantsoa, Thompson, \&
  Hopkins}]{dave2017mufasa}
Dav{\'e}, R., Rafieferantsoa, M.~H., Thompson, R.~J., \& Hopkins, P.~F. 2017,
  Monthly Notices of the Royal Astronomical Society, 467, 115

\bibitem[{Davies {et~al.}(2018)Davies, Schreiber, {\"U}bler, Genzel, Lutz,
  Renzini, Tacchella, Tacconi, Belli, Burkert, {et~al.}}]{davies2018kiloparsec}
Davies, R.~L., Schreiber, N. M.~F., {\"U}bler, H., {et~al.} 2018, arXiv
  preprint arXiv:1808.10700

\bibitem[{D'Elia {et~al.}(2007)D'Elia, Fiore, Meurs, Chincarini, Melandri,
  Norci, Pellizza, Perna, Piranomonte, Sbordone, {et~al.}}]{d2007uves}
D'Elia, V., Fiore, F., Meurs, E., {et~al.} 2007, Astronomy \& Astrophysics,
  467, 629

\bibitem[{D'Elia {et~al.}(2010)D'Elia, Fynbo, Covino, Goldoni, Jakobsson,
  Matteucci, Piranomonte, Sollerman, Th{\"o}ne, Vergani, {et~al.}}]{d2010vlt}
D'Elia, V., Fynbo, J. P.~U., Covino, S., {et~al.} 2010, Astronomy \&
  Astrophysics, 523, A36

\bibitem[{Dessauges-Zavadsky {et~al.}(2006)Dessauges-Zavadsky, Chen, Prochaska,
  Bloom, \& Barth}]{dessauges2006temporal}
Dessauges-Zavadsky, M., Chen, H.-W., Prochaska, J.~X., Bloom, J.~S., \& Barth,
  A.~J. 2006, The Astrophysical Journal Letters, 648, L89

\bibitem[{D’elia {et~al.}(2014)D’elia, Fynbo, Goldoni, Covino,
  de~Ugarte~Postigo, Ledoux, Calura, Gorosabel, Malesani, Matteucci,
  {et~al.}}]{d2014vlt}
D’elia, V., Fynbo, J. P.~U., Goldoni, P., {et~al.} 2014, Astronomy \&
  Astrophysics, 564, A38

\bibitem[{Erb {et~al.}(2006)Erb, Steidel, Shapley, Pettini, Reddy, \&
  Adelberger}]{erb2006stellar}
Erb, D.~K., Steidel, C.~C., Shapley, A.~E., {et~al.} 2006, The Astrophysical
  Journal, 646, 107

\bibitem[{Fiore {et~al.}(2005)Fiore, d’Elia, Lazzati, Perna, Sbordone,
  Stratta, Meurs, Ward, Antonelli, Chincarini, {et~al.}}]{fiore2005flash}
Fiore, F., d’Elia, V., Lazzati, D., {et~al.} 2005, The Astrophysical Journal,
  624, 853

\bibitem[{Ford {et~al.}(2014)Ford, Dav{\'e}, Oppenheimer, Katz, Kollmeier,
  Thompson, \& Weinberg}]{ford2014tracing}
Ford, A.~B., Dav{\'e}, R., Oppenheimer, B.~D., {et~al.} 2014, Monthly Notices
  of the Royal Astronomical Society, 444, 1260

\bibitem[{Foreman-Mackey {et~al.}(2013)Foreman-Mackey, Hogg, Lang, \&
  Goodman}]{foreman2013emcee}
Foreman-Mackey, D., Hogg, D.~W., Lang, D., \& Goodman, J. 2013, Publications of
  the Astronomical Society of the Pacific, 125, 306

\bibitem[{Fox {et~al.}(2008)Fox, Ledoux, Vreeswijk, Smette, \&
  Jaunsen}]{fox2008high}
Fox, A.~J., Ledoux, C., Vreeswijk, P.~M., Smette, A., \& Jaunsen, A.~O. 2008,
  Astronomy \& Astrophysics, 491, 189

\bibitem[{Fox {et~al.}(2007)Fox, Petitjean, Ledoux, \& Srianand}]{fox2007hot}
Fox, A.~J., Petitjean, P., Ledoux, C., \& Srianand, R. 2007, Astronomy \&
  Astrophysics, 465, 171

\bibitem[{Fox {et~al.}(2014)Fox, Wakker, Barger, Hernandez, Richter, Lehner,
  Bland-Hawthorn, Charlton, Westmeier, Thom, {et~al.}}]{fox2014cos}
Fox, A.~J., Wakker, B.~P., Barger, K.~A., {et~al.} 2014, The Astrophysical
  Journal, 787, 147

\bibitem[{Fox {et~al.}(2015)Fox, Bordoloi, Savage, Lockman, Jenkins, Wakker,
  Bland-Hawthorn, Hernandez, Kim, Benjamin, {et~al.}}]{fox2015probing}
Fox, A.~J., Bordoloi, R., Savage, B.~D., {et~al.} 2015, The Astrophysical
  Journal Letters, 799, L7

\bibitem[{Friis {et~al.}(2015)Friis, De~Cia, Kr{\"u}hler, Fynbo, Ledoux,
  Vreeswijk, Watson, Malesani, Gorosabel, Starling, {et~al.}}]{friis2015warm}
Friis, M., De~Cia, A., Kr{\"u}hler, T., {et~al.} 2015, Monthly Notices of the
  Royal Astronomical Society, 451, 167

\bibitem[{Fumagalli {et~al.}(2011)Fumagalli, Prochaska, Kasen, Dekel, Ceverino,
  \& Primack}]{fumagalli2011absorption}
Fumagalli, M., Prochaska, J.~X., Kasen, D., {et~al.} 2011, Monthly Notices of
  the Royal Astronomical Society, 418, 1796

\bibitem[{Fynbo {et~al.}(2009)Fynbo, Jakobsson, Prochaska, Malesani, Ledoux,
  de~Ugarte~Postigo, Nardini, Vreeswijk, Wiersema, Hjorth,
  {et~al.}}]{fynbo2009low}
Fynbo, J. P.~U., Jakobsson, P., Prochaska, J., {et~al.} 2009, The Astrophysical
  Journal Supplement Series, 185, 526

\bibitem[{Gilmore {et~al.}(2009)Gilmore, Madau, Primack, Somerville, \&
  Haardt}]{gilmore2009gev}
Gilmore, R.~C., Madau, P., Primack, J.~R., Somerville, R.~S., \& Haardt, F.
  2009, Monthly Notices of the Royal Astronomical Society, 399, 1694

\bibitem[{Girichidis {et~al.}(2016)Girichidis, Naab, Walch, Hanasz, Mac~Low,
  Ostriker, Gatto, Peters, W{\"u}nsch, Glover,
  {et~al.}}]{girichidis2016launching}
Girichidis, P., Naab, T., Walch, S., {et~al.} 2016, The Astrophysical Journal
  Letters, 816, L19

\bibitem[{Glidden {et~al.}(2016)Glidden, Cooper, Cooksey, Simcoe, \&
  O’Meara}]{glidden2016predominantly}
Glidden, A., Cooper, T.~J., Cooksey, K.~L., Simcoe, R.~A., \& O’Meara, J.~M.
  2016, The Astrophysical Journal, 833, 270

\bibitem[{Goerdt {et~al.}(2012)Goerdt, Dekel, Sternberg, Gnat, \&
  Ceverino}]{goerdt2012detectability}
Goerdt, T., Dekel, A., Sternberg, A., Gnat, O., \& Ceverino, D. 2012, Monthly
  Notices of the Royal Astronomical Society, 424, 2292

\bibitem[{Goodman \& Weare(2010)}]{goodman2010ensemble}
Goodman, J., \& Weare, J. 2010, Communications in applied mathematics and
  computational science, 5, 65

\bibitem[{Greiner {et~al.}(2015)Greiner, Fox, Schady, Kr{\"u}hler, Trenti,
  Cikota, Bolmer, Elliott, Delvaux, Perna, {et~al.}}]{greiner2015gamma}
Greiner, J., Fox, D.~B., Schady, P., {et~al.} 2015, The Astrophysical Journal,
  809, 76

\bibitem[{Hartoog {et~al.}(2015)Hartoog, Malesani, Fynbo, Goto, Kr{\"u}hler,
  Vreeswijk, De~Cia, Xu, M{\o}ller, Covino, {et~al.}}]{hartoog2015vlt}
Hartoog, O., Malesani, D., Fynbo, J., {et~al.} 2015, Astronomy \& Astrophysics,
  580, A139

\bibitem[{Hartoog {et~al.}(2013)Hartoog, Wiersema, Vreeswijk, Kaper, Tanvir,
  Savaglio, Berger, Chornock, Covino, D’Elia, {et~al.}}]{hartoog2013host}
Hartoog, O.~E., Wiersema, K., Vreeswijk, P.~M., {et~al.} 2013, Monthly Notices
  of the Royal Astronomical Society, 430, 2739

\bibitem[{Hayward \& Hopkins(2016)}]{hayward2016stellar}
Hayward, C.~C., \& Hopkins, P.~F. 2016, Monthly Notices of the Royal
  Astronomical Society, stw2888

\bibitem[{Heckman {et~al.}(2017)Heckman, Borthakur, Wild, Schiminovich, \&
  Bordoloi}]{heckman2017cos}
Heckman, T., Borthakur, S., Wild, V., Schiminovich, D., \& Bordoloi, R. 2017,
  The Astrophysical Journal, 846, 151

\bibitem[{Heckman {et~al.}(2015)Heckman, Alexandroff, Borthakur, Overzier, \&
  Leitherer}]{heckman2015systematic}
Heckman, T.~M., Alexandroff, R.~M., Borthakur, S., Overzier, R., \& Leitherer,
  C. 2015, The Astrophysical Journal, 809, 147

\bibitem[{Heintz {et~al.}(2018)Heintz, Watson, Jakobsson, Fynbo, Bolmer,
  Arabsalmani, Cano, Covino, D’Elia, Gomboc, {et~al.}}]{heintz2018highly}
Heintz, K., Watson, D., Jakobsson, P., {et~al.} 2018, Monthly Notices of the
  Royal Astronomical Society

\bibitem[{Hennawi {et~al.}(2006)Hennawi, Prochaska, Burles, Strauss, Richards,
  Schlegel, Fan, Schneider, Zakamska, Oguri, {et~al.}}]{hennawi2006quasars}
Hennawi, J.~F., Prochaska, J.~X., Burles, S., {et~al.} 2006, The Astrophysical
  Journal, 651, 61

\bibitem[{Hopkins {et~al.}(2014)Hopkins, Kere{\v{s}}, O{\~n}orbe,
  Faucher-Gigu{\`e}re, Quataert, Murray, \& Bullock}]{hopkins2014galaxies}
Hopkins, P.~F., Kere{\v{s}}, D., O{\~n}orbe, J., {et~al.} 2014, Monthly Notices
  of the Royal Astronomical Society, 445, 581

\bibitem[{Karim {et~al.}(2018)Karim, Fox, Jenkins, Bordoloi, Wakker, Savage,
  Lockman, Crawford, Jorgenson, \& Bland-Hawthorn}]{karim2018probing}
Karim, M.~T., Fox, A.~J., Jenkins, E.~B., {et~al.} 2018, The Astrophysical
  Journal, 860, 98

\bibitem[{Kornei {et~al.}(2012)Kornei, Shapley, Martin, Coil, Lotz,
  Schiminovich, Bundy, \& Noeske}]{kornei2012properties}
Kornei, K.~A., Shapley, A.~E., Martin, C.~L., {et~al.} 2012, The Astrophysical
  Journal, 758, 135

\bibitem[{Kr{\"u}hler {et~al.}(2011)Kr{\"u}hler, Greiner, Schady, Savaglio,
  Afonso, Clemens, Elliott, Filgas, Gruber, Kann, {et~al.}}]{kruhler2011seds}
Kr{\"u}hler, T., Greiner, J., Schady, P., {et~al.} 2011, Astronomy \&
  Astrophysics, 534, A108

\bibitem[{Kr{\"u}hler {et~al.}(2013)Kr{\"u}hler, Ledoux, Fynbo, Vreeswijk,
  Schmidl, Malesani, Christensen, De~Cia, Hjorth, Jakobsson,
  {et~al.}}]{kruhler2013molecular}
Kr{\"u}hler, T., Ledoux, C., Fynbo, J., {et~al.} 2013, Astronomy \&
  Astrophysics, 557, A18

\bibitem[{Kr{\"u}hler {et~al.}(2015)Kr{\"u}hler, Malesani, Fynbo, Hartoog,
  Hjorth, Jakobsson, Perley, Rossi, Schady, Schulze, {et~al.}}]{kruhler2015grb}
Kr{\"u}hler, T., Malesani, D., Fynbo, J., {et~al.} 2015, Astronomy \&
  Astrophysics, 581, A125

\bibitem[{Lagos {et~al.}(2014)Lagos, Baugh, Zwaan, Lacey, Gonzalez-Perez,
  Power, Swinbank, \& van Kampen}]{lagos2014galaxies}
Lagos, C. d.~P., Baugh, C.~M., Zwaan, M., {et~al.} 2014, Monthly Notices of the
  Royal Astronomical Society, 440, 920

\bibitem[{Lan \& Mo(2018)}]{lan2018circumgalactic}
Lan, T.-W., \& Mo, H. 2018, arXiv preprint arXiv:1806.05786

\bibitem[{Lehner {et~al.}(2015)Lehner, Howk, \& Wakker}]{lehner2015evidence}
Lehner, N., Howk, J.~C., \& Wakker, B.~P. 2015, The Astrophysical Journal, 804,
  79

\bibitem[{Lehner {et~al.}(2014)Lehner, O'Meara, Fox, Howk, Prochaska, Burns, \&
  Armstrong}]{lehner2014galactic}
Lehner, N., O'Meara, J.~M., Fox, A.~J., {et~al.} 2014, The Astrophysical
  Journal, 788, 119

\bibitem[{Levesque {et~al.}(2010)Levesque, Berger, Kewley, \&
  Bagley}]{levesque2010host}
Levesque, E.~M., Berger, E., Kewley, L.~J., \& Bagley, M.~M. 2010, The
  Astronomical Journal, 139, 694

\bibitem[{Lopez {et~al.}(2018)Lopez, Tejos, Ledoux, Barrientos, Sharon, Rigby,
  Gladders, Bayliss, \& Pessa}]{lopez2018clumpy}
Lopez, S., Tejos, N., Ledoux, C., {et~al.} 2018, Nature, 554, 493

\bibitem[{Markwardt(2009)}]{markwardt2009non}
Markwardt, C.~B. 2009, arXiv preprint arXiv:0902.2850

\bibitem[{Martin(2005)}]{martin2005mapping}
Martin, C.~L. 2005, The Astrophysical Journal, 621, 227

\bibitem[{Martin {et~al.}(2012)Martin, Shapley, Coil, Kornei, Bundy, Weiner,
  Noeske, \& Schiminovich}]{martin2012demographics}
Martin, C.~L., Shapley, A.~E., Coil, A.~L., {et~al.} 2012, The Astrophysical
  Journal, 760, 127

\bibitem[{McLean {et~al.}(1994)McLean, Mitchell, \&
  Swanston}]{mclean1994implementation}
McLean, A., Mitchell, C., \& Swanston, D. 1994, Journal of Electron
  Spectroscopy and Related Phenomena, 69, 125

\bibitem[{Morton(2003)}]{morton2003atomic}
Morton, D.~C. 2003, The Astrophysical Journal Supplement Series, 149, 205

\bibitem[{Moster {et~al.}(2012)Moster, Naab, \& White}]{moster2012galactic}
Moster, B.~P., Naab, T., \& White, S.~D. 2012, Monthly Notices of the Royal
  Astronomical Society, 428, 3121

\bibitem[{Muratov {et~al.}(2015)Muratov, Kere{\v{s}}, Faucher-Gigu{\`e}re,
  Hopkins, Quataert, \& Murray}]{muratov2015gusty}
Muratov, A.~L., Kere{\v{s}}, D., Faucher-Gigu{\`e}re, C.-A., {et~al.} 2015,
  Monthly Notices of the Royal Astronomical Society, 454, 2691

\bibitem[{Muratov {et~al.}(2017)Muratov, Kere{\v{s}}, Faucher-Gigu{\`e}re,
  Hopkins, Ma, Angl{\'e}s-Alc{\'a}zar, Chan, Torrey, Hafen, Quataert,
  {et~al.}}]{muratov2017metal}
---. 2017, Monthly Notices of the Royal Astronomical Society, 468, 4170

\bibitem[{Murray {et~al.}(2011)Murray, M{\'e}nard, \&
  Thompson}]{murray2011radiation}
Murray, N., M{\'e}nard, B., \& Thompson, T.~A. 2011, The Astrophysical Journal,
  735, 66

\bibitem[{Murray {et~al.}(2005)Murray, Quataert, \&
  Thompson}]{murray2005galaxies}
Murray, N., Quataert, E., \& Thompson, T.~A. 2005, The Astrophysical Journal,
  618, 569

\bibitem[{Nelson {et~al.}(2019)Nelson, Pillepich, Springel, Pakmor, Weinberger,
  Genel, Torrey, Vogelsberger, Marinacci, \& Hernquist}]{Nelson:2019jkf}
Nelson, D., Pillepich, A., Springel, V., {et~al.} 2019, arXiv:1902.05554

\bibitem[{Oppenheimer \& Schaye(2013)}]{oppenheimer2013non}
Oppenheimer, B.~D., \& Schaye, J. 2013, Monthly Notices of the Royal
  Astronomical Society, 434, 1043

\bibitem[{Page {et~al.}(2009)Page, Willingale, Bissaldi, Postigo, Holland,
  McBreen, O'Brien, Osborne, Prochaska, Rol,
  {et~al.}}]{page2009multiwavelength}
Page, K.~L., Willingale, R., Bissaldi, E., {et~al.} 2009, Monthly Notices of
  the Royal Astronomical Society, 400, 134

\bibitem[{Pallottini {et~al.}(2014)Pallottini, Gallerani, \&
  Ferrara}]{pallottini2014circumgalactic}
Pallottini, A., Gallerani, S., \& Ferrara, A. 2014, Monthly Notices of the
  Royal Astronomical Society: Letters, 444, L105

\bibitem[{Peeples {et~al.}(2014)Peeples, Werk, Tumlinson, Oppenheimer,
  Prochaska, Katz, \& Weinberg}]{peeples2014budget}
Peeples, M.~S., Werk, J.~K., Tumlinson, J., {et~al.} 2014, The Astrophysical
  Journal, 786, 54

\bibitem[{Peeples {et~al.}(2018)Peeples, Corlies, Tumlinson, O'Shea, Lehner,
  O'Meara, Howk, Smith, Wise, \& Hummels}]{peeples2018figuring}
Peeples, M.~S., Corlies, L., Tumlinson, J., {et~al.} 2018, arXiv preprint
  arXiv:1810.06566

\bibitem[{Perley {et~al.}(2016)Perley, Tanvir, Hjorth, Laskar, Berger, Chary,
  de~Ugarte~Postigo, Fynbo, Kr{\"u}hler, Levan, {et~al.}}]{perley2016swift_2}
Perley, D., Tanvir, N.~R., Hjorth, J., {et~al.} 2016, The Astrophysical
  Journal, 817, 8

\bibitem[{Perley {et~al.}(2009)Perley, Cenko, Bloom, Chen, Butler, Kocevski,
  Prochaska, Brodwin, Glazebrook, Kasliwal, {et~al.}}]{perley2009host}
Perley, D.~A., Cenko, S., Bloom, J., {et~al.} 2009, The Astronomical Journal,
  138, 1690

\bibitem[{Petitjean(1995)}]{petitjean1995qso}
Petitjean, P. 1995, in Science with the VLT (Springer), 339--348

\bibitem[{Prochaska {et~al.}(2007)Prochaska, Chen, Bloom, Dessauges-Zavadsky,
  O’Meara, Foley, Bernstein, Burles, Dupree, Falco,
  {et~al.}}]{prochaska2007interstellar}
Prochaska, J., Chen, H.-W., Bloom, J., {et~al.} 2007, The Astrophysical Journal
  Supplement Series, 168, 231

\bibitem[{Prochaska {et~al.}(2006)Prochaska, Chen, \&
  Bloom}]{prochaska2006dissecting}
Prochaska, J.~X., Chen, H.-W., \& Bloom, J.~S. 2006, The Astrophysical Journal,
  648, 95

\bibitem[{Prochaska {et~al.}(2008{\natexlab{a}})Prochaska, Chen,
  Dessauges-Zavadsky, Bloom, Galassi, Palmer, \&
  Fenimore}]{prochaska2008resolving}
Prochaska, J.~X., Chen, H.-W., Dessauges-Zavadsky, M., {et~al.}
  2008{\natexlab{a}}in , AIP, 479--485

\bibitem[{Prochaska {et~al.}(2008{\natexlab{b}})Prochaska, Dessauges-Zavadsky,
  Ramirez-Ruiz, \& Chen}]{prochaska2008survey}
Prochaska, J.~X., Dessauges-Zavadsky, M., Ramirez-Ruiz, E., \& Chen, H.-W.
  2008{\natexlab{b}}, The Astrophysical Journal, 685, 344

\bibitem[{Prochaska {et~al.}(2014)Prochaska, Lau, \&
  Hennawi}]{prochaska2014quasars}
Prochaska, J.~X., Lau, M.~W., \& Hennawi, J.~F. 2014, The Astrophysical
  Journal, 796, 140

\bibitem[{Prochaska {et~al.}(2017)Prochaska, Werk, Worseck, Tripp, Tumlinson,
  Burchett, Fox, Fumagalli, Lehner, Peeples, {et~al.}}]{prochaska2017cos}
Prochaska, J.~X., Werk, J.~K., Worseck, G., {et~al.} 2017, The Astrophysical
  Journal, 837, 169

\bibitem[{Rahmati {et~al.}(2016)Rahmati, Schaye, Crain, Oppenheimer, Schaller,
  \& Theuns}]{rahmati2016cosmic}
Rahmati, A., Schaye, J., Crain, R.~A., {et~al.} 2016, Monthly Notices of the
  Royal Astronomical Society, 459, 310

\bibitem[{Reddy {et~al.}(2012)Reddy, Pettini, Steidel, Shapley, Erb, \&
  Law}]{reddy2012characteristic}
Reddy, N.~A., Pettini, M., Steidel, C.~C., {et~al.} 2012, The Astrophysical
  Journal, 754, 25

\bibitem[{Rigby {et~al.}(2018)Rigby, Bayliss, Sharon, Gladders, Chisholm,
  Dahle, Johnson, Paterno-Mahler, Wuyts, \& Kelson}]{rigby2018magellan}
Rigby, J., Bayliss, M., Sharon, K., {et~al.} 2018, The Astronomical Journal,
  155, 104

\bibitem[{Robitaille {et~al.}(2013)Robitaille, Tollerud, Greenfield,
  Droettboom, Bray, Aldcroft, Davis, Ginsburg, Price-Whelan, Kerzendorf,
  {et~al.}}]{robitaille2013astropy}
Robitaille, T.~P., Tollerud, E.~J., Greenfield, P., {et~al.} 2013, Astronomy \&
  Astrophysics, 558, A33

\bibitem[{Rubin {et~al.}(2012)Rubin, Prochaska, Koo, \&
  Phillips}]{rubin2012direct}
Rubin, K.~H., Prochaska, J.~X., Koo, D.~C., \& Phillips, A.~C. 2012, The
  Astrophysical Journal Letters, 747, L26

\bibitem[{Rubin {et~al.}(2014)Rubin, Prochaska, Koo, Phillips, Martin, \&
  Winstrom}]{rubin2014evidence}
Rubin, K.~H., Prochaska, J.~X., Koo, D.~C., {et~al.} 2014, The Astrophysical
  Journal, 794, 156

\bibitem[{Rudie {et~al.}(2019)Rudie, Steidel, Pettini, Trainor, Strom, Hummels,
  Reddy, \& Shapley}]{rudie2019column}
Rudie, G.~C., Steidel, C.~C., Pettini, M., {et~al.} 2019, arXiv preprint
  arXiv:1903.00004

\bibitem[{Salvaterra {et~al.}(2009)Salvaterra, Della~Valle, Campana,
  Chincarini, Covino, D’avanzo, Fern{\'a}ndez-Soto, Guidorzi, Mannucci,
  Margutti, {et~al.}}]{salvaterra2009grb}
Salvaterra, R., Della~Valle, M., Campana, S., {et~al.} 2009, Nature, 461, 1258

\bibitem[{Savage \& Sembach(1991)}]{savage1991analysis}
Savage, B.~D., \& Sembach, K.~R. 1991, The Astrophysical Journal, 379, 245

\bibitem[{Savage {et~al.}(1997)Savage, Sembach, \& Lu}]{savage1997absorption}
Savage, B.~D., Sembach, K.~R., \& Lu, L. 1997, The Astronomical Journal, 113,
  2158

\bibitem[{Schady(2017)}]{schady2017gamma}
Schady, P. 2017, Royal Society open science, 4, 170304

\bibitem[{Schaye {et~al.}(2014)Schaye, Crain, Bower, Furlong, Schaller, Theuns,
  Dalla~Vecchia, Frenk, McCarthy, Helly, {et~al.}}]{schaye2014eagle}
Schaye, J., Crain, R.~A., Bower, R.~G., {et~al.} 2014, Monthly Notices of the
  Royal Astronomical Society, 446, 521

\bibitem[{Selsing {et~al.}(2018)Selsing, Malesani, Goldoni, Fynbo, Kr{\"u}hler,
  Antonelli, Arabsalmani, Bolmer, Cano, Christensen, {et~al.}}]{selsing2018x}
Selsing, J., Malesani, D., Goldoni, P., {et~al.} 2018, arXiv preprint
  arXiv:1802.07727

\bibitem[{Shen {et~al.}(2012)Shen, Madau, Aguirre, Guedes, Mayer, \&
  Wadsley}]{shen2012origin}
Shen, S., Madau, P., Aguirre, A., {et~al.} 2012, The Astrophysical Journal,
  760, 50

\bibitem[{Shen {et~al.}(2013)Shen, Madau, Guedes, Mayer, Prochaska, \&
  Wadsley}]{shen2013circumgalactic}
Shen, S., Madau, P., Guedes, J., {et~al.} 2013, The Astrophysical Journal, 765,
  89

\bibitem[{Simcoe {et~al.}(2011)Simcoe, Cooksey, Matejek, Burgasser, Bochanski,
  Lovegrove, Bernstein, Pipher, Forrest, McMurtry,
  {et~al.}}]{simcoe2011constraints}
Simcoe, R.~A., Cooksey, K.~L., Matejek, M., {et~al.} 2011, The Astrophysical
  Journal, 743, 21

\bibitem[{{Smette} {et~al.}(2013){Smette}, {Ledoux}, {Vreeswijk}, {De Cia},
  {Petitjean}, {Fynbo}, {Malesani}, \& {Fox}}]{2013GCN.14848....1S}
{Smette}, A., {Ledoux}, C., {Vreeswijk}, P., {et~al.} 2013, GRB Coordinates
  Network, Circular Service, No.~14848, \#1 (2013), 14848

\bibitem[{Sparre {et~al.}(2014)Sparre, Hartoog, Kr{\"u}hler, Fynbo, Watson,
  Wiersema, D'Elia, Zafar, Afonso, Covino, {et~al.}}]{sparre2014metallicity}
Sparre, M., Hartoog, O., Kr{\"u}hler, T., {et~al.} 2014, The Astrophysical
  Journal, 785, 150

\bibitem[{Steidel {et~al.}(2010)Steidel, Erb, Shapley, Pettini, Reddy,
  Bogosavljevi{\'c}, Rudie, \& Rakic}]{steidel2010structure}
Steidel, C.~C., Erb, D.~K., Shapley, A.~E., {et~al.} 2010, The Astrophysical
  Journal, 717, 289

\bibitem[{Stern {et~al.}(2016)Stern, Hennawi, Prochaska, \&
  Werk}]{stern2016universal}
Stern, J., Hennawi, J.~F., Prochaska, J.~X., \& Werk, J.~K. 2016, The
  Astrophysical Journal, 830, 87

\bibitem[{Stocke {et~al.}(2013)Stocke, Keeney, Danforth, Shull, Froning, Green,
  Penton, \& Savage}]{stocke2013characterizing}
Stocke, J.~T., Keeney, B.~A., Danforth, C.~W., {et~al.} 2013, The Astrophysical
  Journal, 763, 148

\bibitem[{Tacconi {et~al.}(2018)Tacconi, Genzel, Saintonge, Combes,
  Garc{\'\i}a-Burillo, Neri, Bolatto, Contini, Schreiber, Lilly,
  {et~al.}}]{tacconi2018phibss}
Tacconi, L.~J., Genzel, R., Saintonge, A., {et~al.} 2018, The Astrophysical
  Journal, 853, 179

\bibitem[{Tanvir {et~al.}(2009)Tanvir, Fox, Levan, Berger, Wiersema, Fynbo,
  Cucchiara, Kr{\"u}hler, Gehrels, Bloom, {et~al.}}]{tanvir2009gamma}
Tanvir, N.~R., Fox, D.~B., Levan, A., {et~al.} 2009, Nature, 461, 1254

\bibitem[{Th{\"o}ne {et~al.}(2012)Th{\"o}ne, Fynbo, Goldoni, de~Ugarte,
  Campana, Vergani, Covino, Kr{\"u}hler, Kaper, Tanvir,
  {et~al.}}]{thone2012grb}
Th{\"o}ne, C., Fynbo, J. P.~U., Goldoni, P., {et~al.} 2012, Monthly Notices of
  the Royal Astronomical Society, 428, 3590

\bibitem[{Toy {et~al.}(2016)Toy, Cucchiara, Veilleux, Fumagalli, Rafelski,
  Rahmati, Cenko, Capone, \& Pasham}]{toy2016exploring}
Toy, V.~L., Cucchiara, A., Veilleux, S., {et~al.} 2016, The Astrophysical
  Journal, 832, 175

\bibitem[{Trenti {et~al.}(2015)Trenti, Perna, \&
  Jimenez}]{trenti2015luminosity}
Trenti, M., Perna, R., \& Jimenez, R. 2015, The Astrophysical Journal, 802, 103

\bibitem[{Tumlinson {et~al.}(2017)Tumlinson, Peeples, \&
  Werk}]{tumlinson2017circumgalactic}
Tumlinson, J., Peeples, M.~S., \& Werk, J.~K. 2017, Annual Review of Astronomy
  and Astrophysics, 55, 389

\bibitem[{Tumlinson {et~al.}(2011)Tumlinson, Thom, Werk, Prochaska, Tripp,
  Weinberg, Peeples, O’Meara, Oppenheimer, Meiring,
  {et~al.}}]{tumlinson2011large}
Tumlinson, J., Thom, C., Werk, J.~K., {et~al.} 2011, Science, 334, 948

\bibitem[{Tumlinson {et~al.}(2013)Tumlinson, Thom, Werk, Prochaska, Tripp,
  Katz, Dav{\'e}, Oppenheimer, Meiring, Ford, {et~al.}}]{tumlinson2013cos}
---. 2013, The Astrophysical Journal, 777, 59

\bibitem[{Turner {et~al.}(2014)Turner, Schaye, Steidel, Rudie, \&
  Strom}]{turner2014metal}
Turner, M.~L., Schaye, J., Steidel, C.~C., Rudie, G.~C., \& Strom, A.~L. 2014,
  Monthly Notices of the Royal Astronomical Society, 445, 794

\bibitem[{Vergani {et~al.}(2017)Vergani, Palmerio, Salvaterra, Japelj,
  Mannucci, Perley, D’Avanzo, Kr{\"u}hler, Puech, Boissier,
  {et~al.}}]{vergani2017chemical}
Vergani, S., Palmerio, J., Salvaterra, R., {et~al.} 2017, Astronomy \&
  Astrophysics, 599, A120

\bibitem[{Voit {et~al.}(2015)Voit, Donahue, Bryan, \&
  McDonald}]{voit2015regulation}
Voit, G., Donahue, M., Bryan, G., \& McDonald, M. 2015, Nature, 519, 203

\bibitem[{Vreeswijk {et~al.}(2013)Vreeswijk, Ledoux, Raassen, Smette, De~Cia,
  Wo{\'z}niak, Fox, Vestrand, \& Jakobsson}]{vreeswijk2013time}
Vreeswijk, P., Ledoux, C., Raassen, A., {et~al.} 2013, Astronomy \&
  Astrophysics, 549, A22

\bibitem[{Vreeswijk {et~al.}(2007)Vreeswijk, Ledoux, Smette, Ellison, Jaunsen,
  Andersen, Fruchter, Fynbo, Hjorth, Kaufer, {et~al.}}]{vreeswijk2007rapid}
Vreeswijk, P.~M., Ledoux, C., Smette, A., {et~al.} 2007, Astronomy \&
  Astrophysics, 468, 83

\bibitem[{Wainwright {et~al.}(2007)Wainwright, Berger, \&
  Penprase}]{wainwright2007morphological}
Wainwright, C., Berger, E., \& Penprase, B. 2007, The Astrophysical Journal,
  657, 367

\bibitem[{Wechsler \& Tinker(2018)}]{wechsler2018connection}
Wechsler, R.~H., \& Tinker, J.~L. 2018, Annual Review of Astronomy and
  Astrophysics, 56, 435

\bibitem[{Weiner {et~al.}(2009)Weiner, Coil, Prochaska, Newman, Cooper, Bundy,
  Conselice, Dutton, Faber, Koo, {et~al.}}]{weiner2009ubiquitous}
Weiner, B.~J., Coil, A.~L., Prochaska, J.~X., {et~al.} 2009, The Astrophysical
  Journal, 692, 187

\bibitem[{Werk {et~al.}(2014)Werk, Prochaska, Tumlinson, Peeples, Tripp, Fox,
  Lehner, Thom, O'Meara, Ford, {et~al.}}]{werk2014cos}
Werk, J.~K., Prochaska, J.~X., Tumlinson, J., {et~al.} 2014, The Astrophysical
  Journal, 792, 8

\bibitem[{Werk {et~al.}(2016)Werk, Prochaska, Cantalupo, Fox, Oppenheimer,
  Tumlinson, Tripp, Lehner, \& McQuinn}]{werk2016cos}
Werk, J.~K., Prochaska, J.~X., Cantalupo, S., {et~al.} 2016, The Astrophysical
  Journal, 833, 54

\bibitem[{Wiklind {et~al.}(2019)Wiklind, Ferguson, Guo, Koo, Kocevski,
  Mobasher, Brammer, Kassin, Koekemoer, Giavalisco,
  {et~al.}}]{wiklind2019evolution}
Wiklind, T., Ferguson, H.~C., Guo, Y., {et~al.} 2019, arXiv preprint
  arXiv:1903.06962

\bibitem[{Wiseman {et~al.}(2017{\natexlab{a}})Wiseman, Perley, Schady,
  Prochaska, de~Ugarte~Postigo, Kr{\"u}hler, Yates, \&
  Greiner}]{wiseman2017gas}
Wiseman, P., Perley, D., Schady, P., {et~al.} 2017{\natexlab{a}}, Astronomy \&
  Astrophysics, 607, A107

\bibitem[{Wiseman {et~al.}(2017{\natexlab{b}})Wiseman, Schady, Bolmer,
  Kr{\"u}hler, Yates, Greiner, \& Fynbo}]{wiseman2017evolution}
Wiseman, P., Schady, P., Bolmer, J., {et~al.} 2017{\natexlab{b}}, Astronomy \&
  Astrophysics, 599, A24

\bibitem[{Zafar {et~al.}(2018)Zafar, M{\o}ller, Watson, Lattanzio, Hopkins,
  Karakas, Fynbo, Tanvir, Selsing, Jakobsson, {et~al.}}]{zafar2018x}
Zafar, T., M{\o}ller, P., Watson, D., {et~al.} 2018, Monthly Notices of the
  Royal Astronomical Society, 480, 108

\bibitem[{Zhu \& M{\'e}nard(2013)}]{zhu2013calcium}
Zhu, G., \& M{\'e}nard, B. 2013, The Astrophysical Journal, 773, 16

\end{thebibliography}






\newcommand{\wfit}{0.8}

\begin{figure*}
\centering
\includegraphics[width=\wfit\textwidth]{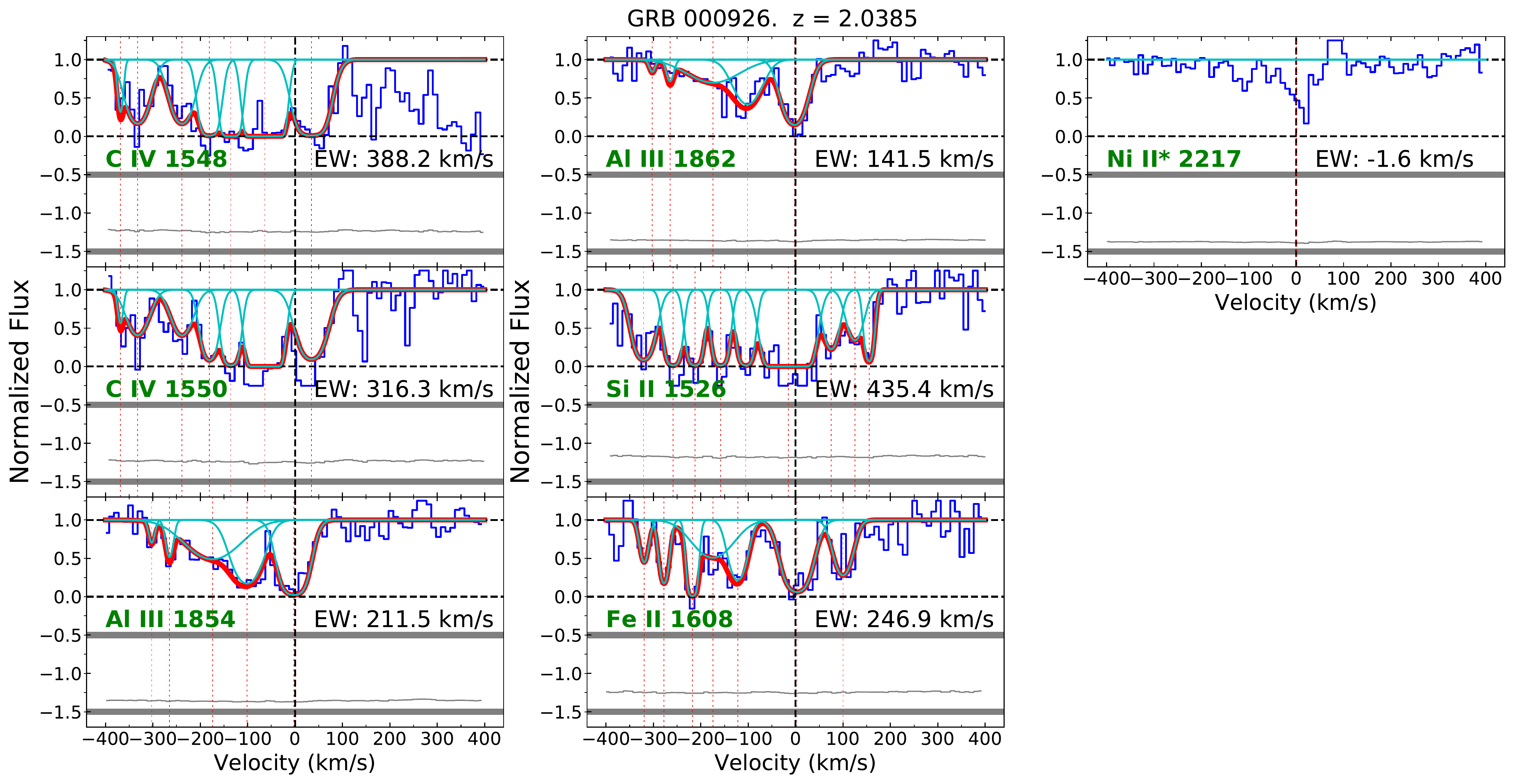}
\figcaption{\label{fig:000926} Voigt profile fit for GRB 000926}
\end{figure*}

\begin{figure*}
\centering
\includegraphics[width=\wfit\textwidth]{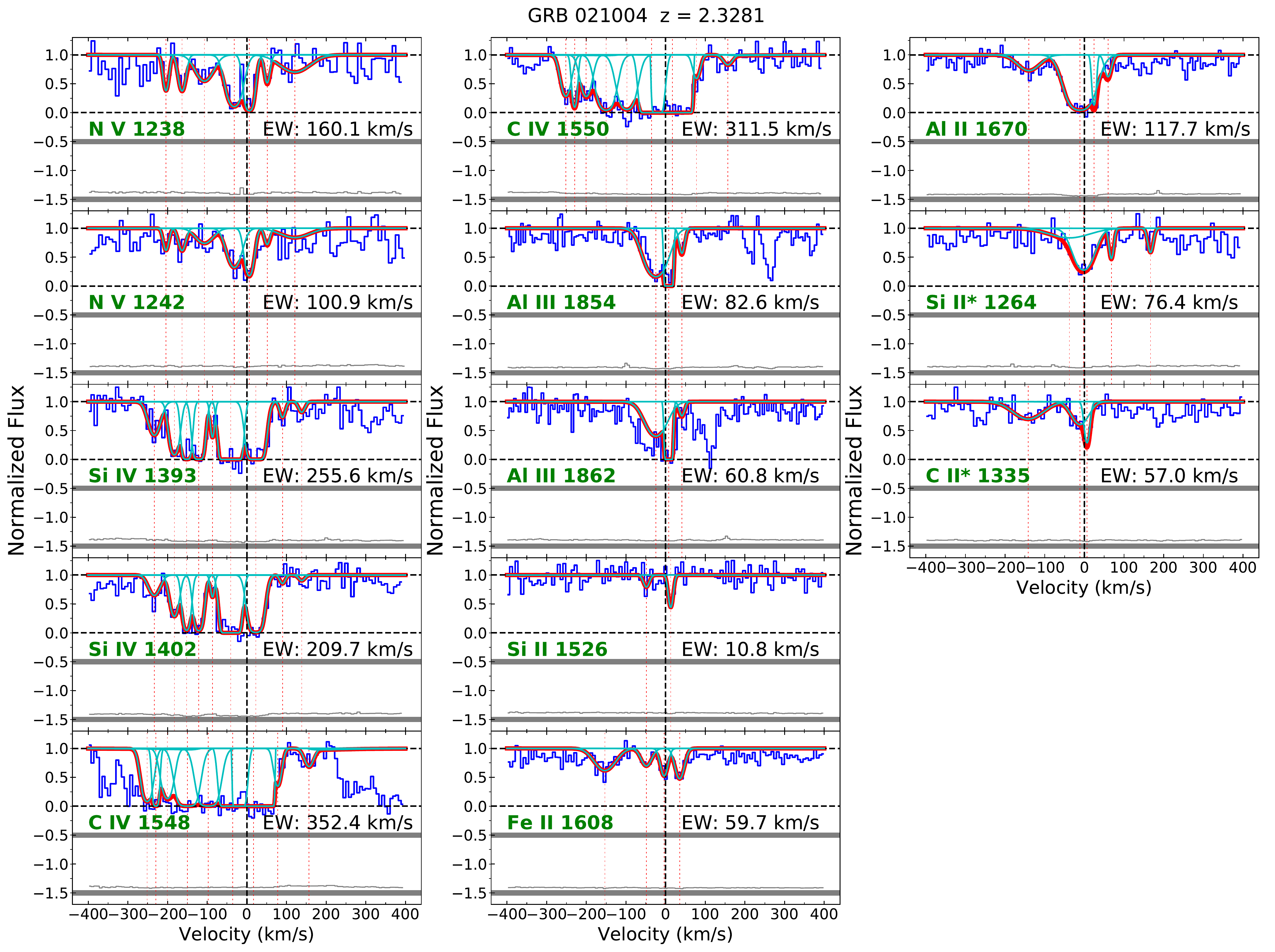}
\figcaption{\label{fig:021004} Voigt profile fit for GRB 021004}
\end{figure*}

\begin{figure*}
\centering
\includegraphics[width=\wfit\textwidth]{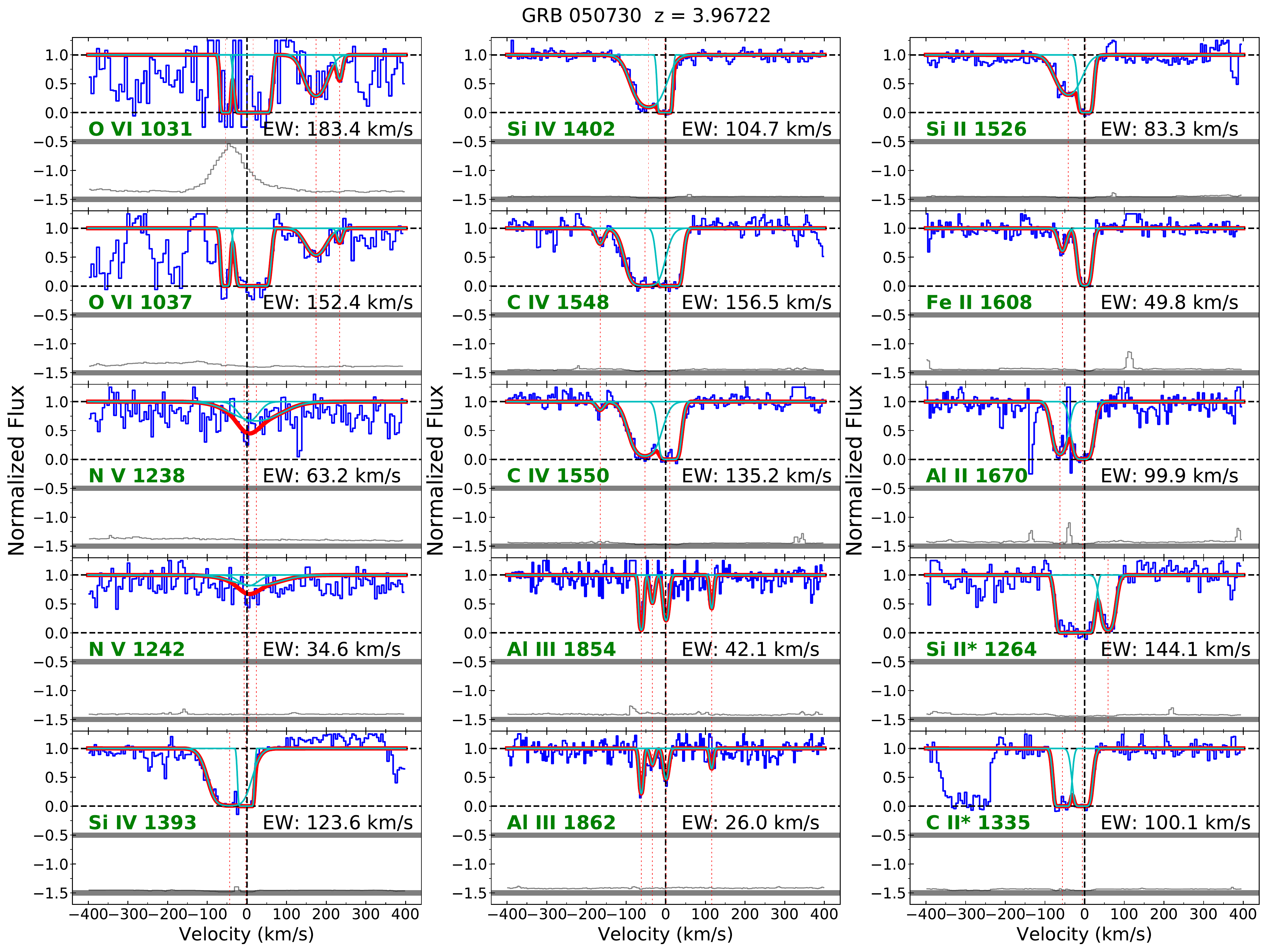}
\figcaption{\label{fig:050730} Voigt profile fit for GRB 050730}
\end{figure*}

\begin{figure*}
\centering
\includegraphics[width=\wfit\textwidth]{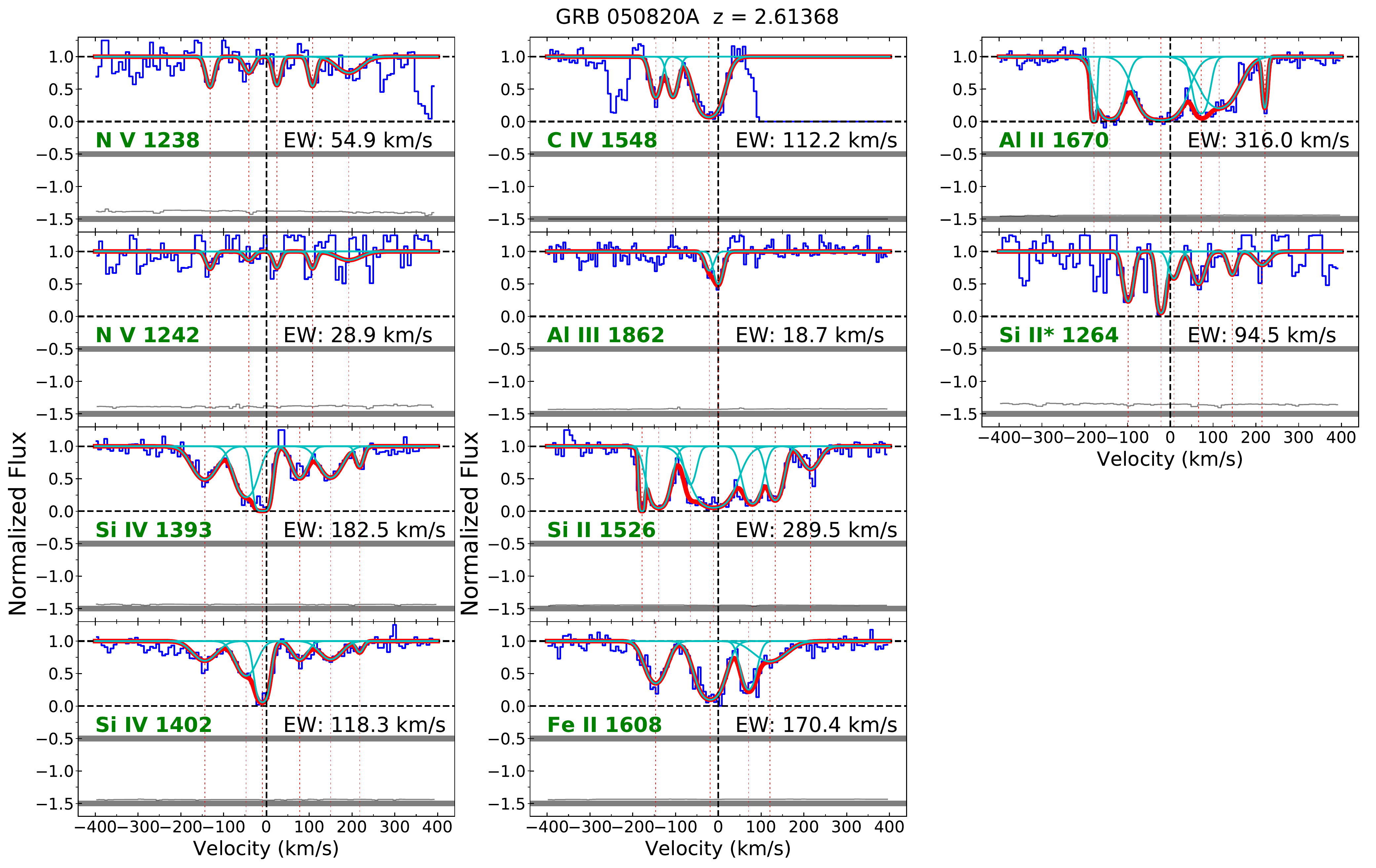}
\figcaption{\label{fig:050820A} Voigt profile fit for GRB 050820A}
\end{figure*}

\begin{figure*}
\centering
\includegraphics[width=\wfit\textwidth]{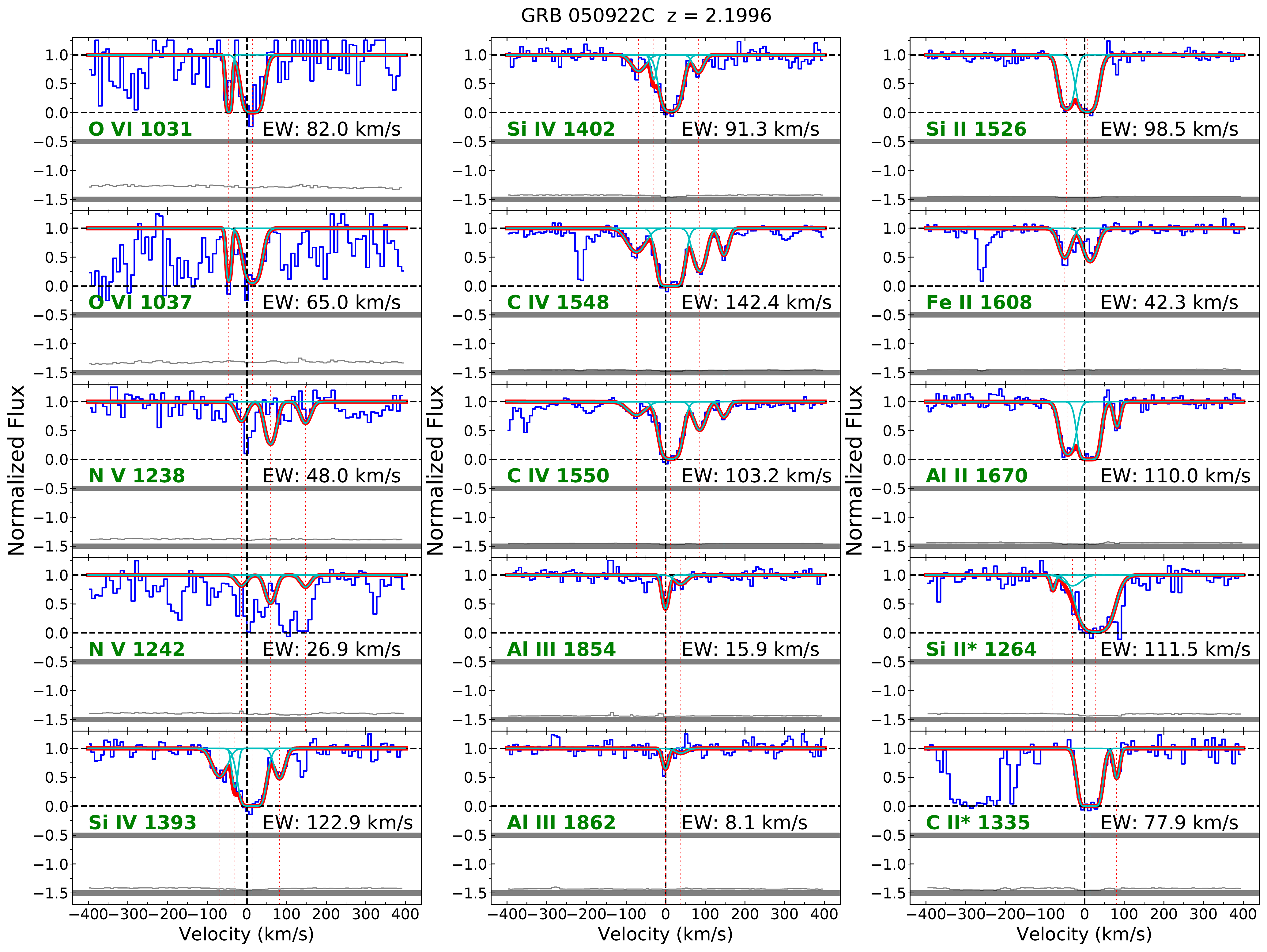}
\figcaption{\label{fig:050922C} Voigt profile fit for GRB 050922C}
\end{figure*}

\begin{figure*}
\centering
\includegraphics[width=\wfit\textwidth]{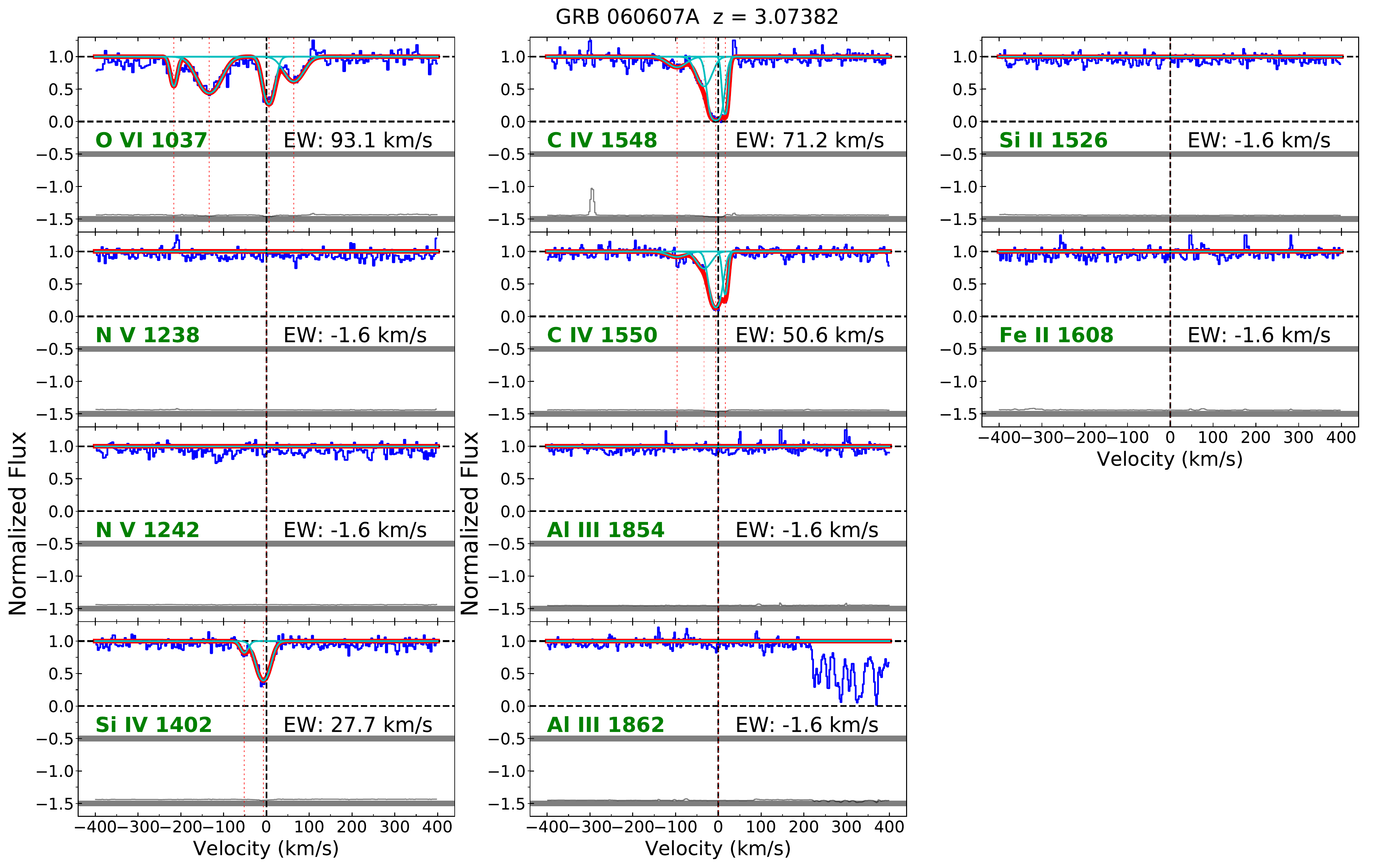}
\figcaption{\label{fig:060607A} Voigt profile fit for GRB 060607A}
\end{figure*}

\begin{figure*}
\centering
\includegraphics[width=\wfit\textwidth]{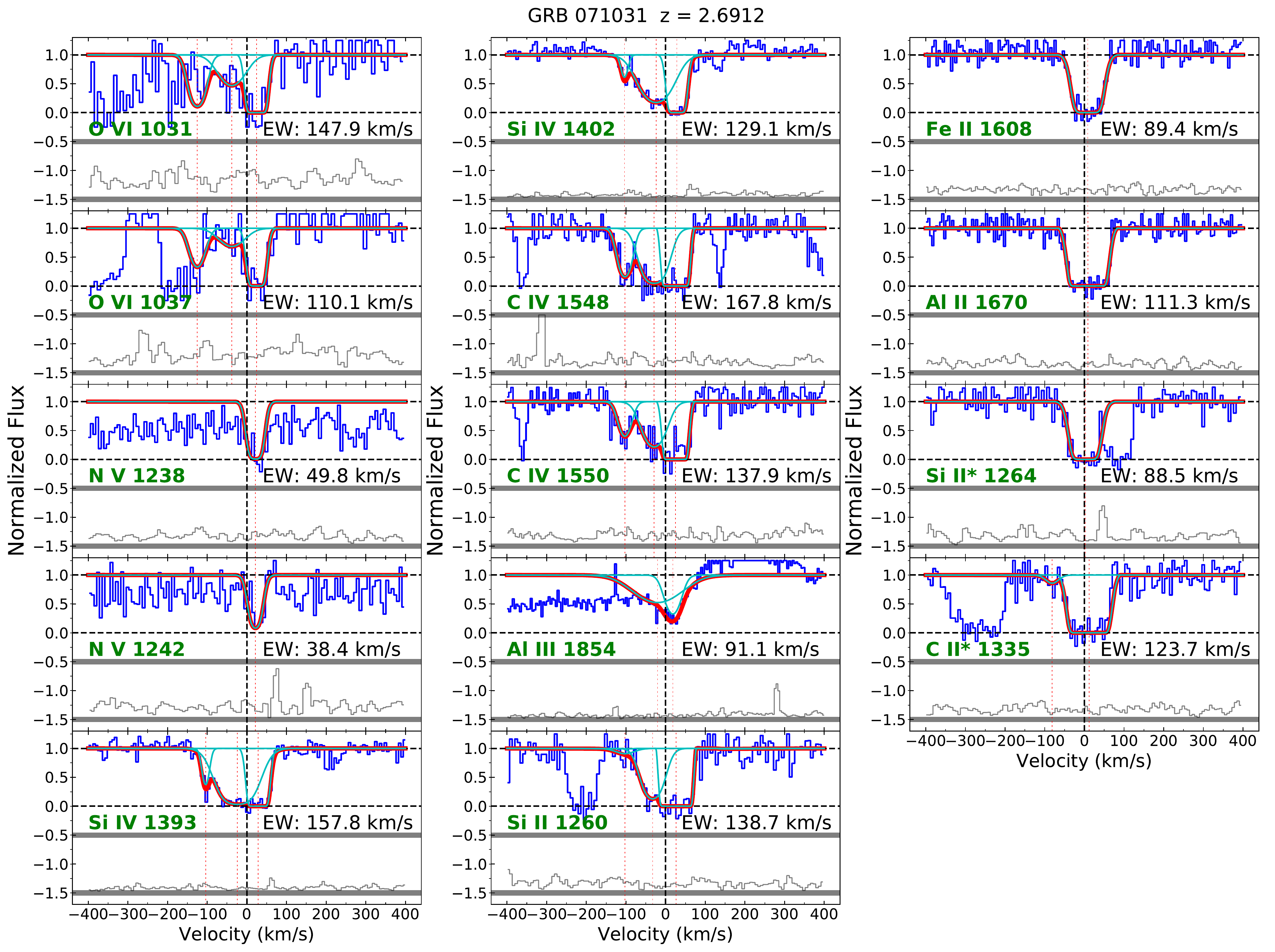}
\figcaption{\label{fig:071031} Voigt profile fit for GRB 071031}
\end{figure*}

\begin{figure*}
\centering
\includegraphics[width=0.85\textwidth]{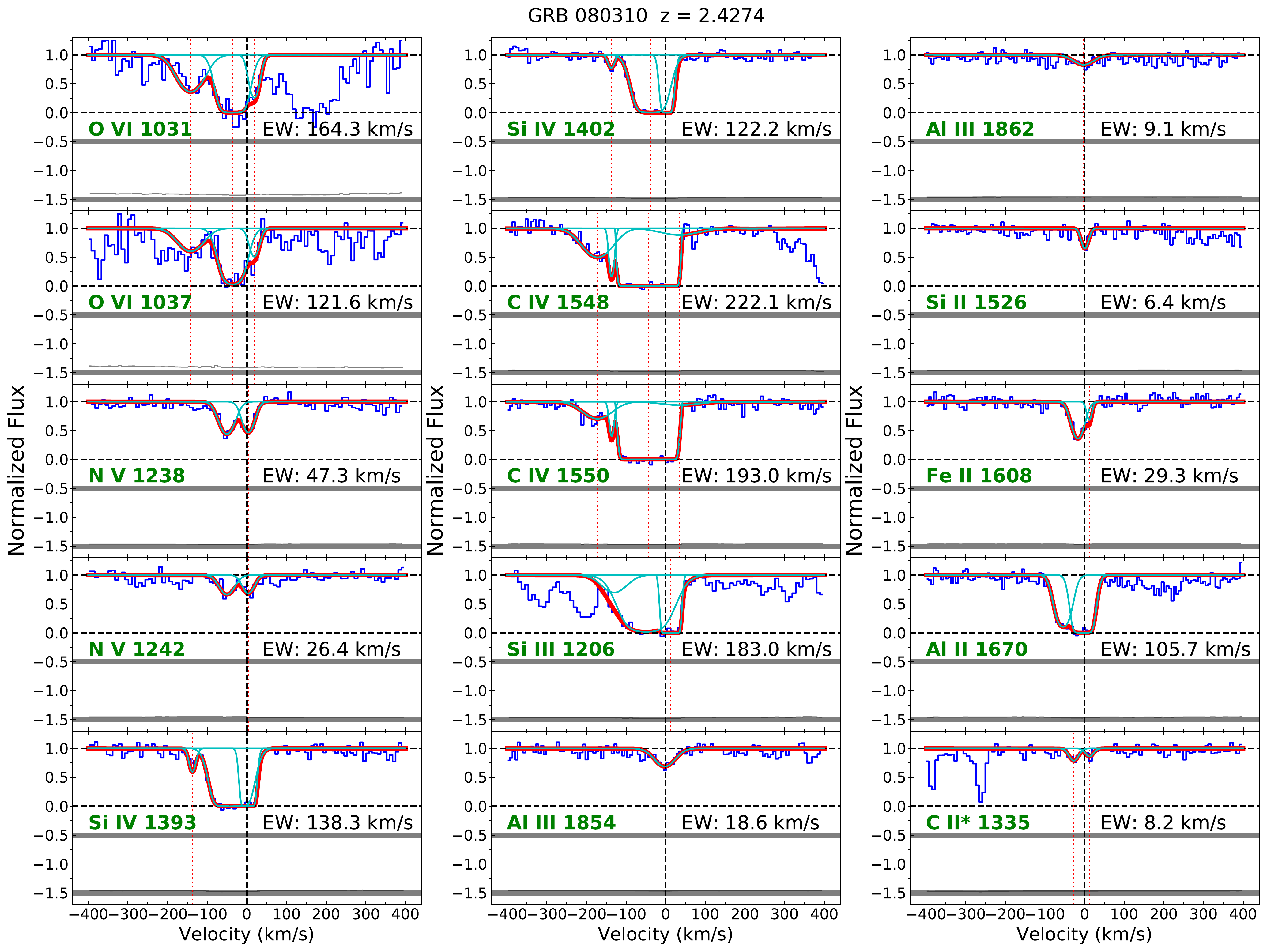}
\figcaption{\label{fig:080310} Voigt profile fit for GRB 080310}
\end{figure*}

\begin{figure*}
\centering
\includegraphics[width=\wfit\textwidth]{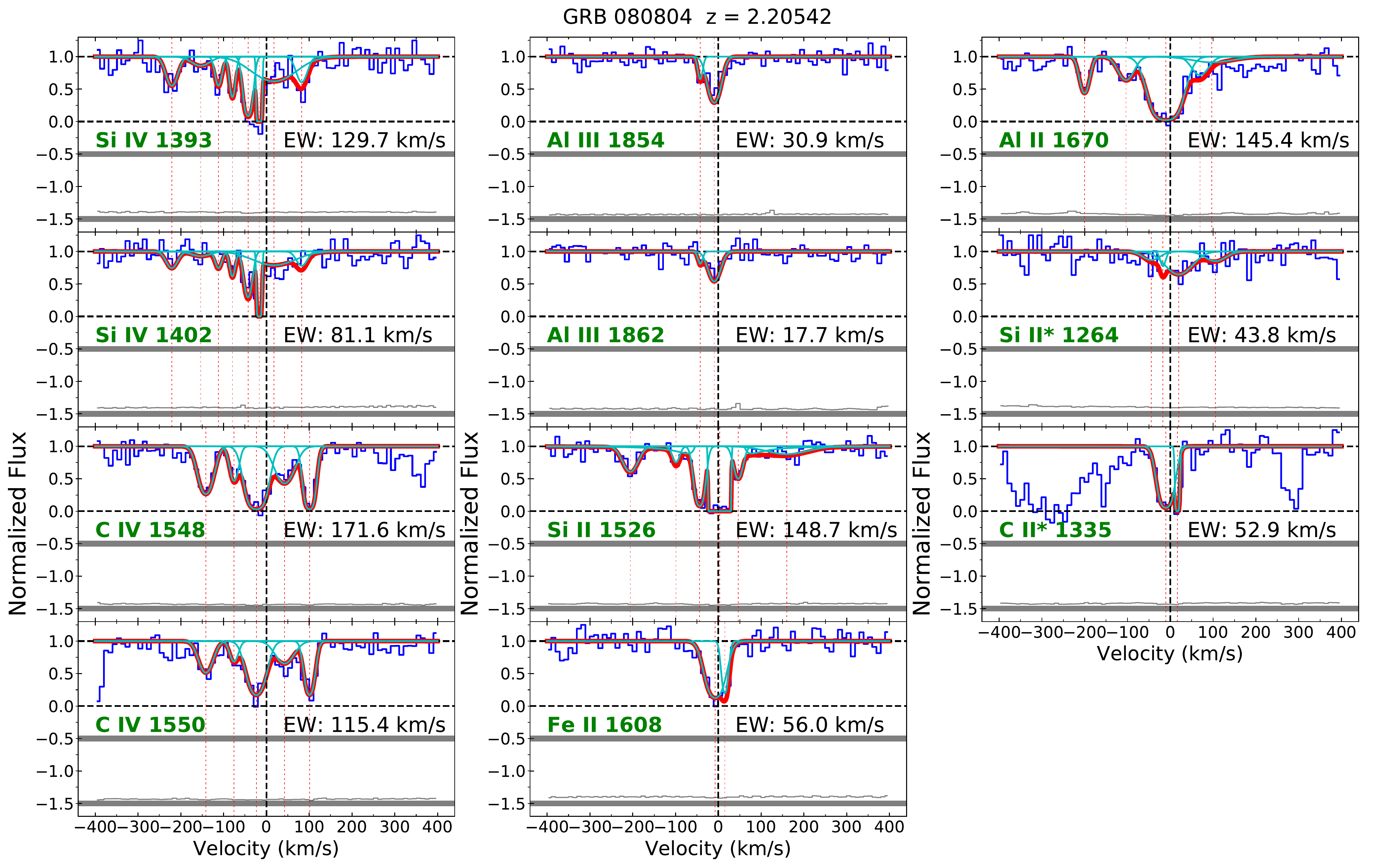}
\figcaption{\label{fig:080804} Voigt profile fit for GRB 080804}
\end{figure*}

\begin{figure*}
\centering
\includegraphics[width=\wfit\textwidth]{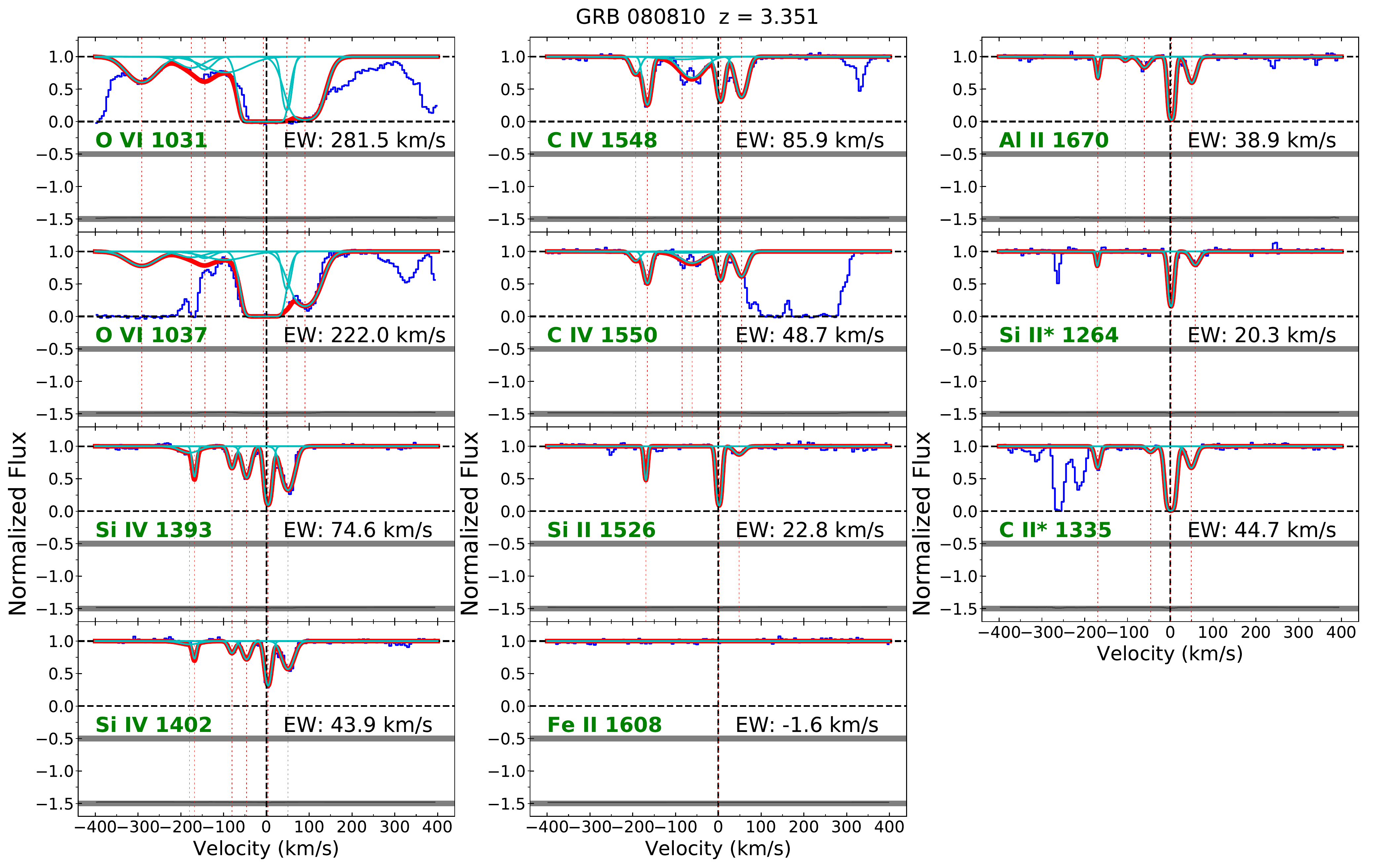}
\figcaption{\label{fig:080810} Voigt profile fit for GRB 080810}
\end{figure*}

\begin{figure*}
\centering
\includegraphics[width=\wfit\textwidth]{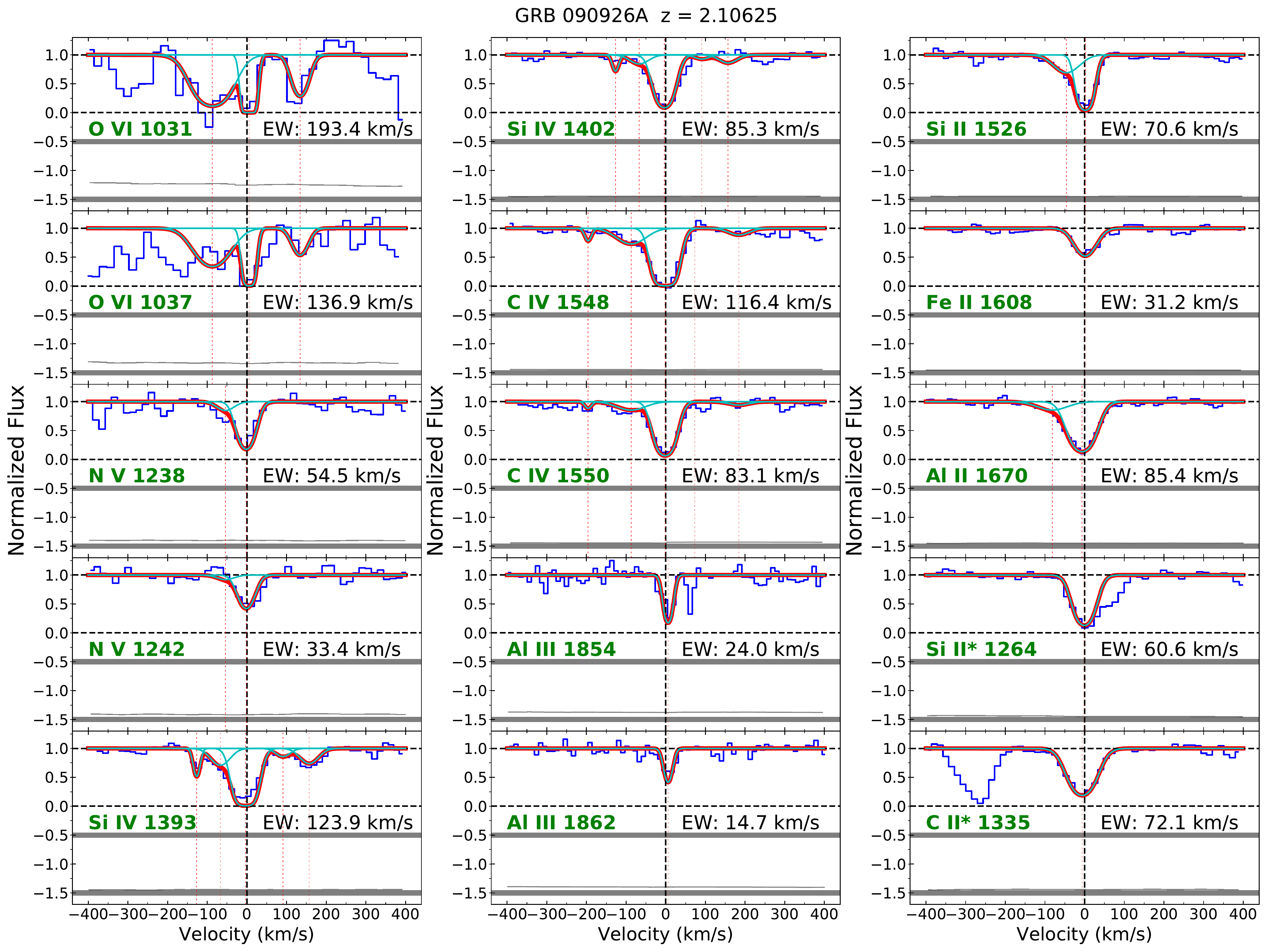}
\figcaption{\label{fig:090926A} Voigt profile fit for GRB 090926A}
\end{figure*}

\begin{figure*}
\centering
\includegraphics[width=\wfit\textwidth]{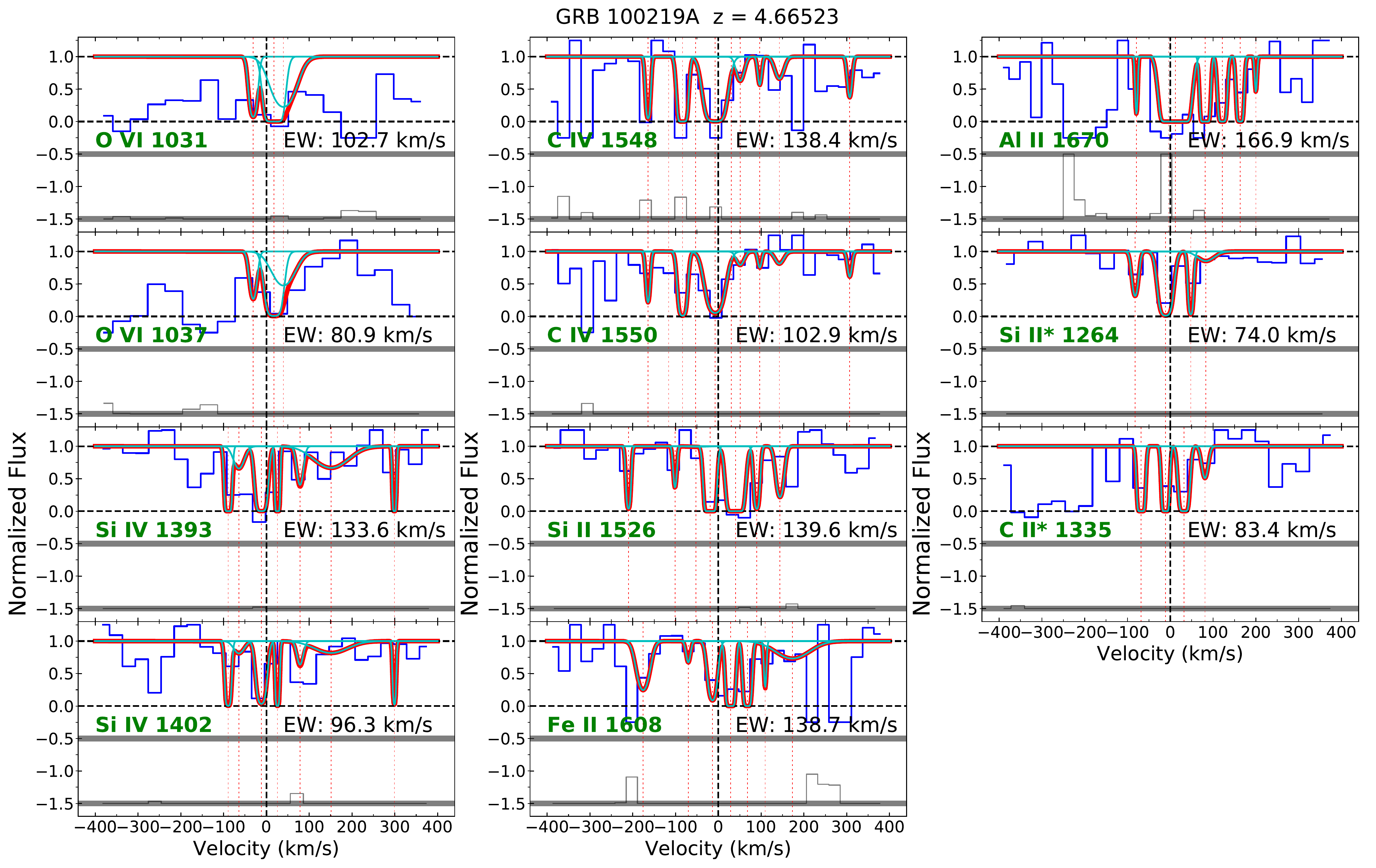}
\figcaption{\label{fig:100219A} Voigt profile fit for GRB 100219A}
\end{figure*}

\begin{figure*}
\centering
\includegraphics[width=\wfit\textwidth]{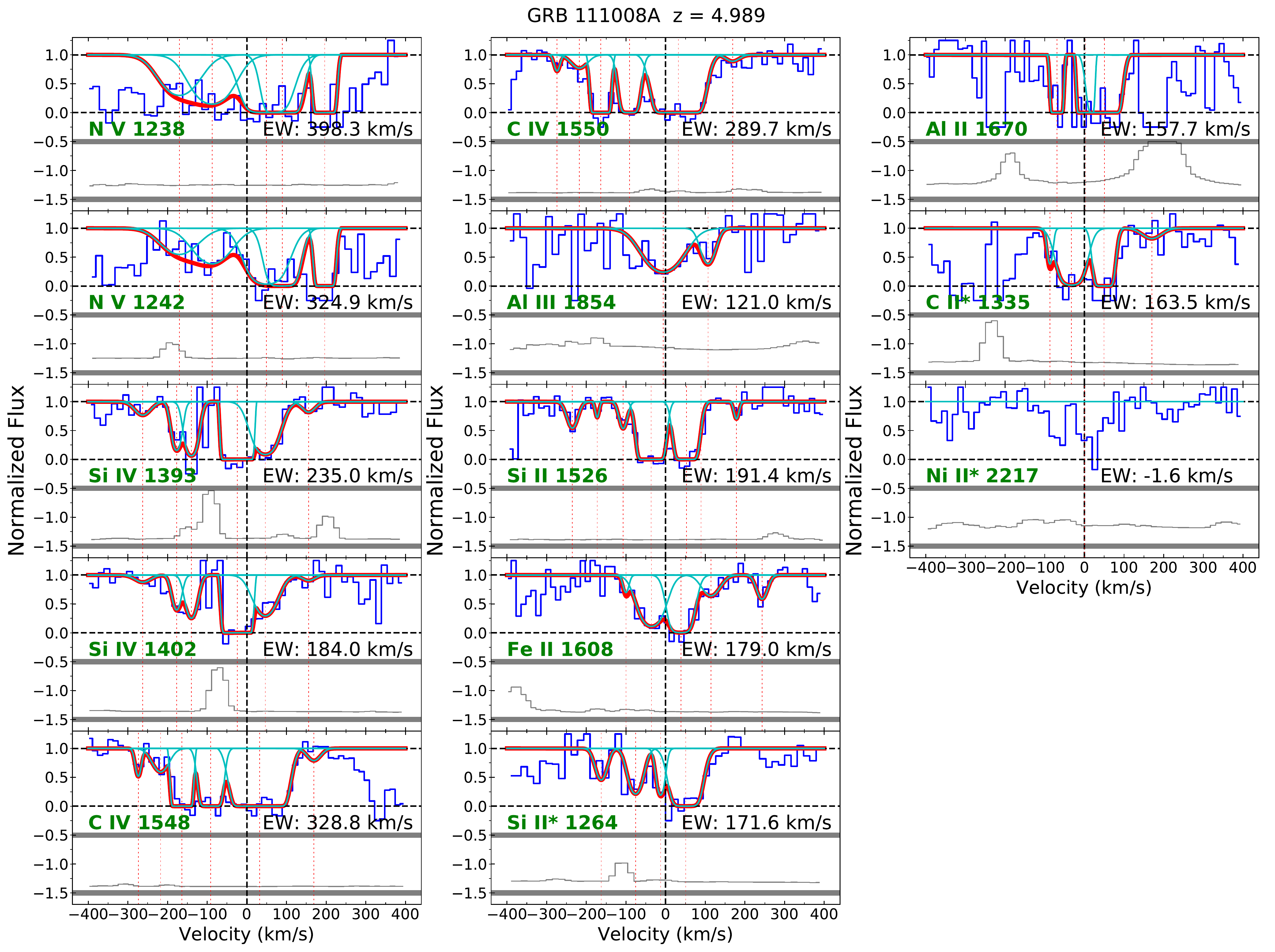}
\figcaption{\label{fig:111008A} Voigt profile fit for GRB 111008A}
\end{figure*}

\begin{figure*}
\centering
\includegraphics[width=\wfit\textwidth]{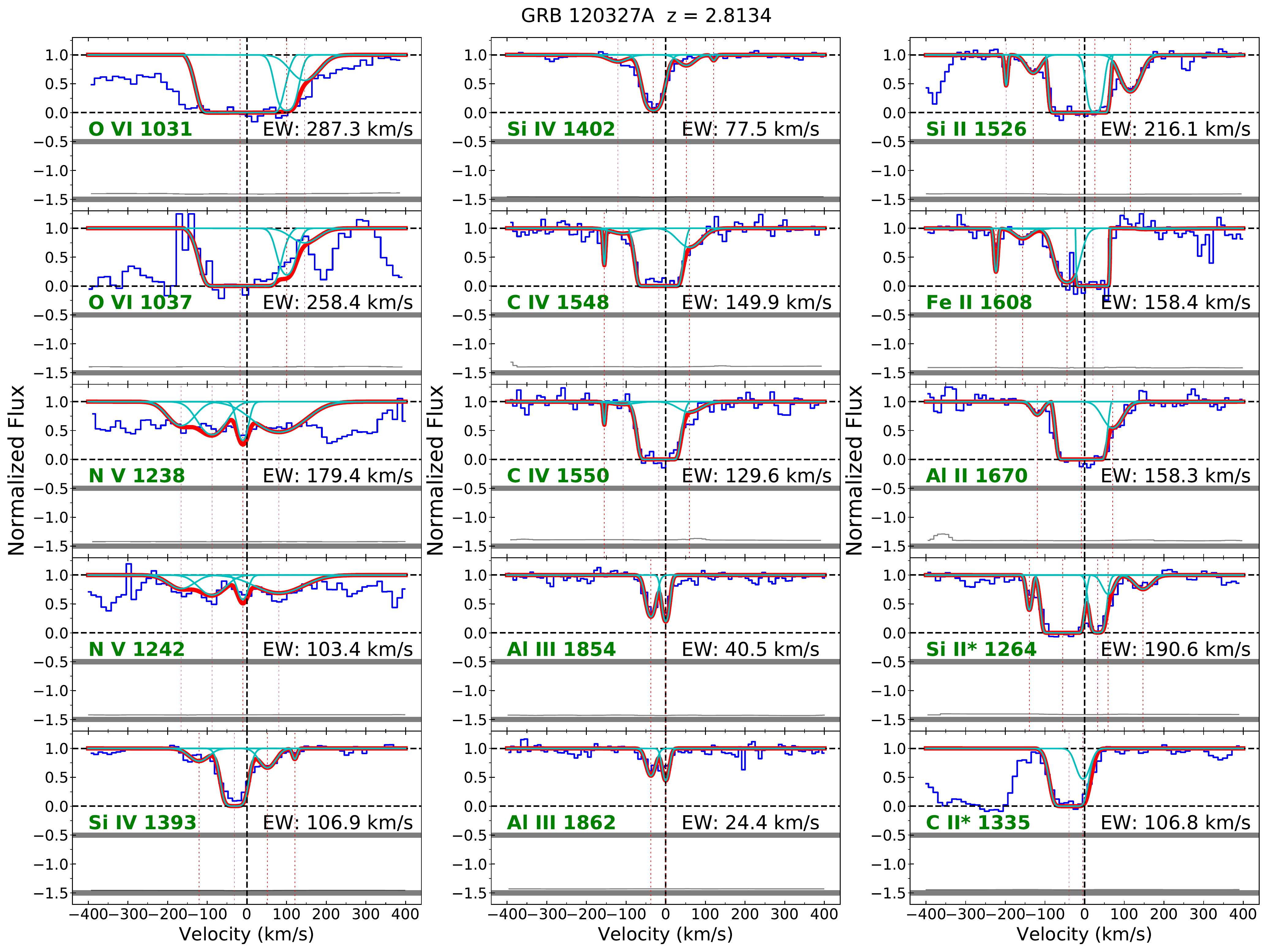}
\figcaption{\label{fig:120815A} Voigt profile fit for GRB 120327A}
\end{figure*}

\begin{figure*}
\centering
\includegraphics[width=\wfit\textwidth]{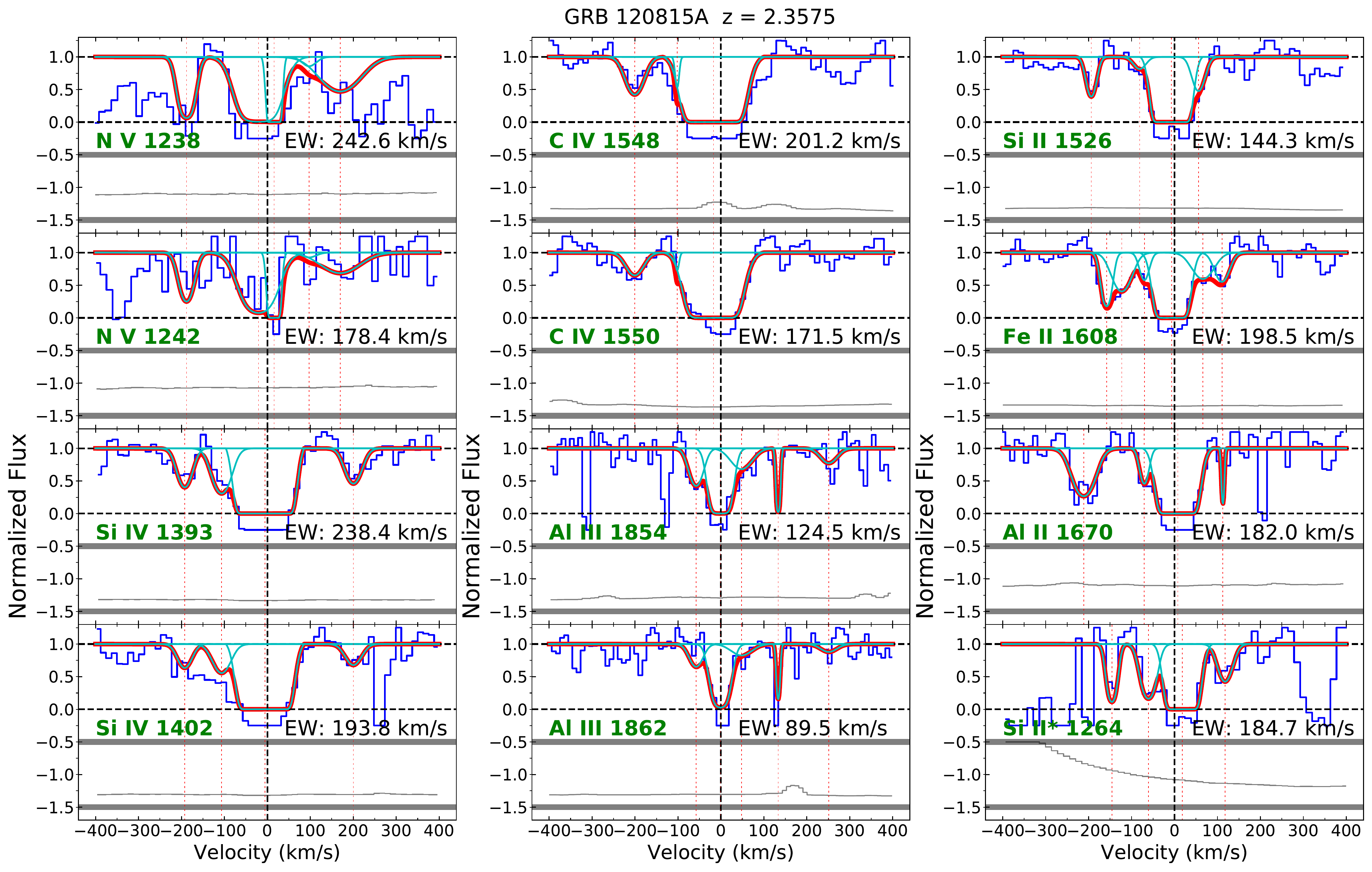}
\figcaption{\label{fig:120815A} Voigt profile fit for GRB 120815A}
\end{figure*}

\begin{figure*}
\centering
\includegraphics[width=\wfit\textwidth]{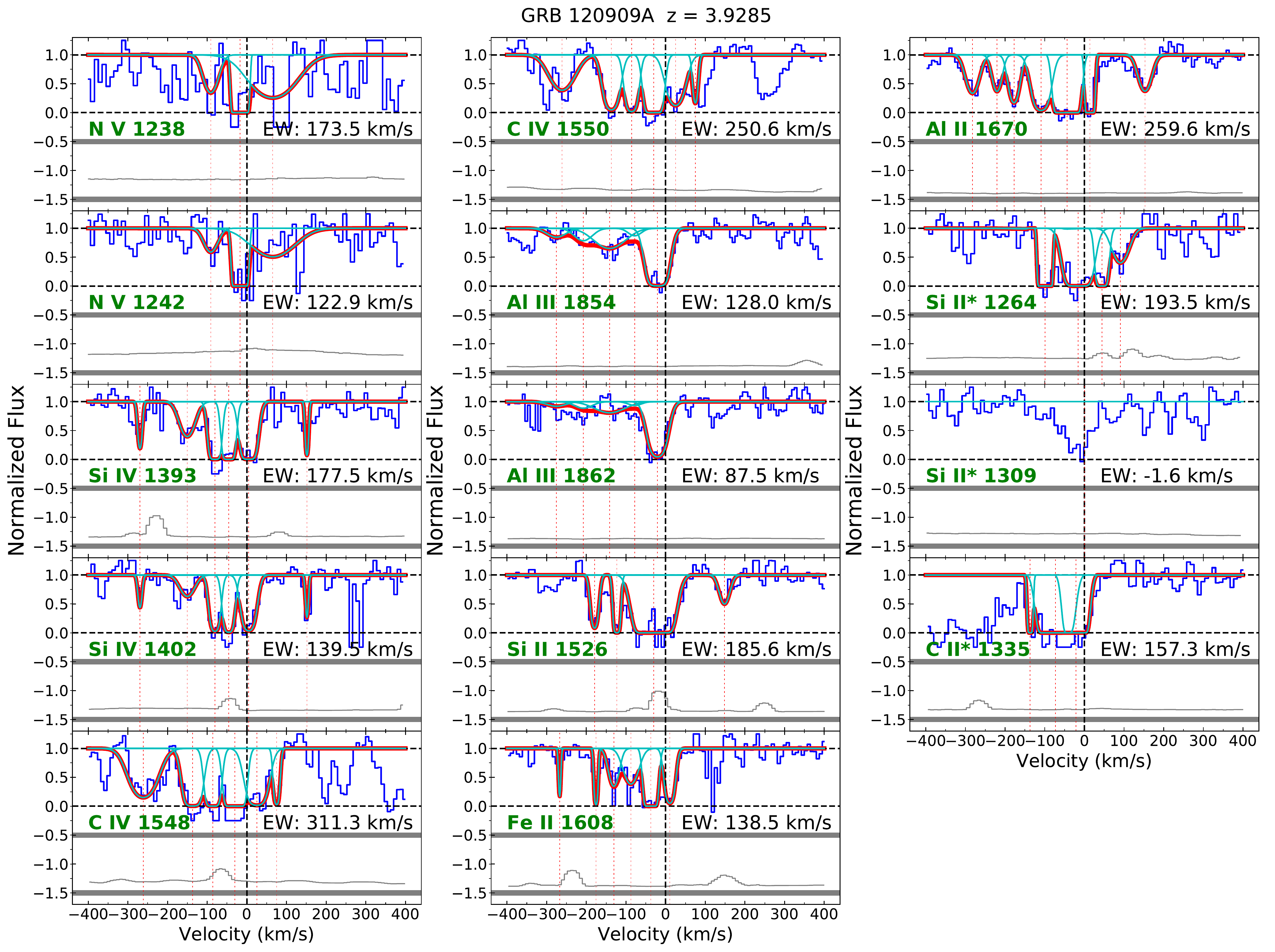}
\figcaption{\label{fig:120909A} Voigt profile fit for GRB 120909A}
\end{figure*}

\begin{figure*}
\centering
\includegraphics[width=\wfit\textwidth]{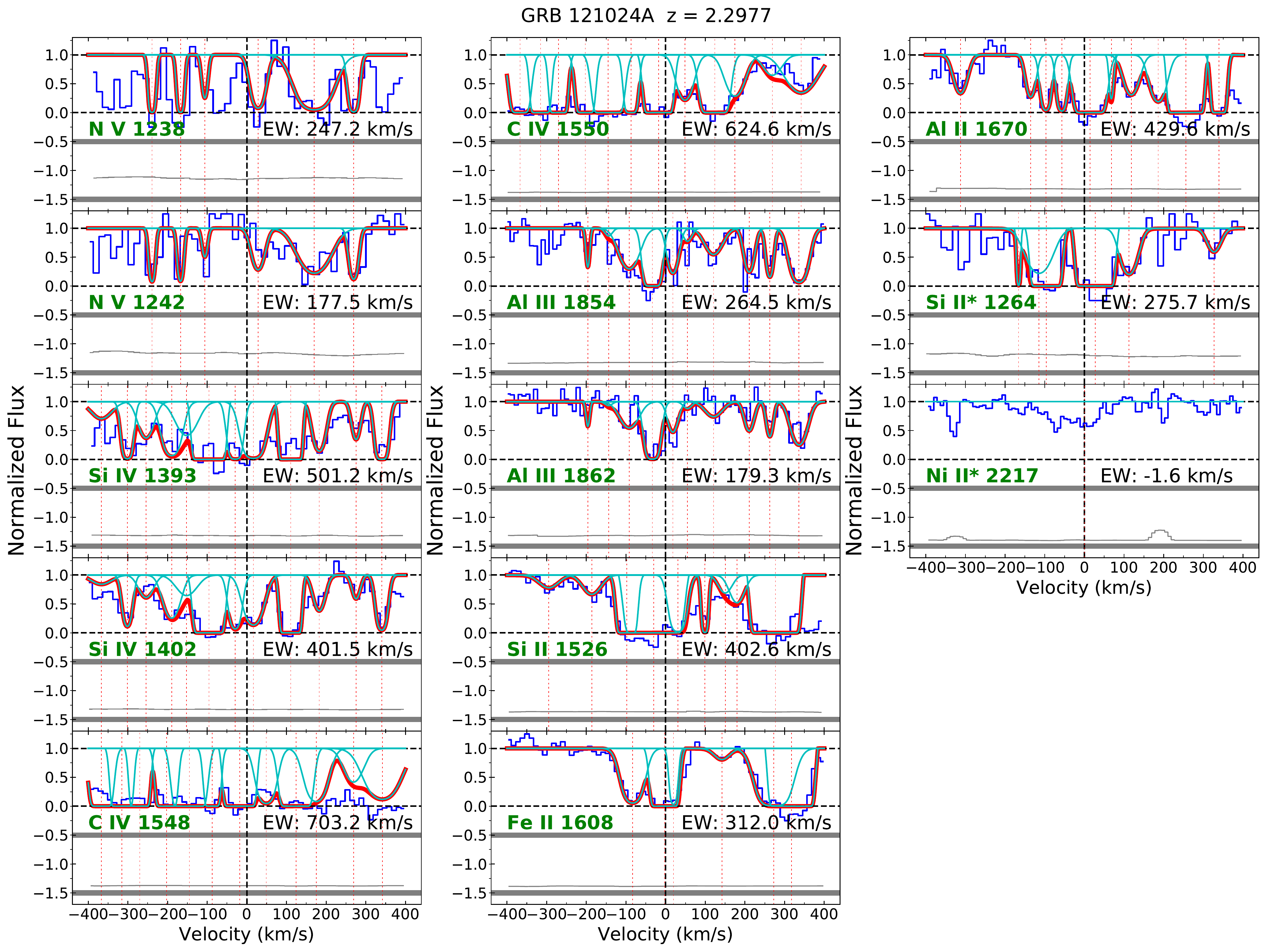}
\figcaption{\label{fig:121024A} Voigt profile fit for GRB 121024A}
\end{figure*}

\begin{figure*}
\centering
\includegraphics[width=\wfit\textwidth]{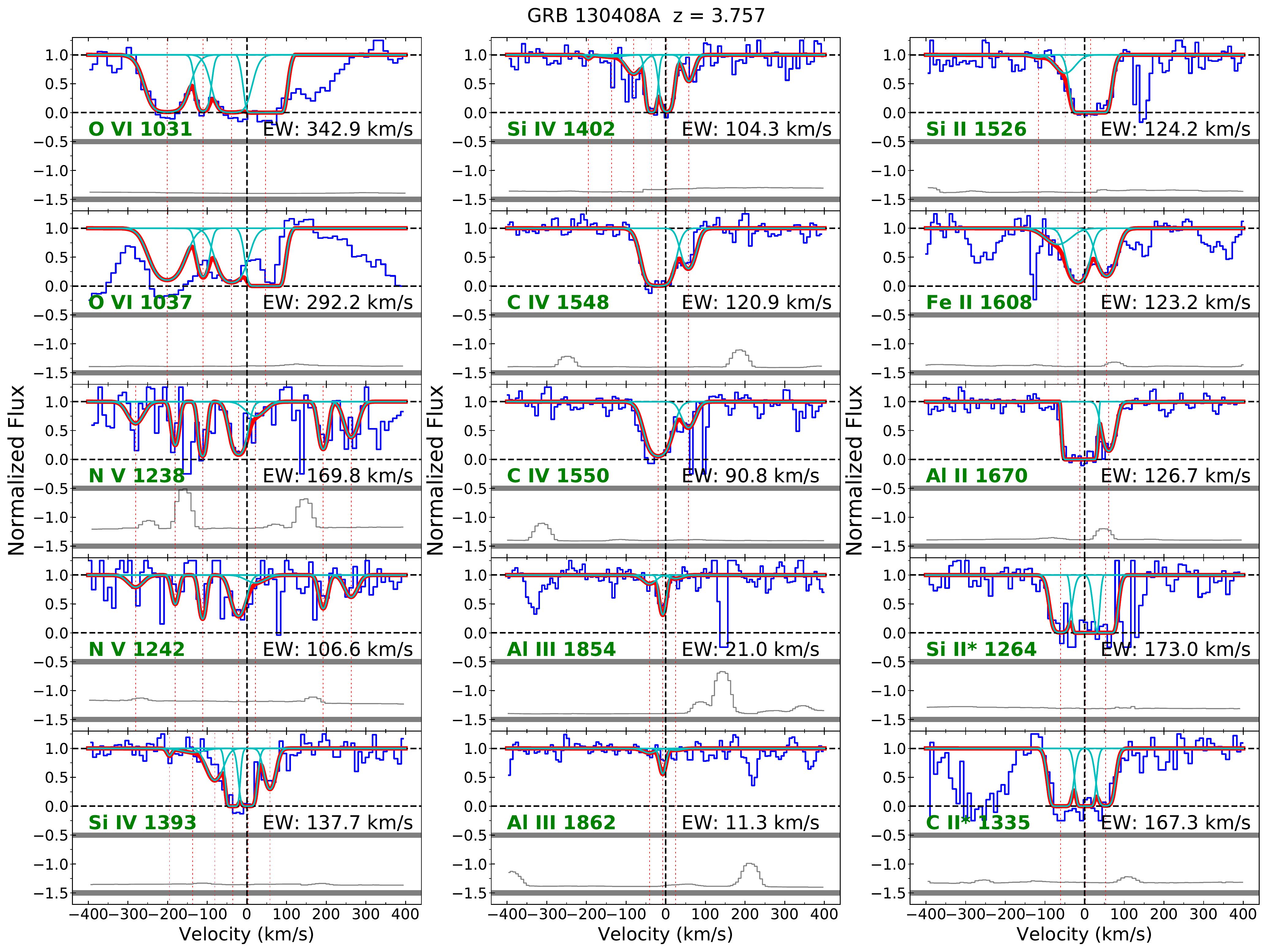}
\figcaption{\label{fig:130408A} Voigt profile fit for GRB 130408A}
\end{figure*}

\begin{figure*}
\centering
\includegraphics[width=\wfit\textwidth]{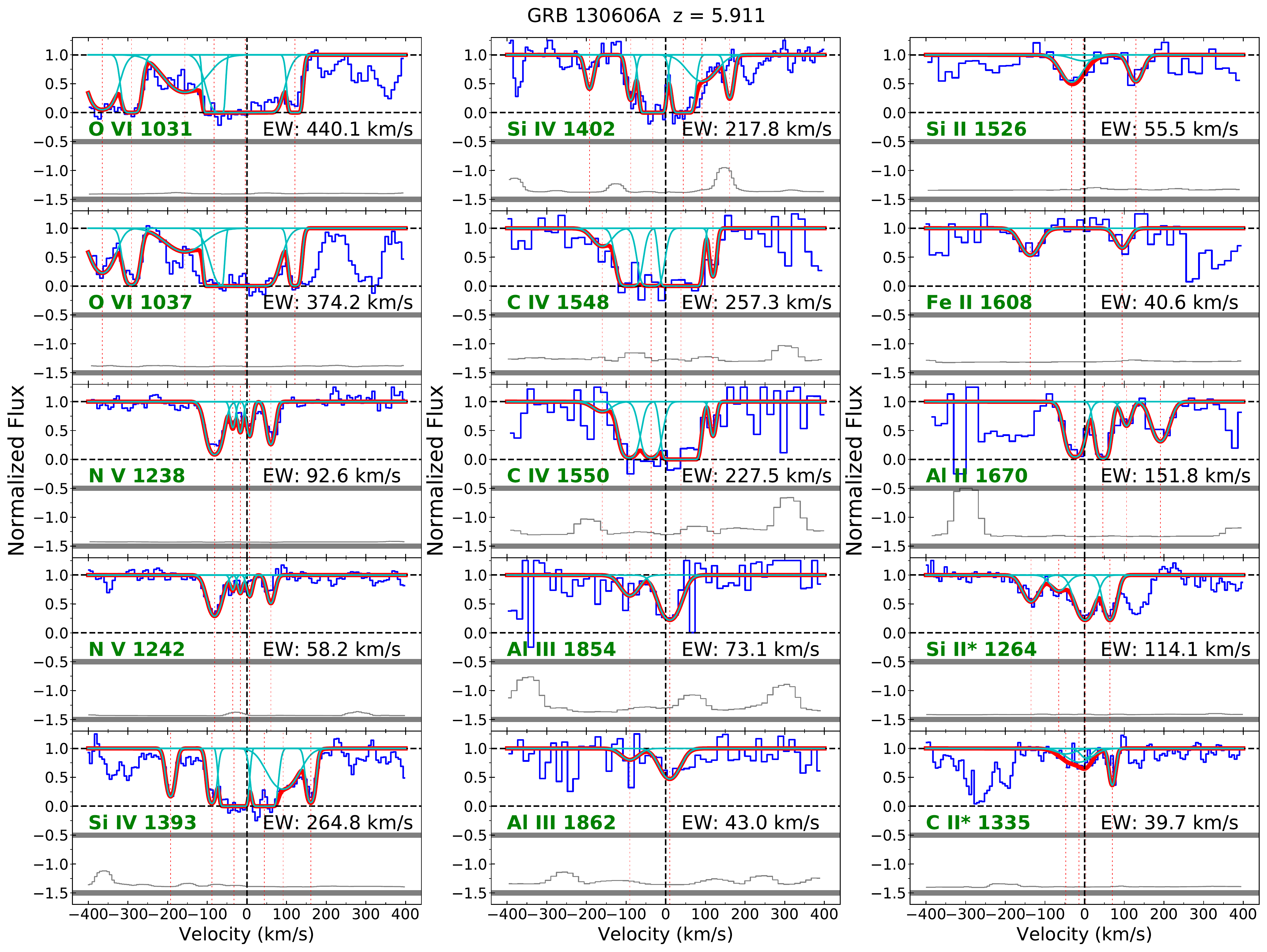}
\figcaption{\label{fig:130606A} Voigt profile fit for GRB 130606A}
\end{figure*}

\begin{figure*}
\centering
\includegraphics[width=\wfit\textwidth]{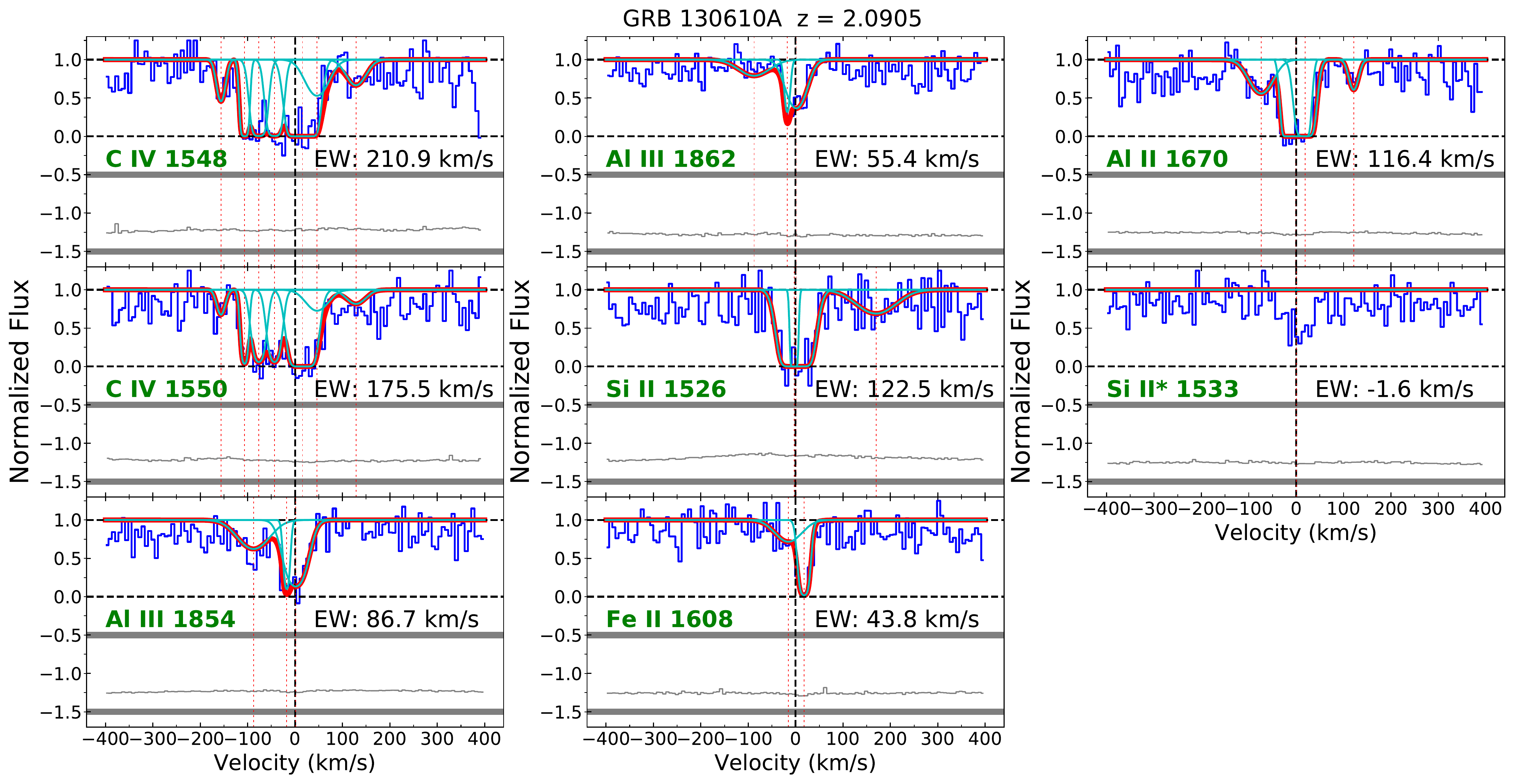}
\figcaption{\label{fig:130610A} Voigt profile fit for GRB 130610A}
\end{figure*}

\begin{figure*}
\centering
\includegraphics[width=\wfit\textwidth]{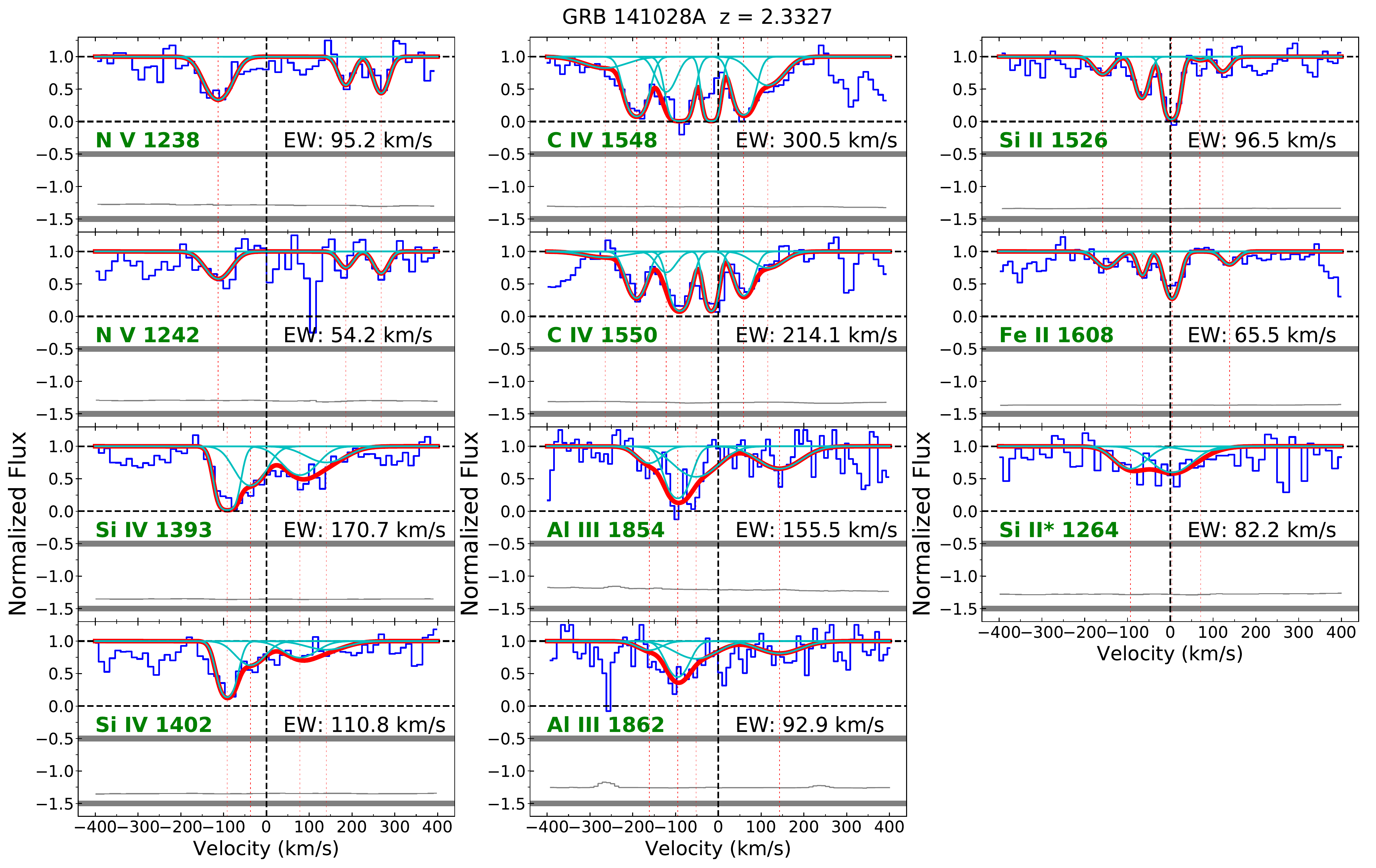}
\figcaption{\label{fig:141028A} Voigt profile fit for GRB 141028A}
\end{figure*}

\begin{figure*}
\centering
\includegraphics[width=\wfit\textwidth]{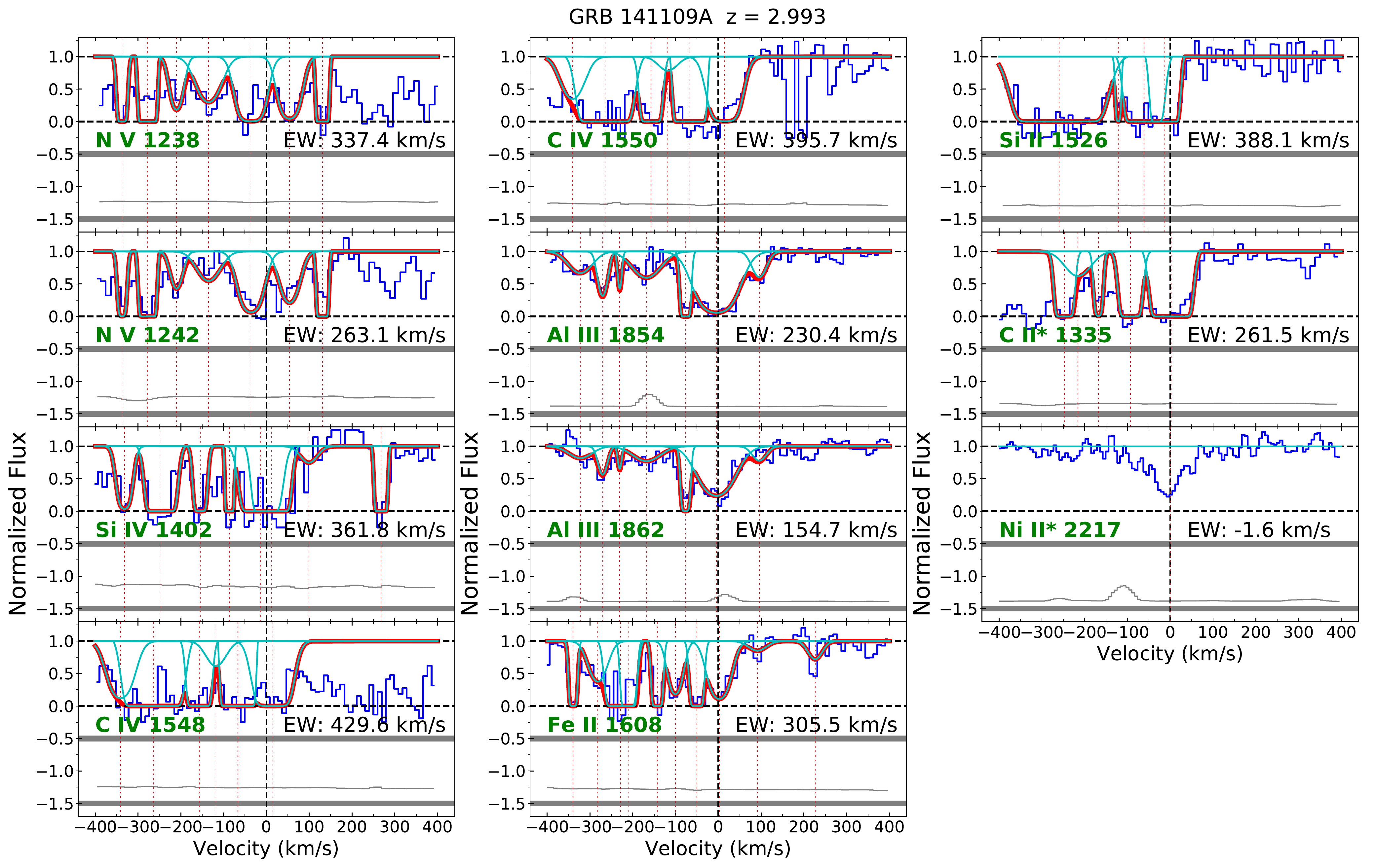}
\figcaption{\label{fig:141109A} Voigt profile fit for GRB 141109A}
\end{figure*}

\begin{figure*}
\centering
\includegraphics[width=\wfit\textwidth]{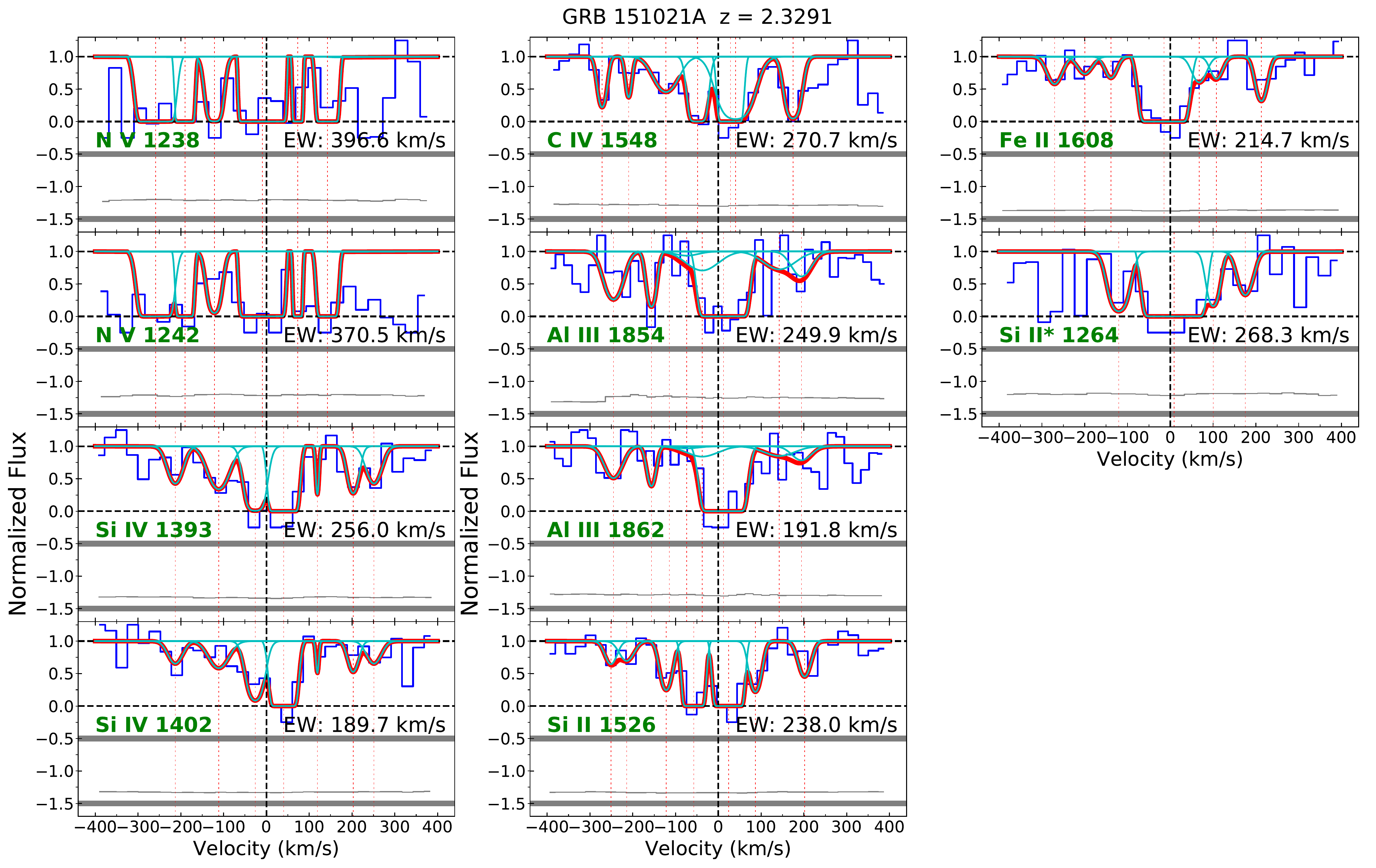}
\figcaption{\label{fig:151021A} Voigt profile fit for GRB 151021A}
\end{figure*}

\begin{figure*}
\centering
\includegraphics[width=\wfit\textwidth]{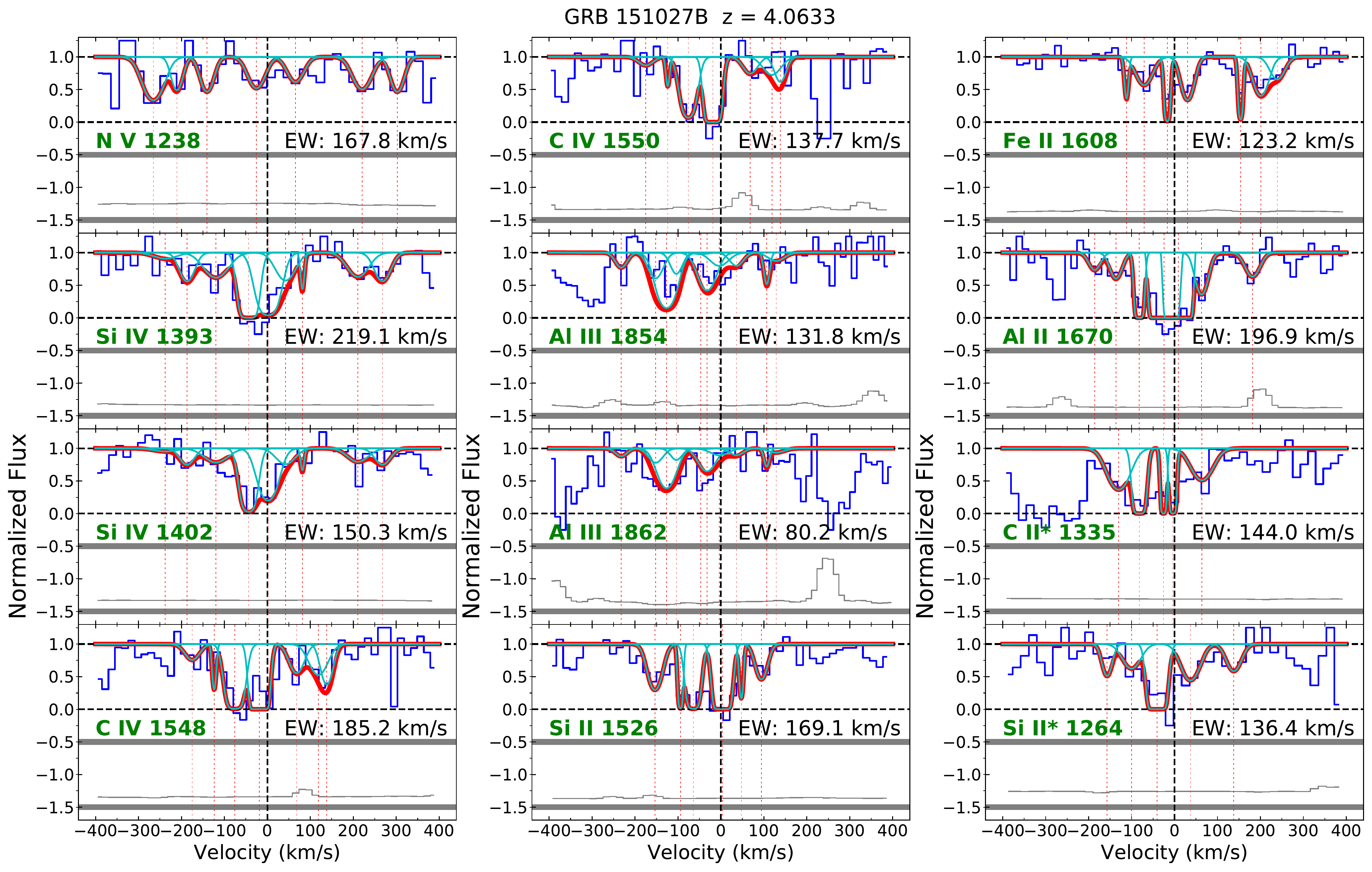}
\figcaption{\label{fig:151027B} Voigt profile fit for GRB 151027B}
\end{figure*}

\begin{figure*}
\centering
\includegraphics[width=\wfit\textwidth]{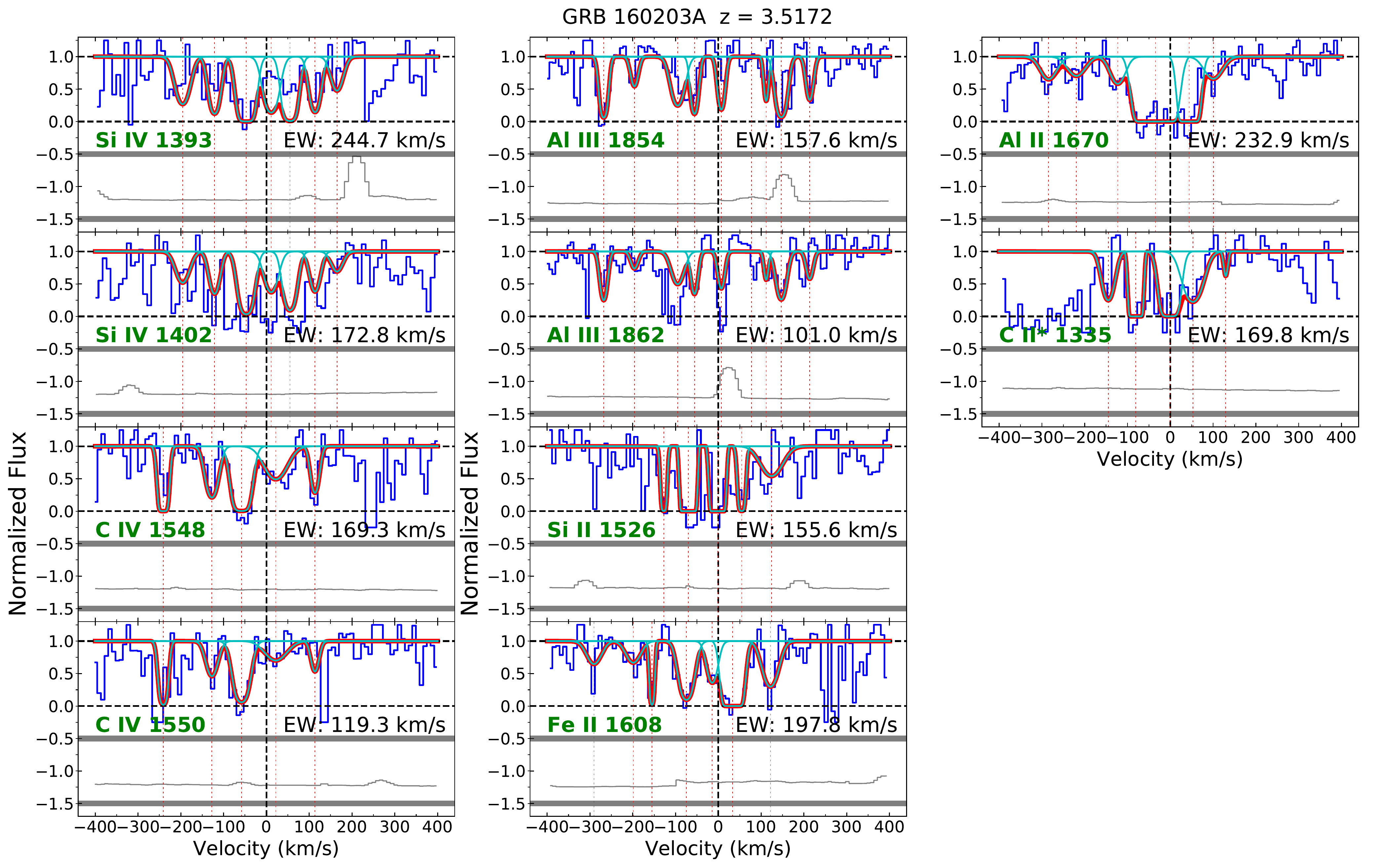}
\figcaption{\label{fig:160203A} Voigt profile fit for GRB 160203A}
\end{figure*}

\begin{figure*}
\centering
\includegraphics[width=\wfit\textwidth]{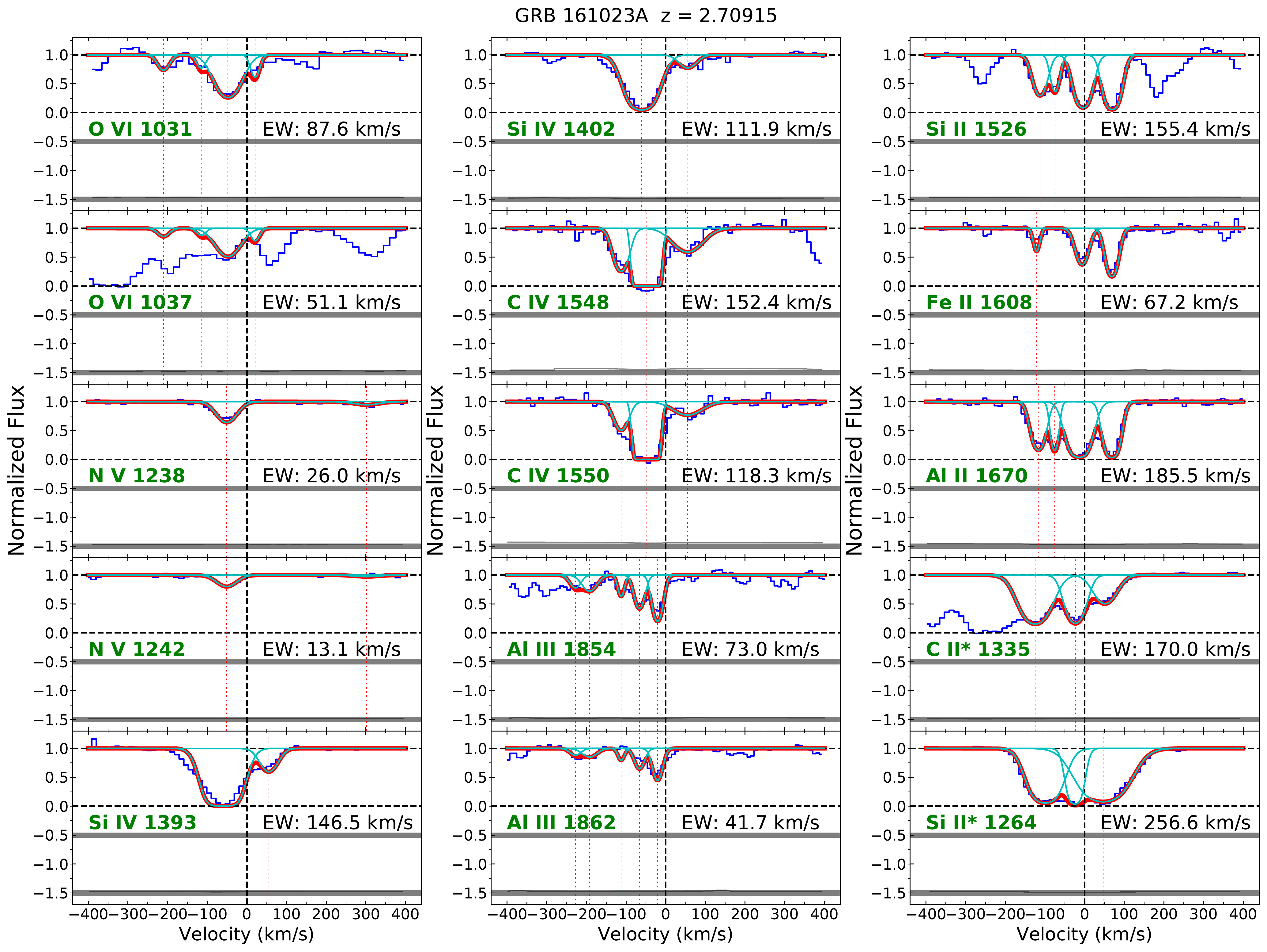}
\figcaption{\label{fig:161023A} Voigt profile fit for GRB 161023A}
\end{figure*}

\begin{figure*}
\centering
\includegraphics[width=\wfit\textwidth]{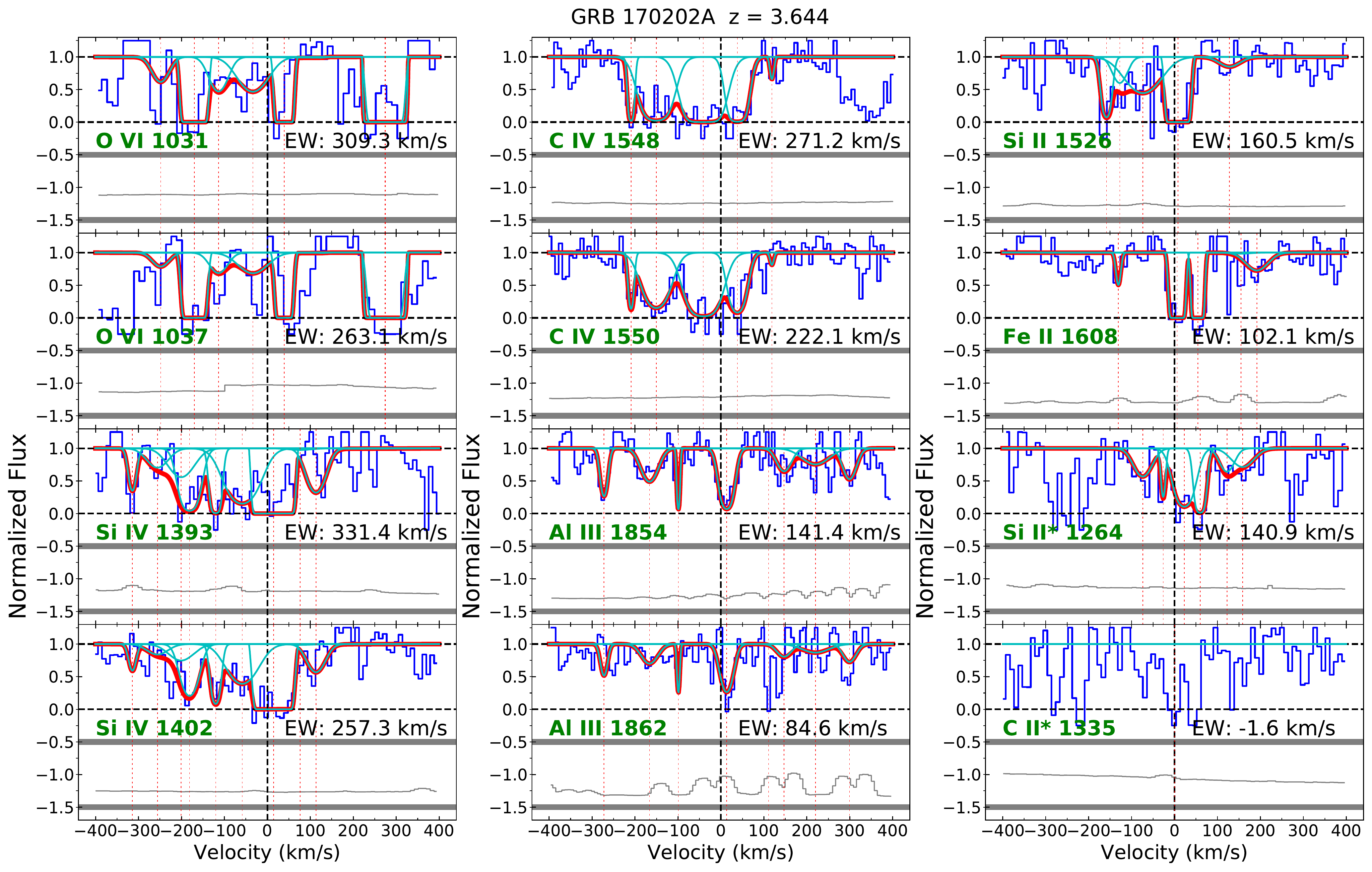}
\figcaption{\label{fig:170202A} Voigt profile fit for GRB 170202A}
\end{figure*}

\end{document}